\documentclass[aps,10pt,prd,amsmath,amsfonts,amssymb,eqsecnum,nofootinbib,
  floatfix,secnumarabic,preprintnumbers,superscriptaddress,nobalancelastpage,
  onecolumn,notitlepage,biblatex]{revtex4-1}
\usepackage{amssymb}
\usepackage{slashed}
\usepackage{makecell}
\usepackage{diagbox}
\usepackage{amsmath}
\usepackage{braket}
\usepackage{float}
\usepackage[utf8]{inputenc}
\usepackage{graphicx}
\usepackage{subfigure}
\usepackage{latexsym}
\usepackage{epic}
\usepackage{eepic}
\usepackage{epsfig}
\usepackage{color}
\usepackage{natbib}
\usepackage[hyperfootnotes=false]{hyperref}
\hypersetup{
 colorlinks=false,
 citecolor=Blue,
 linkcolor=Blue,
 urlcolor=Blue}
\usepackage[normalem]{ulem}

\usepackage{longtable}
\usepackage{pbox}
\usepackage{placeins}
\usepackage[title]{appendix}
\usepackage{pifont}
\usepackage{multirow}
\usepackage{cancel}
\usepackage[normalem]{ulem}

\definecolor{darkred}{rgb}{0.6,0,0}

\newcommand{\AddrAHEP}{Korea Institute for Advanced Study,
	Seoul 02455, Korea}
\newcommand{\AddrHBNI}{
	Homi Bhabha National Institute, BARC Training School Complex, Anushakti Nagar, Mumbai 400094, India }
\newcommand{\AddrIOP}{
	Institute of Physics, Sachivalaya Marg, Bhubaneswar 751005, India}

\newcommand{\Adddur}{
	Institute for Particle Physics Phenomenology, Department of Physics, Durham University\\
	South Road, Durham DH1 3LE, United Kingdom}

\newcommand{\Addpits}{
	Pittsburgh Particle Physics, Astrophysics, and Cosmology Center,
  Department of Physics and Astronomy, \\University of Pittsburgh, Pittsburgh, USA}

\definecolor{linkcolor}{rgb}{0,0,0.5}
\label{key}
\begin{document}

\title{\boldmath Re-examining $N_{R}$-EFT Upto Dimension Six}

\author{Manimala Mitra}\email{manimala@iopb.res.in}
\affiliation{\AddrIOP}
\affiliation{\AddrHBNI}

\author{Sanjoy Mandal}\email{ smandal@kias.re.kr}
\affiliation{\AddrAHEP}

\author{Rojalin Padhan}\email{rojalin.p@iopb.res.in}
\affiliation{\AddrIOP}
\affiliation{\AddrHBNI}
\affiliation{\Addpits}

\author{Agnivo Sarkar}\email{agnivo.sarkar@iopb.res.in}
\affiliation{\AddrIOP}
\affiliation{\AddrHBNI}

\author{Michael Spannowsky}\email{michael.spannowsky@durham.ac.uk}
\affiliation{\Adddur}

\begin{abstract}

The gauge singlet right-handed neutrinos~(RHNs) are essential fields in several neutrino mass models that explain the observed eV scale neutrino mass. We assume RHN field to be present in the vicinity of the electroweak scale and all the other possible beyond the standard model~(BSM) fields arise at high energy scale $\ge\Lambda$. In this scenario, the BSM physics can be described using effective field theory~(EFT) where the set of canonical degrees of freedoms consists of both RHN and SM fields. EFT of this kind is usually dubbed as $N_{R}$-EFT. We systematically construct relevant operators that can arise at dimension five and six while respecting underlying symmetry. To quantify the phenomenological implication of these EFT operators we calculate different couplings that involve RHN fields. We discuss the constraints on these EFT operators coming from different energy and precision frontier experiments. For $pp$, $e^{-}p$ and $e^{+}e^{-}$ colliders, we identify various channels which crucially depends on these operators. We analytically evaluate the decay widths of RHN considering all relevant operators and highlight the differences that arise because of the EFT framework. Based upon the signal cross-section we propose different multi-lepton channels to search for the RHN at 14 TeV LHC as well as \emph{future} particle colliders.
\end{abstract}

\maketitle
\section{Introduction} 
\label{Introduction}
  The tremendous achievement of the Standard Model~(SM) is that it can make precise numerical predictions about the particle dynamics up to the TeV scale. The Higgs boson’s discovery~\cite{ATLAS:2012yve,CMS:2012qbp} at the Large Hadron Collider~(LHC) as well as precision frontier experiments favour the theoretical claims of this model with significant precision. Despite these experimental success, there are many compelling reasons correspond to non-zero neutrino mass, dark matter or the natural explanation behind the electroweak symmetry breaking \emph{etc.} motivate us to construct \emph{Beyond Standard Model} (BSM) theories which can satisfactorily explain these questions. These BSM theories typically contain new degrees of freedom~(\emph{d.o.f}) which interact with the SM particles. Different experimental collaborations have extensively looked for these BSM particles decaying into various SM final states. The results obtained from these searches so far fail to provide any conclusive evidence in support of their existence or their corresponding properties. One of the plausible explanations behind these null results is that these BSM states are situated at a very large energy scale $\Lambda$ and the centre of mass energy of the present day colliders is not sufficient enough to produce them on-shell. However the indirect effects of these particles can be detected while analysing different low-energy observables~\cite{Buchmuller:1985jz}. In view of this, one can consider the effective field theory~(EFT)~\cite{Callan:1969sn,Weinberg:1978kz} approach which can serve as an efficient pathway to parametrise these indirect effects that can help us uncover the nature of BSM.
  
 The construction of any EFT~\cite{Brivio:2017vri,Dobado:1997jx} typically requires two ingredients,  the canonical degrees of freedom~(\emph{d.o.f}) that are present in low energy theory and the symmetries which manifestly dictate the interactions between these fundamental~\emph{d.o.f}. The Lagrangian corresponds to the EFT framework~\cite{Preskill:1990fr} is sum of both the $d = 4$ renormalisable part as well as different higher dimensional operators which are allowed by the symmetry. We assume at the scale $\Lambda$, their exists a gauge theory which contains extra massive~\emph{d.o.f}. At this scale these fields get decouple from the low-energy theory. The effects of these heavy states can be reinstated in forms of a tower of effective operators at each order of mass dimensions $n > 4$. These higher dimensional operators $\{\mathcal{O}_{n}\}$ \footnote{The $n$ stands for the mass dimension of these operators.} are built upon canonical \emph{d.o.f} of low energy theory while respecting space-time as well as the gauge and discrete symmetries. The~\emph{decoupling theorem}~\cite{Appelquist:1974tg,Weinberg:1980wa} guarantees that all measurable observables corresponding to the heavy scale physics are suppressed by inverse powers of cut-off scale $\Lambda$. As a corollary of the decoupling theorem one can establish the hierarchy between the operators that arise at each dimension. As a consequence, the measurable effects of the operators at dimension $n$ in general dominant over the operators arise at dimension $n + 1$. One can optimally use this framework to investigate the physics associated with neutrinos and establish their connection with the SM physics.
  
 The absence of RHNs ($N$) in the SM field content, forbids us to generate neutrino mass similar to other SM fermions. However, various neutrino oscillations experimental measurements~\cite{Super-Kamiokande:1998kpq,Super-Kamiokande:2001bfk,Toshito:2001dk,SNO:2002tuh,deSalas:2020pgw} strongly suggest non zero masses for neutrinos thus encourages us to modify the existing SM. The simplest way to encounter this issue is to add RHNs to the SM particle contents and write down a Yukawa interaction for neutrinos similar to other SM charged fermions. As these RHN fields are charge neutral and singlet under the SM gauge group $SU(3)_{c}\times SU(2)_{L}\times U(1)_{Y}$, one can include a Lepton-number violating Majorana type mass term $M_{N}\overline{N}^{c}_{R}N_{R}$ in the Lagrangian in addition to the previously mentioned Yukawa interaction. The smallness of the neutrino mass can therefore be explained as the hierarchy between the electroweak scale $v$ and the RHN mass scale $M_{N}$ which can be expressed as $ M_{\nu} \sim \frac{y^{2}_{\nu}v^{2}}{M_{N}}$. 
 Here, $y_{\nu}$ stands for Yukawa coupling correspond to neutrinos. If we assume the value of $y_{\nu}$ to be $\mathcal{O}(1)$, one can see that the requirement for tiny neutrino mass set the value of $M_{N}$ in the vicinity of Grand Unification regime (roughly around $10^{14} \sim 10^{15}$ GeV). This simplistic set up for neutrino mass is in general known as \emph{Type-I Seesaw} mechanism~\cite{Gell-Mann:1979vob,Minkowski:1977sc,Mohapatra:1979ia,Yanagida:1979as}. The interaction strength between these heavy neutrinos and the SM particles is controlled by the active sterile mixing parameter $\theta$ which is defined as $\theta \propto \frac{y_{\nu}v}{M_{N}}$.
 The above relation implies a small value of $\theta$ and leads to a small production cross-section for the RHNs at different collider experiments.
 
 The major disadvantage of the Type-I set up is that the physics associated with the RHN fields become relevant at around GUT scale which the current experimental facilities fail to probe. One can alter this situation while assuming that at least one of these RHN fields is within the regime of electroweak scale~\cite{Pilaftsis:1991ug,Borzumati:2000mc,Kersten:2007vk} while satisfying the existing experimental constraints. In this context one can describe the dynamics involving RHN using EFT. The EFT of this kind is denoted as $N_{R}$-EFT. 

 
 There are many works which encompass different aspects of $N_{R}$-EFT. The Ref.~\cite{delAguila:2008ir,PhysRevD.80.013010,Bhattacharya:2015vja,Liao:2016qyd} and Ref.~\cite{Li:2021tsq} presents the non-redundant operator basis upto dimension seven and dimension nine of $N_{R}$-EFT respectively. The Ref.~\cite{Barducci:2020icf,Caputo:2017pit,Delgado:2022fea,Jones-Perez:2019plk} discuss the collider phenomenology of the dimension five $N_{R}$-EFT at \emph{future Higgs} factories as well as LHC. Other studies~\cite{delAguila:2008ir,Alcaide:2019pnf,DeVries:2020jbs,Beltran:2022ast,Duarte:2016miz} also looked into various subset of these higher dimensional operators and presented their phenomenological implication at LHC. If the total decay width of the light RHN is small then it can give rise to interesting displaced decay signatures and detailed study regarding this can be found in Ref.~\cite{Drewes:2019fou,Cottin:2021lrq,Liu:2019ayx,Abada:2018sfh}. The Ref.~\cite{Barducci:2022hll,Cottin:2021lzz,Beltran:2021hpq} focused on the interesting production modes which is invoked by the different four fermi operators that one construct at dimension six. The study assume relevant decay modes for the $N$ field to be $N \to \nu \gamma$ and $N \to 3f$ (where $f$ is SM fermions).  The Ref.~\cite{Butterworth:2019iff} discuss the theoretical aspects of the dimension 6 operators that involve the Higgs doublet and discuss their sensitivity under various Higgs mediated processes. In addition to that, Ref.~\cite{Duarte:2018xst,Zapata:2022qwo} study the sensitivity of different dimension six operators at LHeC and lepton colliders.
 
 In this work we present the complete phenomenological description of the $N_{R}$-EFT upto dimension six. In section~\ref{Sec:Model} we begin with the general set up and systemically construct different dimension five (see sub-section \ref{Sec:dim5}) as well as dimension six (see sub-section \ref{Sec:dim6}) operators along with highlighting their physics aspects. In section~\ref{Sec:bound}, we evaluate the constraints on different operators coming from precision frontier as well as direct search experiments. In section~\ref{prodxs}, we calculate the cross section for RHN production at $pp$, $e^{-}p$ and $e^{+}e^{-}$ colliders. Depending on the RHN mass, the $N$ field can decay either to two body or to three body decay modes respectively. In section~\ref{Sec:NRdecay}, we present the detailed analytic calculations correspond to each of these decay modes and evaluate the branching ratios for different benchmark scenarios. We also present expected number of signal events with multi-lepton final state for above mentioned colliders in section.~\ref{Sec:multilep}. We summarise our findings along with few concluding remarks in section~\ref{Sec:conc}.   
\section{General Set Up}
\label{Sec:Model}
We begin with a \emph{phenomenological} Lagrangian which can be expressed as 
\begin{equation}
\mathcal{L} \equiv \mathcal{L}_{\text{SM}} + \bar{N}_{R}\cancel{\partial}N_{R} - \bar{L}_{\ell}Y_{\nu}\tilde{H}N_{R} - \frac{1}{2}\tilde{M}_{N}\bar{N}^{C}_{R}N_{R} + \sum_{n > 4}\frac{\mathcal{O}_{n}}{\Lambda^{n - 4}} + h.c.
\label{Eq:Model}
\end{equation}
\noindent
where $\tilde{H} = i\sigma^{2}H^{*}$, $N^{C}_{R}$ = $C\bar{N}^{T}_{R}$ with charge conjugation matrix $C = i\gamma^{2}\gamma^{0}$. The term $\tilde{M}_{N}$ stands for the Majorana bare mass term which is a $\mathcal{N}\times\mathcal{N}$ matrix in the flavour space. $L_{\ell}$ is the SM lepton doublet and $Y_\nu$ is the Dirac-type Yukawa coupling. The terms $\bar{L}_{\ell}Y_{\nu}\tilde{H}N_{R}$ and $\frac{1}{2}\tilde{M}_{N}\bar{N}^{C}_{R}N_{R}$ contributes to the neutrino mass matrix. The $\mathcal{O}_{n}$ are the higher dimensional operators, which one can build at each dimension. The effect of these operators are suppressed by cut-off scale $\Lambda$ with appropriate power.  
\subsection{$N_{R}$-EFT Operators at Dimension Five} 
\label{Sec:dim5}
With this general set-up in mind, one can write down three possible $N_{R}$-EFT operators at dimension five. In Table.\ref{Tab:Ope5}, we present the explicit form of these operators where $\alpha^{(5)}_{i}$ ($ i =$ 1 to 3 ) represent the Wilson coefficients correspond to each of these operators. 
\begin{table}[h!]
\centering
\begin{tabular}{|c||c|}
\hline
~~~~~$\mathcal{O}^{\left(5\right)}_{1}$~~~~~ & ~~~~~ $\frac{\alpha^{(5)}_{1}}{\Lambda}\left(\overline{L}^{c}\tilde{H}^{\dagger}\tilde{H}L\right) $ ~~~~~ \\
\hline
~~~~~$\mathcal{O}^{\left(5\right)}_{2}$~~~~~ & ~~~~~ $\frac{\alpha^{(5)}_{2}}{\Lambda}\left(\overline{N_{R}}^{c}N_{R}\right)\left(H^{\dagger}H\right)$ ~~~~~ \\
\hline 
~~~~~$\mathcal{O}^{\left(5\right)}_{3}$~~~~~ & ~~~~~ $\frac{\alpha^{(5)}_{3}}{\Lambda}\left(\overline{N_{R}}^{c}\sigma_{\mu\nu}N_{R}\right)B_{\mu\nu}$~~~~~ \\
\hline
\end{tabular}
\caption{All Possible $N_{R}$-EFT operators that appear at dimension five. The $\sigma^{\mu\nu}$ is defined as, $\sigma^{\mu\nu} = \frac{i}{2}[\gamma^{\mu}, \gamma^{\nu}]$ and $B_{\mu\nu}$ is the field strength tensor corresponds to $U(1)_{Y}$ gauge group. $\Lambda$ is the cut-off scale of underlying $N_{R}$-EFT.}
\label{Tab:Ope5}
\end{table}
\noindent
Considering the space time transformation rules, one can realise that the $\alpha^{(5)}_{1}$ and $\alpha^{(5)}_{2}$ are symmetric matrices in flavour space. In contrast to that, $\alpha^{(5)}_{3}$ is an antisymmetric matrix which arises if we only consider more than one $N_{R}$ fields. The $\mathcal{O}^{\left(5\right)}_{1}$ which is famously known as the \emph{Weinberg} operator~ \cite{Weinberg:1979sa} primarily contributes to active neutrino masses. This is the only operator one can construct in this dimension solely using SM fields. The renormalisable realisation of this operator can be found in Ref.~\cite{Ma:1998dn,Gu:2006dc,CentellesChulia:2018gwr} and its phenomenological implications have been studied in Ref.~\cite{Bonnet:2012kz,Fuks:2020zbm}. On the other hand, operator $\mathcal{O}^{\left(5\right)}_{2}$ provides additional contributions to the Majorana mass term which is mentioned in Eq.~\ref{Eq:MNu5}. However the operator $\mathcal{O}^{\left(5\right)}_{3}$ does not play any role in the neutrino mass matrix but the presence of $B_{\mu\nu}$ in that term brings out non trivial vertices between neutrinos and SM neutral vector boson fields. Assuming the full theory is a gauge theory one may predict that out of these three operators, $\mathcal{O}^{\left(5\right)}_{1}$ and $\mathcal{O}^{\left(5\right)}_{2}$ may be generated in tree level but the $\mathcal{O}^{\left(5\right)}_{3}$ would only appear via loop mediated processes. As a consequence, one can estimate a further $\frac{1}{16\pi^{2}}$ suppression to the $\alpha^{(5)}_{3}$ coefficient ~\cite{Arzt:1994gp}. For a detailed discussion on this aspect the interested reader may follow Ref.~\cite{Aparici:2009fh}. \\
\begin{itemize}  
\item \textbf{Neutrino Mass In Dimension Five}
\end{itemize}        
 We will now define the neutrino mass matrix while considering all the relevant terms upto dimension five. In the basis $\{\nu_{L}, N^{c}_{R}\}$, the neutrino mass matrix will take the following form   
\begin{equation}
\mathcal{M}^{(5)}_{\nu N} = 
\begin{bmatrix}
\frac{\alpha^{(5)}_{1}v^{2}}{\Lambda} & \frac{Y_{\nu}v}{\sqrt{2}} \\
\frac{Y^{T}_{\nu}v}{\sqrt{2}} & \left(\tilde{M}_{N} + \frac{\alpha^{(5)}_{2}v^{2}}{\Lambda}\right)
\end{bmatrix}
\label{Eq:MNu5}
\end{equation}
\noindent
In the seesaw approximation~(when $\nu-N$ blocks are smaller than the ones in the $N-N$ one), this leads to the following light and heavy neutrino mass matrix
\begin{align}
m^{(5)}_{\text{light}} & \approx \frac{\alpha^{(5)}_{1}v^{2}}{\Lambda} - \frac{Y^{T}_{\nu}M^{-1}_{N}v^{2}Y_{\nu}}{2},  \\
m^{(5)}_{\text{heavy}} & \approx M_{N} = \tilde{M}_{N} + \frac{\alpha^{(5)}_{2}v^{2}}{\Lambda}. 
\label{Eq:EigenM5}
\end{align}
\noindent
The mass matrix in Eq.~\ref{Eq:MNu5} can be diagonalized by a unitary matrix as
\begin{align}
V^T \mathcal{M}^{(5)}_{\nu N} V = (\mathcal{M}^{(5)}_{\nu N})^{\text{diag}}.
\end{align}
Following the standard procedure of two step diagonalization $V$ can be expressed as
\begin{align}
V=\mathcal{U}W\,\,\, \text{with}\,\,\, \mathcal{U}^T \mathcal{M}^{(5)}_{\nu N} \mathcal{U}=
\begin{pmatrix}
m^{(5)}_{\text{light}} &  0 \\
0  &   m^{(5)}_{\text{heavy}}
\end{pmatrix}
\end{align}
Hence, $\mathcal{U}$ is the matrix which brings the neutrino mass matrix in the block diagonalized form and further $W=\text{Diag}(U_{\rm PMNS},\kappa)$ diagonalizes the mass matrices in the light and heavy sector. One can approximately write the matrix $V$ as follows
\begin{align}
V=\mathcal{U} W\approx \begin{pmatrix}
1+\mathcal{O}(M_{N}^{-2}) & \theta \\
-\theta^T & 1 + \mathcal{O}(M_{N}^{-2})
\end{pmatrix} \begin{pmatrix}
U_{\text{PMNS}} &  0 \\
0   &   \kappa
\end{pmatrix} \approx \begin{pmatrix}
U_{\text{PMNS}} & \theta \\
- \theta^{T} & \kappa
\end{pmatrix},
\label{Eq:Vdim5}
\end{align}
where $\theta = M^{-1}_{N}\frac{Y_{\nu}v}{\sqrt{2}}$ is the mixing angle between the active and sterile neutrinos, $U_{\text{PMNS}}$ is the PMNS matrix and $\kappa$ is $\mathcal{O}(1)$~(For details see Ref.~\cite{Hashida:1999wh}). Following is the mixing relations between the gauge and mass eigenstates
\begin{align}
\nu_{L} & \simeq U_{\text{PMNS}}\nu_{L, m} + \theta N^{c}_{R, m}, \label{Eq:EigenVV5}\\
N^{c}_{R} & \simeq - \theta^{T}\nu_{L, m} + \kappa N^{c}_{R, m}, \nonumber
\label{Eq:EigenV5}
\end{align}
\noindent
where the subscript ``m" signifies the mass eigenstate.
\begin{itemize}
\item \textbf{Interesting Facets of the Dimension Five Operators}
\end{itemize}
In Eq.~\ref{Eq:EigenVV5}, we show the relation between flavour and mass eigenstates between light (active) and heavy (sterile) neutrinos. In the subsequent discussion, we denote the Majorana mass eigenstate of RHN fields as $N=N_{R,m}+N_{R,m}^c$, while we use similar notation for light neutrino mass basis, $\nu=\nu_{L,m}+\nu_{L,m}^c$. With these definitions we now present various three point vertices that involve neutrino fields which are coming from renormalizable Lagrangian and dimension five operators. The details of the calculations have been included in Appendix~\ref{App:Dim5Couplings}. In Table.\ref{Tab:vertex5}, we illustrate the explicit form of all these couplings. One can notice that the coupling between the $W_{\mu}$ boson and neutrinos does not get any additional contributions from the dimension five operators. However, the situation alters in case of Higgs as well as neutral gauge boson operators.   
\begin{table}[h!]
\centering
\begin{tabular}{|c||c|c|}
\hline
Couplings & Explicit Form & Operator  \\
\hline
$\mathcal{C}^{W_{\mu}}_{\ell\nu}$ & $\frac{g\gamma_{\mu}U}{\sqrt{2}}P_{L}$ + h.c. &~~ RT ~~  \\
\hline
$\mathcal{C}^{W_{\mu}}_{\ell N}$ & $\frac{g\gamma_{\mu}\theta}{\sqrt{2}}P_{L}$ + h.c. &~~ RT ~~  \\
\hline
$\mathcal{C}^{h}_{\overline{\nu}\nu}$ &~~~ $\frac{Y_{\nu}}{\sqrt{2}}U^{\dagger}\theta^{\dagger}P_{R} + \frac{\alpha^{(5)}_{1}v}{\Lambda}U^{T}UP_{L} + \frac{\alpha^{(5)}_{2}v}{\Lambda}\theta^{*}\theta^{\dagger}P_{R}$ + h.c. ~~~ & ~~RT, $\mathcal{O}^{\left(5\right)}_{1}$, $\mathcal{O}^{\left(5\right)}_{2}$~~ \\
\hline 
$\mathcal{C}^{h}_{\overline{N}N}$ &~~~ $- \frac{Y_{\nu}}{\sqrt{2}}\theta^{\dagger}\kappa^{*}P_{R} + \frac{\alpha^{(5)}_{1}v}{\Lambda}\theta^{\dagger}\theta P_{L} + \frac{\alpha^{(5)}_{2}v}{\Lambda}\kappa^{\dagger}\kappa^{*}P_{R}$ + h.c. ~~~ & ~~RT, $\mathcal{O}^{\left(5\right)}_{1}$, $\mathcal{O}^{\left(5\right)}_{2}$~~  \\
\hline
$\mathcal{C}^{h}_{\overline{\nu}N + \overline{N}\nu}$ & ~~~ $\{- \frac{Y_{\nu}}{\sqrt{2}}U^{\dagger}\kappa^{*}P_{R} + \frac{\alpha^{(5)}_{1}v}{\Lambda}U^{\dagger}\theta P_{L} - \frac{\alpha^{(5)}_{2}v}{\Lambda}\theta^{*}\kappa^{*}P_{R}\}$ &   \\
										& ~ + $\{\frac{Y_{\nu}}{\sqrt{2}}\theta^{\dagger}\theta^{\dagger}P_{R} + \frac{\alpha^{(5)}_{1}v}{\Lambda}\theta^{\dagger}U P_{L} - \frac{\alpha^{(5)}_{2}v}{\Lambda}\kappa^{\dagger}\theta^{\dagger}P_{R}\}$ + h.c. & ~~RT, $\mathcal{O}^{\left(5\right)}_{1}$, $\mathcal{O}^{\left(5\right)}_{2}$~~  \\
\hline
$\mathcal{C}^{Z_{\mu}}_{\overline{\nu}\nu}$ & ~~~ $\frac{g\gamma_{\mu}}{2c_{w}}U^{\dagger}UP_{L} - 2i\frac{\alpha^{(5)}_{3}s_{w}}{\Lambda}\theta^{*}\theta p_{\nu}\sigma_{\mu\nu}P_{R}$ + h.c. & ~~ RT, $\mathcal{O}^{\left(5\right)}_{3}$ ~~  \\
\hline
$\mathcal{C}^{Z_{\mu}}_{\overline{N}N}$ & ~~~ $\frac{g\gamma_{\mu}}{2c_{w}}\theta^{\dagger}\theta P_{L} - 2i\frac{\alpha^{(5)}_{3}s_{w}}{\Lambda}\kappa^{\dagger}\kappa^{*} p_{\nu}\sigma_{\mu\nu}P_{R}$ + h.c. & ~~ RT, $\mathcal{O}^{\left(5\right)}_{3}$ ~~  \\
\hline
$\mathcal{C}^{Z_{\mu}}_{\overline{\nu}N + \overline{N}\nu}$ & ~~~ $\{\frac{g\gamma_{\mu}}{2c_{w}}U^{\dagger}\theta P_{L} + 2i\frac{\alpha^{(5)}_{3}s_{w}}{\Lambda}\theta^{*}\kappa^{*} p_{\nu}\sigma_{\mu\nu}P_{R}\}$ &  \\
								& ~ + $\{\frac{g\gamma_{\mu}}{2c_{w}}U\theta^{\dagger}P_{L} + 2i\frac{\alpha^{(5)}_{3}s_{w}}{\Lambda}\kappa^{\dagger}\theta^{\dagger} p_{\nu}\sigma_{\mu\nu}P_{R}\}$ + h.c. & ~~ RT, $\mathcal{O}^{\left(5\right)}_{3}$ ~~ \\
\hline
$\mathcal{C}^{A_{\mu}}_{\overline{\nu}\nu}$ & ~~~ $2i\frac{\alpha^{(5)}_{3}c_{w}}{\Lambda}\theta^{*}\theta^{\dagger} p_{\nu}\sigma_{\mu\nu}P_{R}$ + h.c. & ~~ $\mathcal{O}^{\left(5\right)}_{3}$ ~~  \\
\hline
$\mathcal{C}^{A_{\mu}}_{\overline{N}N}$ & ~~~ $2i\frac{\alpha^{(5)}_{3}c_{w}}{\Lambda}\kappa^{\dagger}\kappa^{*} p_{\nu}\sigma_{\mu\nu}P_{R}$ + h.c. & ~~ $\mathcal{O}^{\left(5\right)}_{3}$ ~~  \\
\hline
$\mathcal{C}^{A_{\mu}}_{\overline{\nu}N + \overline{N}\nu}$ & ~~~ $\{-2i\frac{\alpha^{(5)}_{3}c_{w}}{\Lambda}\theta^{*}\kappa^{*}p_{\nu}\sigma_{\mu\nu}P_{R}\}$ $- \{2i\frac{\alpha^{(5)}_{3}c_{w}}{\Lambda}\kappa^{\dagger}\theta^{\dagger}p_{\nu}\sigma_{\mu\nu}P_{R}\}$ + h.c. & ~~ $\mathcal{O}^{\left(5\right)}_{3}$ ~~  \\
\hline
\end{tabular}
\caption{Coupling from the three-point vertices that arise after taking into account both the dimension four and dimension five terms of the Lagrangian. Here $U$ signifies $U_{\text{PMNS}}$ matrix. The abbreviation ``RT" stands for renormalisable term which includes charge current, neutral current as well as Yukawa term. The chirality projection matrix is denoted by $P_{L}$ and $P_{R}$. The momentum factor $p_{\nu}$ in different vertices arise from field strength tensor $B_{\mu\nu}$ after transforming it into momentum space.}
\label{Tab:vertex5}
\end{table}
\noindent
\begin{itemize}
\item The tree-level vertices that involve Higgs field do get modified due to the presence of $\mathcal{O}^{\left(5\right)}_{1}$, $\mathcal{O}^{\left(5\right)}_{2}$ operators. In view of Eq.~\ref{Eq:EigenM5}, one can see that the operator $\mathcal{O}^{(5)}_{1}$ regulates the SM neutrino masses. The smallness of these mass values forces us to choose a tiny magnitude for $\frac{\alpha^{(5)}_{1}}{\Lambda}$, which is below the order of $\mathcal{O}\left(10^{-11}\right)$ GeV for $\Lambda$ to be in the order TeV. This is why, the effects coming from this operator can not be studied in the present day experimental set up. Due to this, for all our analysis we will set $\alpha^{(5)}_{1}$ to be zero. 
\item In contrast to that, a similar conclusion can not be made for $\frac{\alpha^{(5)}_{2}}{\Lambda}$ coefficient. Hence, one should critically analyse it's role on a case by case basis. 
\item The operator $\mathcal{O}^{(5)}_{3}$ changes the couplings that involve both massless and massive vector boson fields. However as we have mentioned before, the structure of this operator contains two important aspects. First,  the $\alpha^{(5)}_{3}$ is an anti-symmetric matrix in the flavour space and can only exist if we consider more than one flavour of $N_{R}$ fields within the EFT framework. In addition to that, from a full theory point of view the vertices coming from this operators can not possibly be realised in tree level graphs. Hence, the effects come from this operator must be further suppressed by the loop factor $\left(\frac{1}{16\pi^{2}}\right)$. 
\item The compelling facet of the operator $\mathcal{O}^{(5)}_{3}$ is to invoke a non-trivial coupling between the photon field and neutrinos which are not present in the SM counterpart. The presence of $B_{\mu\nu}$ tensor in the $\mathcal{O}^{(3)}_{5}$ operator introduce interaction term between the RHN fields and hyper-charge gauge boson $B_{\mu}$. After the symmetry breaking the $B_{\mu}$ field can be written as the linear combination of $Z$ boson and photon ($B_{\mu} = -s_{w}Z_{\mu} +c_{w} A_{\mu}$). As a consequence of the field redefinition, $N_{R}$ fields would couple to photon. These couplings would have a direct impact to neutrino magnetic moment \cite{Aparici:2009fh}. Using XENON data~\cite{Miranda:2020kwy} one can determine the size of the associated Wilson coefficient $\frac{\alpha^{(5)}_{3}}{\Lambda}$. Here we conclude our discussion on dimension five operators and in the subsequent section, we discuss $d =$ 6 operators.  
\end{itemize}    

\subsection{$N_{R}$-EFT Operators at Dimension Six} 
\label{Sec:dim6}
In the last section we have presented various aspects of dimension five operators. We will now turn our attention to the details of the dimension six operators. In Table.~\ref{tab:Ope6}, we enlist all possible operators in systematic manner. For a methodical construction of these operators, one may read through Ref.~\cite{Liao:2016qyd}. 
\begin{itemize}  
\item \textbf{Neutrino Mass In Dimension Six}
\end{itemize} 
\noindent
Before engaging ourselves into an extensive discussion on these operators, we like to point out possible modification happens in the neutrino mass matrix when one consider dimension six operators. The operator that falls under the class of $\psi^{2}H^{3}$, where $\psi^{2}$ represents two fermionic fields, contributes towards the neutrino mass matrix as this operator would give additional contribution towards the off-diagonal Dirac elements of the matrix mentioned in Eq.~\ref{Eq:MNu5}. The updated form of this matrix can be illustrated in the following fashion -            
 \begin{equation}
\mathcal{M}^{(6)}_{\nu N} = 
\begin{bmatrix} 
\frac{\alpha^{(5)}_{1}v^{2}}{\Lambda} &~~~~ \frac{Y_{\nu}v}{\sqrt{2}} + \left(\frac{\alpha_{LNH}v^{3}}{2\sqrt{2}\Lambda^{2}}\right) \\
\frac{Y^{T}_{\nu}v}{\sqrt{2}} + \left(\frac{\alpha^{T}_{LNH}v^{3}}{2\sqrt{2}\Lambda^{2}}\right) &~~~~ \left(\tilde{M}_{N} + \frac{\alpha^{(5)}_{2}v^{2}}{\Lambda}\right)
\end{bmatrix}.
\end{equation}
\noindent
Our next task is to obtain the correct form of eigenvalues and eigenvectors corresponds to the light and heavy neutrinos respectively. To do so, we would consider the following re-definition of the off-diagonal element of the above matrix.  
\begin{equation}
 \tilde{Y}_{\nu} = Y_{\nu} + \left(\frac{\alpha_{LNH}v^{2}}{2\Lambda^{2}}\right)
 \label{Eq:theta6}
 \end{equation}
 \noindent
We use this parametrisation, to write down the light neutrino mass. To evaluate the eigenvalues of the above matrix we choose the limit $M_{N} \gg \frac{\alpha_{LNH}v^{3}}{\Lambda^{2}}, \frac{\alpha^{(5)}_{1}v^{2}}{\Lambda}$. In this limit, the light and heavy mass eigenvalue will take the following matrix form 
\begin{align}
m^{(6)}_{light} & \approx \frac{\alpha^{(5)}_{1}v^{2}}{\Lambda} - \frac{\tilde{Y}^{T}_{\nu}(\tilde{M}^{-1}_{N})v^{2}\tilde{Y}_{\nu}}{2}  \nonumber \\
m^{(6)}_{heavy} & \approx M_{N}. 
\end{align}
\noindent
Looking at the above form of the neutrino mass matrices one can appreciate the rational behind the parametrisation mentioned in Eq.~\ref{Eq:theta6}. The inclusion of the dimension six contribution does not alter the form of the mass eigenvalues as compare to Eq.~\ref{Eq:EigenM5}. The mass matrix in the dimension six set up can also be diagonalised using the prescription discussed in the last section. To do so we need to re-define the mixing angle between active and sterile neutrino. The matrix $V$ of Eq.~\ref{Eq:Vdim5} will take the following form

\begin{align}
V \approx \begin{pmatrix}
U_{\text{PMNS}} & \tilde{\theta} \\
- \tilde{\theta}^{T} & \kappa
\end{pmatrix},
\label{Eq:Vdim6}
\end{align}
\noindent
where $\tilde{\theta}$ is -
\[ \tilde{\theta} = \theta^{(5)} + \theta^{(6)} = M^{-1}_{N}\frac{\tilde{Y}_{\nu}v}{\sqrt{2}},\,\,\, \text{and} \, \, \, \theta^{(5)} = M^{-1}_{N}\frac{Y_{\nu}v}{\sqrt{2}}, \,\, \, \theta^{(6)} = M^{-1}_{N}\frac{\alpha_{LNH}v^{3}}{2\sqrt{2}\Lambda^{2}}   \]
With this new definition of the mixing angle one can obtain the corresponding mass eigenstates for the neutrinos. We illustrate this results in Eq.~\ref{Eq:EigenV6}    
\begin{align}
\nu_{L} & \simeq U_{\text{PMNS}}\nu_{L, m} + \tilde{\theta}N^{c}_{m, R} \nonumber \\
N^{c}_{R} & \simeq - \tilde{\theta}^{T}\nu_{L, m} + \kappa N^{c}_{m, R} 
\label{Eq:EigenV6}
\end{align}
\noindent
where, the orthonormality between these two states dictate $U_{\text{PMNS}} \simeq \kappa $. 

\begin{itemize}
\item \textbf{Interesting Facets of the Dimension Six Operators}
\end{itemize}
\noindent
With this set up, we now point out various aspect of the dimension six operators that are tabulated in Table.~\ref{tab:Ope6}. We have categorised these operators based upon their Lorentz structure as well as the field contents that are present. 
\begin{itemize}
\item $\mathbf{\mathcal{O}_{LNH}:}$ In dimension six, the only possible operator can come under the class of $\psi^{2}H^{3}$ is $\mathcal{O}_{LNH}$ where $\psi$ denotes the charged as well as neutral fermionic fields present in that operator. Apart from modifying the neutrino mass matrix, this operator also contributes toward the Higgs neutrino couplings. The modification of these couplings are explicitly presented in Table.~\ref{tab:vertex6} that we will discuss in detail in the later part of this section. Before that we like to highlight interesting aspects of other classes of dimension six operators.  
\item $\mathbf{\mathcal{O}_{HN} ~\&~\mathcal{O}_{HNe}:}$ Two operators $\mathcal{O}_{HN}$ and $\mathcal{O}_{HNe}$ falls in the class of $\psi^{2}H^{2}D$, where $D$ represents the covariant derivative corresponds to $SU(2)_{L}\times U(1)_{Y}$ gauge group. Upon expanding $H^{\dagger}i\overleftrightarrow{D}_{\mu}H$\footnote{where the explicit form of $H^{\dagger}i\overleftrightarrow{D}_{\mu}H$ is $i\left(H^{\dagger}D_{\mu}H - (D_{\mu}H)^{\dagger}H\right)$. Replacing $H$ field with the SM Higgs doublet one can see the terms which contain $Z$ boson field would only survive.} and $\tilde{H}^{\dagger}iD_{\mu}H$, one can see both active and sterile neutrinos couple to $Z$ boson via $\mathcal{O}_{HN}$ operator, and not via $\mathcal{O}_{HNe}$. On the other hand, $\mathcal{O}_{HNe}$ only contributes to the $\mathcal{C}^{W_{\mu}}_{\ell\nu}$ and $\mathcal{C}^{W_{\mu}}_{\ell N}$ couplings. If we expand this operator explicitly, one can see the $W$ boson couple to right-handed chiral leptons via this operator. This is indeed a striking departure from the existing SM theory. The SM is a $SU(2)_{L}$ theory which forbids the coupling between charged gauge bosons and right-handed fermions. The present experimental bounds on the respective coupling can be translated to estimate the current limit on this operator. 
\item $\mathbf{\mathcal{O}_{LNW}~\&~\mathbf{\mathcal{O}_{LNB}}:}$ The dimension six allows us to write two operators that involve the field-strength tensor i.e. $W^{I}_{\mu\nu}$ and $B_{\mu\nu}$ corresponding to $SU(2)_{L}$ and $U(1)_{Y}$ respectively. Similar to the operator $\mathcal{O}^{(5)}_{3}$, both these operators would be suppressed by an extra $\frac{1}{16\pi^{2}}$ as these can only be realised in a full theory via loop mediated processes. In case of dimension five, the operator $\mathcal{O}^{\left(3\right)}_{5}$ is an antisymmetric tensor in the flavour space which is not the same for dimension six case.  
\item \textbf{4-Fermi:} The dimension six case allows us to construct various kind of four-Fermi operators. Considering the chirality of different fermions these operators can be categorised into four separate classes - $\left(\overline{L}R\right)\left(\overline{R}L\right)$, $\left(\overline{R}R\right)\left(\overline{R}R\right)$, $\left(\overline{L}L\right)\left(\overline{R}R\right)$ and $\left(\overline{L}R\right)\left(\overline{L}R\right)$. In addition to that, one can write three more operators that violate either lepton number or lepton-baryon number. The operator $\mathcal{O}_{QuNL}$ that comes under the $\left(\overline{L}R\right)\left(\overline{R}L\right)$ class would contributes toward the neutral as well as charged four-point contact interaction terms. Considering the underlying Lorentz structure, one can see in the full theory this operator can be realised via the charged scalar mediated process. In contrast to that, the operator, $\mathcal{O}_{duNe}$ would also give rise to the charged four-point contact interaction. Moreover, the presence of $\gamma^{\mu}$ matrices in this interaction vertex suggest that it can be incorporated in a possible non-abelian gauge extended theory. These operators lead to single production of $N_{R}$ which is not mixing suppressed. The other operators $\mathcal{O}_{eN}$, $\mathcal{O}_{uN}$ and $\mathcal{O}_{dN}$  would also invoke the pair production of the $N_{R}$ fields in the lepton and hadron colliders respectively. The corresponding production cross section is independent of active-sterile mixing angle. The operators that come under $\left(\overline{L}L\right)\left(\overline{R}R\right)$ class would also give rise to pair production processes similar to the operators $\mathcal{O}_{eN}$, $\mathcal{O}_{uN}$ and $\mathcal{O}_{dN}$. The operator $\mathcal{O}_{LN}$, would also invoke additional contact interaction term $(\overline{\nu}_{L}\gamma^{\mu}\nu_{L})(\overline{N}_{R}\gamma_{\mu}N_{R})$.  Nevertheless, one can neglect this term for further discussion as it is phenomenologically imprudent.                              
\end{itemize} 

\begin{table}[h!]
\centering
\begin{tabular}{ |c|c|c|c| }
\hline
\multicolumn{4}{ |c| }{Relevant Operators in dim-6} \\
\hline
& $\mathcal{O}_{6}$ & Explicit Form & $n_{f}$ \\
\hline
\multirow{1}{*}{\hspace{0.5cm}$\psi^{2}H^{3}$\hspace{0.5cm}} & \hspace{0.5cm} $\mathcal{O}_{LNH} := $ \hspace{0.5cm} & \hspace{0.5cm} $\frac{\alpha_{LNH}}{\Lambda^{2}}\left(\overline{L}N_{R}\right)\tilde{H}\left(H^{\dagger}H\right) +$ h.c. \hspace{0.5cm} & \hspace{0.5cm} $2n^{2}_{f}$ \hspace{0.5cm} \\ \cline{1-4} 
\hline \hline
\multirow{2}{*}{\hspace{0.5cm}$\psi^{2}H^{2}D$\hspace{0.5cm}} & \hspace{0.5cm} $\mathcal{O}_{HN} :=$ \hspace{0.5cm} & \hspace{0.25cm} $\frac{\alpha_{HN}}{\Lambda^{2}}\left(\overline{N}_{R}\gamma^{\mu}N_{R}\right)\left(H^{\dagger}i\overleftrightarrow{D}_{\mu}H\right)$ &\hspace{0.5cm} $n^{2}_{f}$ \hspace{0.5cm} \\ \cline{2-4}
	& \hspace{0.25cm} $\mathcal{O}_{HNe} :=$ \hspace{0.5cm} & \hspace{0.5cm} $\frac{\alpha_{HNe}}{\Lambda^{2}}\left(\overline{N}_{R}\gamma^{\mu}e_{R}\right)\left(\tilde{H}^{\dagger}iD_{\mu}H\right)$ + h.c. \hspace{0.5cm} &\hspace{0.5cm} $2n^{2}_{f}$ \hspace{0.5cm} \\ \cline{2-4}
	\hline \hline
\multirow{2}{*}{\hspace{0.5cm}$\psi^{2}H^{2}X$\hspace{0.5cm}} & \hspace{0.5cm}  $\mathcal{O}_{LNB} :=$ \hspace{0.5cm} & $\frac{\alpha_{LNB}}{\Lambda^{2}}\left(\overline{L}\sigma_{\mu\nu}N_{R}\right)\tilde{H}B_{\mu\nu} +$ h.c. & $2n^{2}_{f}$ \\ \cline{2-4}
	& $\mathcal{O}_{LNW} :=$ & $\frac{\alpha_{LNW}}{\Lambda^{2}}\left(\overline{L}\sigma_{\mu\nu}N_{R}\right)\tau^{I}\tilde{H}W^{I\mu\nu}$ + h.c. & $2n^{2}_{f}$ \\ \cline{2-4}
	\hline \hline
\multirow{1}{*}{\hspace{0.5cm}$\left(\overline{L}R\right)\left(\overline{R}L\right)$\hspace{0.5cm}} & $\mathcal{O}_{QuNL} :=$ & $\frac{\alpha_{QuNL}}{\Lambda^{2}}\left(\overline{Q}u_{R}\right)\left(\overline{N}_{R}L\right)$ + h.c. & $2n^{4}_{f}$ \\ \cline{1-4}
\hline \hline
\multirow{5}{*}{\hspace{0.5cm}$\left(\overline{R}R\right)\left(\overline{R}R\right)$\hspace{0.5cm}} & $\mathcal{O}_{NN} :=$ & $\frac{\alpha_{NN}}{\Lambda^{2}}\left(\overline{N}_{R}\gamma^{\mu}N_{R}\right)\left(\overline{N}_{R}\gamma_{\mu}N_{R}\right)$ & $\frac{n^{2}_{f}(n_{f} + 1)^{2}}{4}$ \\ \cline{2-4}
	& $\mathcal{O}_{eN} :=$ &  $\frac{\alpha_{eN}}{\Lambda^{2}}\left(\overline{e}_{R}\gamma^{\mu}e_{R}\right)\left(\overline{N}_{R}\gamma_{\mu}N_{R}\right)$ & $n^{4}_{f}$ \\ \cline{2-4}
	& $\mathcal{O}_{uN} :=$ &  $\frac{\alpha_{uN}}{\Lambda^{2}}\left(\overline{u}_{R}\gamma^{\mu}u_{R}\right)\left(\overline{N}_{R}\gamma_{\mu}N_{R}\right)$ & $n^{4}_{f}$ \\ \cline{2-4}
	& $\mathcal{O}_{dN} :=$ &  $\frac{\alpha_{dN}}{\Lambda^{2}}\left(\overline{d}_{R}\gamma^{\mu}d_{R}\right)\left(\overline{N}_{R}\gamma_{\mu}N_{R}\right)$ & $n^{4}_{f}$ \\ \cline{2-4}
	& $\mathcal{O}_{duNe} :=$ & $\frac{\alpha_{duNe}}{\Lambda^{2}}\left(\overline{d}_{R}\gamma^{\mu}u_{R}\right)\left(\overline{N}_{R}\gamma_{\mu}e_{R}\right)$ + h.c. & $2n^{2}_{f}$ \\ \cline{2-4} \hline \hline
\multirow{2}{*}{\hspace{0.5cm}$\left(\overline{L}L\right)\left(\overline{R}R\right)$\hspace{0.5cm}} & $\mathcal{O}_{LN} :=$ & $\frac{\alpha_{LN}}{\Lambda^{2}}\left(\overline{L}\gamma^{\mu}L\right)\left(\overline{N}_{R}\gamma_{\mu}N_{R}\right)$ & $n^{4}_{f}$ \\ \cline{2-4}
	& $\mathcal{O}_{QN} :=$ & $\frac{\alpha_{QN}}{\Lambda^{2}}\left(\overline{Q}\gamma^{\mu}Q\right)\left(\overline{N}_{R}\gamma N_{R}\right)$ & $n^{4}_{f}$ \\ \cline{2-4} \hline \hline
\multirow{3}{*}{\hspace{0.5cm}$\left(\overline{L}R\right)\left(\overline{L}R\right)$\hspace{0.5cm}} & $\mathcal{O}_{LNLe} :=$ & $\frac{\alpha_{LNLe}}{\Lambda^{2}}\left(\overline{L}N_{R}\right)\epsilon\left(\overline{L}e_{R}\right)$ h.c. & $2n^{4}_{f}$ \\ \cline{2-4}
	& $\mathcal{O}_{LNQd} :=$ & $\frac{\alpha_{LNQd}}{\Lambda^{2}}\left(\overline{L}N_{R}\right)\epsilon\left(\overline{Q}d_{R}\right)$ + h.c. & $2n^{4}_{f}$ \\ \cline{2-4} 
	& $\mathcal{O}_{LdQN} :=$ & $\frac{\alpha_{LdQN}}{\Lambda^{2}}\left(\overline{L}d_{R}\right)\epsilon\left(\overline{Q}N_{R}\right)$ + h.c. & $2n^{4}_{f}$ \\ \cline{2-4} \hline \hline
\multirow{1}{*}{\hspace{0.5cm}$\sout{L}\cap B$\hspace{0.5cm}} & $\mathcal{O}_{NNNN} :=$ & $\frac{\alpha_{NNNN}}{\Lambda^{2}}\left(N_{R}CN_{R}\right)\left(N_{R}CN_{R}\right)$ + h.c. & $\frac{n^{2}_{f}\left(n^{2}_{f} - 1\right)}{6}$ \\ \cline{2-4} \hline \hline
\multirow{2}{*}{\hspace{0.5cm}$\sout{L}\cap \sout{B}$\hspace{0.5cm}} & $\mathcal{O}_{QQdN} :=$ & $\frac{\alpha_{QQdN}}{\Lambda^{2}}\epsilon_{ij}\epsilon_{\alpha\beta\sigma}\left(Q^{i}_{\alpha}CQ^{j}_{\beta}\right)\left(d_{R\sigma}CN_{R}\right)$ + h.c. & $n^{3}_{f}\left(n_{f} + 1\right)$ \\ \cline{2-4} 
		& $\mathcal{O}_{uddN} :=$ & $\frac{\alpha_{uddN}}{\Lambda^{2}}\epsilon_{\alpha\beta\sigma}\left(u^{\alpha}_{R}Cd^{\beta}_{R}\right)\left(d^{\sigma}_{R}CN_{R}\right)$ + h.c. & $2n^{4}_{f}$ \\ \cline{2-4} \hline \hline
\end{tabular}
\caption{List of all possible operators that appear in dimension six construction. The four-Fermi operators can arise in this order as oppose to dimension five $N_{R}-$EFT. In this paper we would refrain ourselves from discussing the phenomenology that arise from operators mentioned in last two rows.}
\label{tab:Ope6}
\end{table}
\noindent
\begin{itemize}
 \item $\mathbf{\left(\overline{L}R\right)\left(\overline{L}R\right) :}$ In case of $\left(\overline{L}R\right)\left(\overline{L}R\right)$ scenario, one can write three operators $\mathcal{O}_{LNLe}$, $\mathcal{O}_{LNQd}$ and $\mathcal{O}_{LdQN}$ where $\epsilon$ stands for $2\times2$ antisymmetric matrices. From the structure of these operators one can interpret their effects with heavy scalar mediated processes. The terms arise from these operators would contribute to both neutral as well as charged four-point vertices. Furthermore, the Lorentz structure of these vertices can possibly be incorporated in a \emph{full} theory which contains both neutral and charged scalar \emph{d.o.f}.  
 \item $\mathbf{\mathcal{O}_{NN}~\&~\mathbf{\mathcal{O}_{NNNN}}:}$ The dimension six also allows us to build two operators $\mathcal{O}_{NN}$, and $\mathcal{O}_{NNNN}$ that involve four $N_{R}$ fields. The differences between these two are many folds. From the stand point of Lorentz structure one can see, $\mathcal{O}_{NN}$ invokes a vector-like process in contrast to scalar-like $\mathcal{O}_{NNNN}$ operator. On the other hand $\mathcal{O}_{NNNN}$ explicitly violates lepton number as oppose to $\mathcal{O}_{NN}$. In addition to that, the Wilson coefficient $\alpha_{NNNN}$ is antisymmetric in flavour space, whereas $\alpha_{NN}$ is symmetric. However, the explicit computation of the operators suggest that the coupling of $N_{R}$ with SM neutrinos coming from these operators would be $\tilde{\theta}^{3}$ suppressed (see Table.~\ref{tab:vertexfermi} for the explicit form). As a result, both these operators remain inaccessible from present day collider experiments.  
 \item $\mathbf{\mathcal{O}_{QQdN}~\&~\mathcal{O}_{uddN}:}$ We conclude our discussion while presenting two Lepton $\oplus$ Baryon numbers violating operators - $\mathcal{O}_{QQdN}$ and $\mathcal{O}_{uddN}$. Both these operators invoke non trivial decay mode of $N$ such as $N_{i} \to d_{\alpha}u_{\alpha}d_{\beta}$. These operators also play an important role in physics involving $B - L$ asymmetry. However in the current paper, we refrain ourselves to discussing the aspects of these operators.                                        
\end{itemize} 
Our next step is to discuss the modifications as well as emergence of various three-point couplings that involve RHNs in the dimension six set up. The detail calculation to determine the explicit form of these couplings are illustrated in Appendix~\ref{App:dim6}. In the beginning of this section we have shown how the mass eigenstates of both the active and sterile neutrinos evolve due to the inclusion of the operator $\mathcal{O}_{LNH}$. This change can be incorporated by redefining the mixing parameter from $\theta$ to $\tilde{\theta}$. 

One can divide all the relevant couplings into two sub categories. In Table.~\ref{tab:vertex6}, we also include those couplings that one can already find in the dimension five $N_{R}$-EFT.  We write different couplings as $\mathcal{C}^{6}_{Sff^{'}}$, where $\mathcal{C}^{6}_{Sff^{'}}$ includes sum of dimension five contribution $C^{S}_{ff^{'}}$ and the $\mathcal{O}\left(\frac{1}{\Lambda^{2}}\right)$ corrections\footnote{In the notation $\mathcal{C}^{6}_{Sff^{'}}$ and $C^{S}_{ff^{'}}$ the $S$ denotes $W/Z/\gamma/h$, $f$ stands for either charged or neutral leptons and $f^{'}$ correspond to either active or sterile neutrinos.}.
\begin{itemize}
\item Unlike dimension five the coupling between $W$ and $\ell\nu/N$ receives additional correction due to both $\mathcal{O}_{HNe}$ and the loop-suppressed $\mathcal{O}_{LNW}$ (which has in general small value) operators. As mentioned earlier, the operator $\mathcal{O}_{HNe}$ invokes right handed coupling between SM leptons and the W boson. The upshot of this coupling is that, it would modify the branching ratios as well as total decay width of $W$ bosons. The precision measurements on this charged gauge boson can be used to place meaningful bounds on the Wilson coefficient $\frac{\alpha_{HNe}}{\Lambda^{2}}$. A recent article~\cite{Xue:2022mde} proposed similar kind of right handed couplings between SM fermions and W boson on the verge of solving the W boson mass tension (see Ref.~\cite{CDF:2022hxs}).   
\item The Z boson coupling gets addition corrections both from tree-level mediated operators $\mathcal{O}_{HN}$ as well as loop-mediated operators $\mathcal{O}_{LNB}$ and $\mathcal{O}_{LNW}$ respectively. The last two operators also generate appropriate alteration to photon neutrino couplings via neutral gauge state mixing. We like to reiterate that the terms coming from these would be $\frac{1}{16\pi^{2}}$ suppressed. Processes induced from these operators receive appropriate constraints from the different precession measurements such as $Z-$boson total width~\cite{ParticleDataGroup:2018ovx} measurement, $\mathcal{BR}(Z \to \nu N)$\cite{OPAL:1990fcc} \emph{etc}. 
\item Apart from the gauge bosons, the Higgs-neutrino couplings also get modified in the underlying EFT setup due to the operator $\mathcal{O}_{LNH}$. In the beginning of this section we have shown how this operator enters into the off-diagonal elements of the mass matrix. From the parametrisation which is displayed in Eq.~[\ref{Eq:theta6}], one can see this operator also redefines the active-sterile mixing angles $\tilde{\theta}$.        
\end{itemize}      
\begin{table}[h!]
\centering
\begin{tabular}{|c||c|c|}
\hline
Couplings & Explicit Form & Operator  \\
\hline
$\mathcal{C}^{6}_{W_{\mu}\ell\nu}$ & $\mathcal{C}^{W_{\mu}}_{\ell\nu} + \{\frac{g\alpha_{HNe}v^{2}}{2\sqrt{2}\Lambda^{2}}\tilde{\theta}^{\dagger}\gamma^{\mu}P_{R} - 2ip_{\nu}\frac{v\alpha_{LNW}}{\sqrt{2}\Lambda^{2}}\tilde{\theta}^{\dagger}\sigma_{\mu\nu}P_{R}\}+$ h.c. &~~ $\mathcal{O}_{HNe}, \mathcal{O}_{LNW}$~~ \\
\hline
$\mathcal{C}^{6}_{W_{\mu}\ell N}$ & $\mathcal{C}^{W_{\mu}}_{\ell N} + \{ -\frac{g\alpha_{HNe}v^{2}}{2\sqrt{2}\Lambda^{2}}\kappa^{*}\gamma^{\mu}P_{R} + 2ip_{\nu}\frac{v\alpha_{LNW}}{\sqrt{2}\Lambda^{2}}\kappa^{*}\sigma_{\mu\nu}P_{R}\}+$ h.c. & $\mathcal{O}_{HNe}, \mathcal{O}_{LNW}$ \\
\hline
$\mathcal{C}^{6}_{h\overline{\nu}\nu}$ & $\mathcal{C}^{h}_{\overline{\nu}\nu} + \{- \frac{3v^{2}\alpha_{LNH}}{2\sqrt{2}\Lambda^{2}}U^{\dagger}\tilde{\theta}^{\dagger}P_{R}\} $ + h.c.&   $\mathcal{O}_{LNH}$    \\
\hline
$\mathcal{C}^{6}_{h\overline{N}N}$ & $\mathcal{C}^{h}_{\overline{N}N} + \{ \frac{3v^{2}\alpha_{LNH}}{2\sqrt{2}\Lambda^{2}}\tilde{\theta}^{\dagger}\kappa^{*}P_{R}\}$ + h.c.                              &    $\mathcal{O}_{LNH}$      \\
\hline
$\mathcal{C}^{6}_{h(\overline{\nu}N + \overline{N}\nu)}$ & $\mathcal{C}^{h}_{(\overline{\nu}N + \overline{N}\nu)} + \{ \frac{3v^{2}\alpha_{LNH}}{2\sqrt{2}\Lambda^{2}}U^{\dagger}\kappa^{*}P_{R} - \frac{3v^{2}\alpha_{LNH}}{2\sqrt{2}\Lambda^{2}}\tilde{\theta}^{\dagger}\tilde{\theta}^{\dagger}P_{R}\}$ + h.c.  &     $\mathcal{O}_{LNH}$     \\
\hline
$\mathcal{C}^{6}_{Z_{\mu}\overline{\nu}\nu}$ &~~~ $\mathcal{C}^{Z_{\mu}}_{\overline{\nu}\nu} + \{ -\frac{\alpha_{HN}vm_{Z}}{\Lambda^{2}}\tilde{\theta}\tilde{\theta}^{\dagger}\gamma^{\mu}P_{R} + 2is_{w}p_{\nu}\frac{v\alpha_{LNB}}{\sqrt{2}\Lambda^{2}}U^{\dagger}\tilde{\theta}^{\dagger}\sigma_{\mu\nu}P_{R}$ ~~~ &  $\mathcal{O}_{HN}, \mathcal{O}_{LNB}$, \\
                     &$- 2ic_{w}p_{\nu}\frac{v\alpha_{LNW}}{\sqrt{2}\Lambda^{2}}U^{\dagger}\tilde{\theta}^{\dagger}\sigma_{\mu\nu}P_{R}\}$ + h.c.  & $\mathcal{O}_{LNW}$ \\
\hline 
$\mathcal{C}^{6}_{Z_{\mu}\overline{N}N}$ &~~~ $\mathcal{C}^{Z_{\mu}}_{\overline{N}N} + \{ -\frac{\alpha_{HN}vm_{Z}}{\Lambda^{2}}\kappa^{\dagger}\kappa^{*}\gamma^{\mu}P_{R} - 2is_{w}p_{\nu}\frac{v\alpha_{LNB}}{\sqrt{2}\Lambda^{2}}\tilde{\theta}^{\dagger}\kappa^{*}\sigma_{\mu\nu}P_{R}$ ~~~ & $\mathcal{O}_{HN}, \mathcal{O}_{LNB}$,  \\
						&$+ 2ic_{w}p_{\nu}\frac{v\alpha_{LNW}}{\sqrt{2}\Lambda^{2}}\tilde{\theta}^{\dagger}\kappa^{*}\sigma_{\mu\nu}P_{R}\}$ + h.c.  & $\mathcal{O}_{LNW}$ \\
\hline 
$\mathcal{C}^{6}_{Z_{\mu}(\overline{\nu}N + \overline{N}\nu)}$ & $\mathcal{C}^{Z_{\mu}}_{(\overline{\nu}N + \overline{N}\nu)} + \{\frac{\alpha_{HN}vm_{Z}}{\Lambda^{2}}\tilde{\theta}\kappa^{*}\gamma^{\mu}P_{R} - 2is_{w}p_{\nu}\frac{\alpha_{LNB}v}{\sqrt{2}\Lambda^{2}}U^{\dagger}\kappa^{\dagger}\sigma_{\mu\nu}P_{R}$ & \\
						& $+ 2ic_{w}p_{\nu}\frac{\alpha_{LNW}v}{\sqrt{2}\Lambda^{2}}U^{\dagger}\kappa^{\dagger}\sigma_{\mu\nu}P_{R}\} + \{\frac{\alpha_{HN}vm_{Z}}{\Lambda^{2}}\kappa^{T}\theta^{\dagger}\gamma^{\mu}P_{R} $ & $\mathcal{O}_{HN}, \mathcal{O}_{LNB}$,   \\
						& $+ 2is_{w}p_{\nu}\frac{\alpha_{LNB}v}{\sqrt{2}\Lambda^{2}}\tilde{\theta}^{\dagger}\tilde{\theta}^{\dagger}\sigma_{\mu\nu}P_{R} - 2ic_{w}p_{\nu}\frac{\alpha_{LNW}v}{\sqrt{2}\Lambda^{2}}\tilde{\theta}^{\dagger}\tilde{\theta}^{\dagger}\sigma_{\mu\nu}P_{R}\}$ + h.c.  & $\mathcal{O}_{LNW}$  \\
\hline
$\mathcal{C}^{6}_{A_{\mu}\overline{\nu}\nu}$ &~~~ $\mathcal{C}^{A_{\mu}}_{\overline{\nu}\nu} + \{-2ic_{w}p_{\nu}\frac{\alpha_{LNB}v}{\sqrt{2}\Lambda^{2}}U^{\dagger}\tilde{\theta}^{\dagger}\sigma_{\mu\nu}P_{R} + 2is_{w}p_{\nu}\frac{\alpha_{LNW}v}{\sqrt{2}\Lambda^{2}}U^{\dagger}\tilde{\theta}^{\dagger}\sigma_{\mu\nu}P_{R} \} $ + h.c. &   $ \mathcal{O}_{LNB}, \mathcal{O}_{LNW}$              \\
\hline
$\mathcal{C}^{6}_{A_{\mu}\overline{N}N}$ &~~~ $\mathcal{C}^{A_{\mu}}_{\overline{N}N} + \{2ic_{w}p_{\nu}\frac{\alpha_{LNB}v}{\sqrt{2}\Lambda^{2}}\tilde{\theta}^{\dagger}\kappa^{*}\sigma_{\mu\nu}P_{R} + 2is_{w}p_{\nu}\frac{\alpha_{LNW}v}{\sqrt{2}\Lambda^{2}}\tilde{\theta}^{\dagger}\kappa^{*}\sigma_{\mu\nu}P_{R} \} $ + h.c. &    $ \mathcal{O}_{LNB}, \mathcal{O}_{LNW}$             \\
\hline
$\mathcal{C}^{6}_{A_{\mu}(\overline{\nu}N + \overline{N}\nu)}$ & $\mathcal{C}^{A_{\mu}}_{(\overline{\nu}N + \overline{N}\nu)} + \{2ic_{w}p_{\nu}\frac{\alpha_{LNB}v}{\sqrt{2}\Lambda^{2}}U^{\dagger}\kappa^{*}\sigma_{\mu\nu}P_{R} + 2is_{w}p_{\nu}\frac{\alpha_{LNW}v}{\sqrt{2}\Lambda^{2}}U^{\dagger}\kappa^{\dagger}\sigma_{\mu\nu}P_{R}$ &   \\
		& $- 2ic_{w}p_{\nu}\frac{\alpha_{LNB}v}{\sqrt{2}\Lambda^{2}}\tilde{\theta}^{\dagger}\tilde{\theta}^{\dagger}\sigma_{\mu\nu}P_{R} - 2is_{w}p_{\nu}\frac{\alpha_{LNB}v}{\sqrt{2}\Lambda^{2}}\tilde{\theta}^{\dagger}\tilde{\theta}^{\dagger}\sigma_{\mu\nu}P_{R}\}$ + h.c. &   $ \mathcal{O}_{LNB}, \mathcal{O}_{LNW}$     \\
\hline
\end{tabular}
\caption{The explicit modification of three point couplings after inclusion of the dimension six operator contributions. The couplings up to dimension five is embedded in $\mathcal{C}^{S}_{ff^{'}}$ term where $S$ is SM bosons and $f$ and $f^{'}$ can be the charged as well as neutral leptons. The column three highlights the dimension six operators which participate in each of these couplings.}
\label{tab:vertex6}
\end{table}
\noindent
The compelling ingredient of the dimension six $N_{R}$-EFT setup is the emergence of various four fermi operators. As an upshot, one can write down the four point contact interactions that involve at least one heavy $N_{R}$ field. Couplings like these play a crucial role in the production as well as decay of the right handed neutrinos. Few of these operators would contribute to various pair production processes in lepton and hadron colliders with appreciable cross section. In parallel to that, these operators have the capacity to participate into the three body decay modes of $N_{R}$. Upto dimension five, the RHN fields can decay to three body leptonic/semi-leptonic final states, where decay is mediated via  off-shell $W/Z$ bosons. These can serve as dominant channels if mass of the $N_{R}$ is below $M_{W}$. The four fermi operators instead give rise to relevant four-point contact interactions which contribute towards these three body decay modes. In Section~\ref{Sec:NRdecay}, we present the analytic expression of these decay modes and the relevance of four Fermi operators in this context. 

 In Table.~\ref{tab:vertexfermi}, we present some of these couplings with its explicit structure. Here we restrict ourselves to the couplings which only involve RHN field and SM fermions as they are relevant for our later discussion on the three body decay modes. In addition to that, we have chosen the mass of various flavour of heavy neutrinos to be same. As a result the decay from $N_{i}$ to $N_{j}$ states with $i \neq j$ are kinematically forbidden. Out of these seven couplings, three of them would be purely leptonic and other three would be an admixture of hadronic and leptonic state. There exist one coupling which involves only the active and sterile neutrinos.    

\begin{itemize}
\item The coupling, $\mathcal{G}^{N}_{\ell_{j}\ell_{k}\nu_{k}} (j \neq k)$, $\mathcal{G}^{N}_{\nu_{k}\ell_{k}\ell_{k}}$ and $\mathcal{G}^{N}_{\nu_{j}\ell_{k}\ell_{k}}$  are controlled by same operators. However, the label of the SM leptons suggest that one can not treat them as an equal footing. The contribution coming from $\mathcal{O}_{LNLe}$ operator only depends on the associated Wilson coefficient and the effect coming from it would be prominent. On the other hand, the operator $\mathcal{O}_{eN}$ and $\mathcal{O}_{LN}$ have an additional dependence on the mixing angle. As a consequence, their phenomenological implications are difficult to probe. 
\item Apart from the leptonic channels, the $N$ can couple to quarks via four point interactions. Noticeably, all the operators which furnish the coupling $\mathcal{G}^{N}_{\ell_{j}u_{\alpha}d_{\beta}}$ are phenomenologically viable as their impact is not suppressed by the smallness of mixing angle. Along with that, two other couplings $\mathcal{G}^{N}_{\nu_{j}u_{\alpha}u_{\alpha}}$ and $\mathcal{G}^{N}_{\nu_{j}d_{\alpha}d_{\alpha}}$ are also possible which can provide addition signatures for the collider study of $N_{R}$-EFT. The operators $\mathcal{O}_{uN}$, $\mathcal{O}_{QN}$ and $\mathcal{O}_{dN}$ contributions in these couplings face an additional $\tilde{\theta}$ suppression, while $\mathcal{O}_{QuNL}$, $\mathcal{O}_{LdQN}$ are unsuppressed. The vertex $\mathcal{G}^{N}_{\ell_{j}u_{\alpha}d_{\beta}}$ is not accompanied with any such suppression.  
\item The coupling $\mathcal{G}^{N}_{\nu_{j}\nu\nu}$ is mediated via $\mathcal{O}_{NNNN}$ and $\mathcal{O}_{NN}$ operators but its magnitude is proportional to the cubic order of $\tilde{\theta}$.     
\end{itemize}        

\begin{table}[h!]
\centering
\begin{tabular}{|c||c|c|}
\hline
Couplings & Explicit Form & Operator  \\
\hline
 & $\frac{\alpha_{LNLe}}{\Lambda^{2}}\{U^{\dagger}\kappa^{*}\left(\overline{\nu}_{m}P_{R}N_{m}\right)\left(\overline{e}_{m}P_{R}e_{m}\right) - \kappa^{*}U^{\dagger}\left(\overline{e}_{m}P_{R}N_{m}\right)\left(\overline{\nu}_{m}P_{R}e_{m}\right)\}$   & $\mathcal{O}_{LNLe}$, \\
$\mathcal{G}^{N}_{\ell_{j}\ell_{k}\nu_{k}}$						& $ + \frac{\alpha_{eN}}{\Lambda^{2}}\left(\overline{e}_{m}\gamma^{\mu}P_{R}e_{m}\right)\{-\kappa^{T}\tilde{\theta}^{\dagger}\left(\overline{N}_{m}\gamma^{\mu}P_{R}\nu_{m}\right) - \tilde{\theta}\kappa^{*}\left(\overline{\nu}_{m}\gamma^{\mu}P_{R}N_{m}\right)\}  $  & $\mathcal{O}_{eN}$, \\
						& $ + \frac{\alpha_{LN}}{\Lambda^{2}}\left(\overline{e}_{m}\gamma^{\mu}P_{L}e_{m}\right)\{-\kappa^{T}\tilde{\theta}^{\dagger}\left(\overline{N}_{m}\gamma^{\mu}P_{R}\nu_{m}\right) - \tilde{\theta}\kappa^{*}\left(\overline{\nu}_{m}\gamma^{\mu}P_{R}N_{m}\right)\}  $ + h.c.   & $\mathcal{O}_{LN}$ \\
\hline
$\mathcal{G}^{N}_{\nu_{k}\ell_{k}\ell_{k}}$ & Same As Above & Same \\
                       &                &          As Above \\
\hline
$\mathcal{G}^{N}_{\nu_{j}\ell_{k}\ell_{k}}$ & Same As Above & Same \\
                        &                 &         As Above \\
\hline
$\mathcal{G}^{N}_{\ell_{j}u_{\alpha}d_{\beta}}$ & ~ $\frac{\alpha_{duNe}}{\Lambda^{2}}\left(\overline{d}_{m}\gamma^{\mu}P_{R}u_{m}\right)\left(\kappa^{T}\overline{N}_{m}\gamma^{\mu}P_{R}e_{m}\right) + \frac{\alpha_{QuNL}}{\Lambda^{2}}\left(\overline{d}_{m}P_{R}u_{m}\right)\left(\kappa^{T}\overline{N}_{m}P_{L}e_{m}\right) $ ~	& $\mathcal{O}_{duNe}, \mathcal{O}_{QuNL}$            \\
				& $- \left[\frac{\alpha_{LNQd}}{\Lambda^{2}} + \frac{\alpha_{LdQN}}{\Lambda^{2}} \right]\kappa^{\dagger}\left(\overline{e}_{m}P_{R}N_{m}\right)\left(\overline{u}_{m}P_{R}d_{m}\right)$ + h.c.        &   $\mathcal{O}_{LNQd}, \mathcal{O}_{LdQN}$           \\
\hline	
$\mathcal{G}^{N}_{\nu_{j}u_{\alpha}u_{\alpha}}$ & $\frac{\alpha_{QuNL}}{\Lambda^{2}}\kappa^{T}U\left(\overline{u}_{m}P_{R}u_{m}\right)\left(\overline{N}_{m}P_{L}\nu_{m}\right) - \frac{\alpha_{uN}}{\Lambda^{2}}\kappa^{T}\tilde{\theta}\left(\overline{u}\gamma^{\mu}P_{R}u_{m}\right)\left(\overline{N}_{m}\gamma^{\mu}P_{R}\nu_{m}\right)$ &   $\mathcal{O}_{QuNL}, \mathcal{O}_{uN}$           \\
                 &      $ - \frac{\alpha_{QN}}{\Lambda^{2}}\kappa^{T}\tilde{\theta}^{\dagger}\left(\overline{u}_{m}\gamma^{\mu}P_{L}u_{m}\right)\left(\overline{N}_{m}\gamma^{\mu}P_{R}\nu_{m}\right)$     + h.c.    & $\mathcal{O}_{QN}$  \\
\hline
$\mathcal{G}^{N}_{\nu_{j}d_{\alpha}d_{\alpha}}$ & $\frac{\alpha_{LdQN}}{\Lambda^{2}}\left(U^{\dagger}\kappa^{*}\overline{\nu}_{m}P_{R}d_{m}\overline{d}_{m}P_{R}N_{m}\right) + \frac{\alpha_{LNQd}}{\Lambda^{2}}\left(U^{\dagger}\kappa^{*}\overline{\nu}_{m}P_{R}N_{m}\overline{d}_{m}P_{R}d_{m}\right)$      & $\mathcal{O}_{LNQd}, \mathcal{O}_{LdQN}$ \\
						&   $ - \frac{\alpha_{dN}}{\Lambda^{2}}\left(\kappa^{T}\tilde{\theta}^{\dagger}\overline{d}_{m}\gamma^{\mu}P_{R}d_{m}\overline{N}_{m}\gamma^{\mu}P_{R}\nu_{m}\right)$                    &      $\mathcal{O}_{dN}, \mathcal{O}_{QN}$            \\
						&      $  - \frac{\alpha_{QN}}{\Lambda^{2}}\left(\kappa^{T}\tilde{\theta}^{\dagger}\overline{d}_{m}\gamma^{\mu}P_{R}d_{m}\overline{N}_{m}\gamma^{\mu}P_{R}\nu_{m}\right)$   + h.c.       &           \\
\hline
$\mathcal{G}^{N}_{\nu_{j}\nu\nu}$                        &  $\frac{\alpha_{NNNN}}{\Lambda^{2}}\left(\tilde{\theta}^{*}\tilde{\theta}^{\dagger}\kappa^{\dagger}\tilde{\theta}^{\dagger}\right)\left(\overline{\nu}_{m}P_{R}\nu_{m}\overline{N}_{m}P_{R}\nu_{m}\right)$    & $\mathcal{O}_{NNNN}$ \\
   & $ - \frac{\alpha_{NN}}{\Lambda^{2}}\left(\tilde{\theta}\tilde{\theta}^{\dagger}\kappa^{T}\tilde{\theta}^{\dagger}\right)\left(\overline{\nu}_{m}P_{R}\nu_{m}\overline{N}_{m}P_{R}\nu_{m}\right)$ + h.c.    &           $\mathcal{O}_{NN}$                \\
\hline
\end{tabular}
\caption{The four point coupling arise from different four fermi operators. Here we have only presented those couplings which are relevant for the three body decay calculation for $N$. The other four point couplings that involve more than one heavy neutrino fields are presented in Appendix.~\ref{App:dim6}. We represent these couplings while adopting a generic structure $\mathcal{G}^{N}_{f_{1}f_{2}f_{3}}$, where $f_{1}, f_{2}$ and $f_{3}$ represent the SM fermions. The greek indices $\alpha, \beta$ and latin indices $i$, $j$, $k$ describe the underlying flavour of the quarks and leptons respectively.}
\label{tab:vertexfermi}
\end{table}

\section{Constraints on relevant $N_{R}$-EFT Parameters}
\label{Sec:bound}
The EFT operators discussed above can contribute to the processes that has already been searched at the LHC and hence receive constraints from these experimental searches. Further these operators leads to BSM decay modes of the SM particles and hence there are constraints from their branching ratios measurements. In the following we briefly discussed that. 
\begin{itemize}
	\item  \textbf{Constraints from decay of $\mathbf{Z}$}: 
	The Operators $\mathcal{O}_{LNB}$ and $ \mathcal{O}_{LNW}$ can enhance the decay width of $Z$ boson for non-zero  $\alpha_{LNB}$ and $ \alpha_{LNW}$ through the decay mode $Z\to \nu N$, decay width for which is 
\begin{align}
\Gamma(Z\to N\nu)=\frac{3M_Z^3 v^2}{12 \pi \Lambda^4} (c_w\alpha_{LNW}-s_w\alpha_{LNB})^2 (1- M_N^2/M_Z^2)^{3/2}.
\end{align}
The decay mode $Z\to \nu N$ and subsequent decay of $N$ to $\nu \gamma$~(which can come from operators such as $\mathcal{O}^{(5)}_{3}, \mathcal{O}_{LNW, LNB}$) leads to $Z\to 2\nu +\gamma$. There exist experimental limit on $\text{BR}(Z\to2\nu +\gamma )<3.2 \times 10^{-6}$~\cite{L3:1997exg} that restricts the values of $\alpha_{LNB}$ and $ \alpha_{LNW}$ for a given $\Lambda$. This limit has been presented in Table.~\ref{limit:decay} assuming $\alpha_{LNB}= \alpha_{LNW}$. To calculate this we consider $\text{BR}(N\to\nu \gamma)=1~(0.1)$ and $\Lambda=4~\text{TeV}, 500~\text{GeV}$. Similarly, $\mathcal{O}_{HN}$ leads to the decay mode $Z\to NN$, with decay width 
\begin{align}
\Gamma(Z\to NN)=\frac{m_Z^3 v^2 \alpha_{HN}^2}{8 \pi \Lambda^4} (1- 4M_N^2/M_Z^2)^{3/2}.
\end{align}
 Subsequent decay of $N\to \nu\gamma$ can lead to the decay mode $Z\to  2\nu+ 2\gamma$, whose branching ratio is bounded as $\text{BR}(Z\to2\nu +2\gamma )<3.1 \times 10^{-6}$~\cite{ParticleDataGroup:2018ovx}. The upper limit on $\alpha_{HN}$ obtained from this observation is presented in Table.~\ref{limit:decay}. 
	\item  \textbf{Constraints from decay of $\mathbf{h}$:} In our framework, SM Higgs can have BSM decay modes such as $h\to \nu N/NN/\nu\gamma N$ where $\nu\gamma$ arises due to $N$ decays. If $N$ is stable at the detector length scale, these decay modes lead to the invisible decay of Higgs which is constrained as $\text{Br}(h\to \text{invisible})\le0.13$~\cite{ATLAS:2020cjb}. This in turn limits the couplings $\alpha_{2}^{(5)}$, $\alpha_{LNH}$ and $\alpha_{LNB/W}$ which is given in Table.~\ref{limit:higBr}. We consider the values of these parameters as per Table.~\ref{limit:higBr} for the evaluation of cross-section in Sec.~\ref{prodxs}. 
	\item $\mathbf{pp\to \ell+N}$: RHN has been searched for at the LHC by both the CMS~\cite{CMS:2018jxx} and ATLAS~\cite{ATLAS:2022atq} collaboration through the process $pp\to W^\pm \to \ell+N$. These searches put bounds on the active-sterile neutrino mixing~($\tilde{\theta}$) as a function of RHN mass. The limit on $\tilde{\theta}$ can be translated into the limit on $\sigma(pp\to W^\pm\to \ell^\pm+N)$. In EFT framework the production of $N+\ell^\pm$  can occur dominantly via 4-fermion interaction~($pp\to \ell+N$) \cite{Beltran:2021hpq} along with the s-channel $W$-mediated process. As the kinematic feature of $pp\to \ell+N$ is not similar to that of $pp\to W^\pm \to \ell^\pm+N$ for all RHN mass, the above-mentioned constraints can not be used in our case, without proper recasting of the CMS analysis. We check that the $p_T(\ell)$ distributions for both the processes are similar for $M_N\simeq800$ GeV. However, for  $M_N<800$ GeV, the distributions are different and hence inappropriate to apply the limit directly. Without going into the detail recasting, we comment on the bounds for $M_N\ge800$ GeV. The $95\%$ C.L. limit on active-sterile mixing $\tilde{\theta}\le0.387$ for $M_N=800$ GeV~\cite{CMS:2018jxx} leads to the limit on cross-section $\sigma(pp \to \ell^\pm+N)\le0.8$ fb, which can be translated to the limit on the relevant couplings. For $M_N\ge800$ GeV, the total cross-section is dominated by the 4-fermion interaction which is mainly controlled by $\alpha_{duNe},\alpha_{LdQN},\alpha_{LNQd} ~\text{and}~\alpha_{QuNL}$. We assume all these couplings equal to be  $\alpha$. Considering $\sigma(pp \to \ell^\pm+N) \le 0.8$ fb for $M_N\simeq800$ GeV, we obtain constraint on  $\alpha \le 0.28(0.04)$ for $\Lambda=4 \ \text{TeV} \ (1.5 \ \text{TeV})$. As this is the strongest constraint on $\alpha_{duNe},\alpha_{LdQN},\alpha_{LNQd} ~\text{and}~\alpha_{QuNL}$, we consider this for the cross-section calculation in Sec.~\ref{prodxs}.	
	\item $\mathbf{pp\to \ell+ \gamma + MET}$: Production of RHN with one associated lepton occurs via two processes, one is the 4-fermion interaction and the other $pp\to W\to N\ell$ via s-channel $W$-mediated process. The decay mode of RHN $N\to \nu \gamma$ leads to the final state $ \ell+ \gamma + MET$ which has been searched by the CMS collaboration~\cite{CMS:2018fon}. The couplings $\alpha_{duNe},\alpha_{LdQN},\alpha_{LNQd},$ $\alpha_{QuNL}$, $\alpha_{LNW}$ and $\alpha_{HNe}$ that are involved in the above process receive constraints from the CMS result. This search has been recasted in~\cite{Biekotter:2020tbd} and limits has been set on the four-fermi operators. 
	Using the $95\%$ C.L. limit on BSM events $\le9.7$ calculated in Ref.~\cite{Biekotter:2020tbd}, we calculate constraints on the  coefficient ($\alpha_{duNe},\alpha_{LdQN},\alpha_{LNQd},\alpha_{QuNL}$) assuming all equal to $\alpha$ and BR($N\to \nu+\gamma$)=0.1. For $M_N=800$ GeV and $\sigma(pp \to \ell N)=9.7$ fb, we obtain $\alpha\le0.5 (0.07)$ for $\Lambda=4 \ \text{TeV} \ (1.5 \ \text{TeV})$. For $M_N=200$ GeV and $\sigma(pp \to \ell N)=30$ fb, we obtain $\alpha\le0.28 (0.04)$ for $\Lambda=4 \ \text{TeV} \ (1.5 \ \text{TeV})$. For these masses total cross-section is dominated by the 4-fermion interaction and the contribution from the $W$-mediated channel can be ignored. 
	\item $\mathbf{pp\to 2\gamma + MET}$: This signature can be obtained via the process $pp\to NN$ and subsequent decay $N\to\nu+\gamma$. Both 4-fermion interaction and higgs production via gluon fusion process lead to pair of RHN. For $M_N\ge100$ GeV, the RHN production occurs primarily via 4-fermion interaction. The \emph{Wilson} co-efficients that are involved in this process are $\alpha_{dN}, \alpha_{QN},\alpha_{uN},\alpha_{LNQd}, \alpha_{LdQN}, \alpha_{QuNL}$, which we assume to be equal. However, one must note that contributions coming from the respective operators can not be treated equally. From Eq.~\ref{Eq:OLNQd}, Eq.~\ref{Eq:OLdQN} and Eq.~\ref{Eq:OQuNL} one can notice that the relevant coupling coming from the operators $\mathcal{O}_{LNQd}, \mathcal{O}_{LdQN}$ and $\mathcal{O}_{QuNL}$ receives additional $\tilde{\theta}^{2}$ suppression. Hence we can ignore their individual effect for present constraint calculation. There exists a search by the CMS collaboration for $2\gamma + MET$ signature~\cite{CMS:2019vzo}, which sets limits on the BSM contribution to the $2\gamma + MET$ events. We adopt the $95\%$ C.L. limit on observed $ 2\gamma + MET$ events $\le9.6$ for $\mathcal{L}=35.9/\text{fb}$, from the Ref.~\cite{Biekotter:2020tbd}, where the CMS analysis has been recasted. Considering this limit we obtain $\alpha\le1.49 (0.209)$ for $\Lambda=4 \ \text{TeV} \ (1.5 \ \text{TeV})$ assuming $M_N\ge800$ GeV and BR($N\to \nu+\gamma$)=0.1.
	\item $\mathbf{pp\to \nu +N}$: Production of one RHN in the process $pp\to \nu +N$ leads to $\gamma+MET$ signature for $N\to\nu+\gamma$ decay mode. This process involves Drell-Yan production~($pp\to \gamma^{*} \to\nu +N$), Higgs production~($pp\to h\to\nu +N$) and via four-fermi~($pp\to \nu +N$) interaction. The process $pp\to h\to\nu +N(\to \nu+\gamma)$ is not constrainted as the photon is comparatively soft than for the process $pp\to \gamma\to\nu +N$~\cite{Butterworth:2019iff}. The relevant couplings $\alpha_{dN},\alpha_{QN},\alpha_{uN},\alpha_{LNQd}, \alpha_{LdQN}, \alpha_{QuNL}$ and $\alpha_{LNW/B}$ are assume to be equal. 
	The CMS analysis~\cite{CMS:2018ffd} for the $\gamma+MET$ signature sets limits on the BSM contribution.
	We consider $95\%$ C.L. limit on observed $ \gamma + MET$ events $\le16$ from Ref.~\cite{Butterworth:2019iff}, where the CMS analysis has been recasted. We obtain $\alpha\le8.67(1.28)$ for $\Lambda=4 \ \text{TeV} \ (1.5 \ \text{TeV})$ assuming $M_N\ge800$ GeV and BR($N\to \nu+\gamma$)=0.1. For the considered mass leading contribution to the $\sigma(pp\to \nu +N)$  is from four-fermi operators that involve only one RHN~($\alpha_{LNQd}, \alpha_{LdQN}, \alpha_{QuNL}$).  
	
\end{itemize}
In view of the above discussion, to estimate the production rate in the next section we consider the values of the Wilsonian coefficient as follows. For $\Lambda=4~(1.5)$ TeV, we assume $\alpha_{HN}=1~(1)$,  $\alpha_{HNe}=1~(1)$,  $\alpha_{LNH}=0.1~(0.04)$,$\alpha_{LNW}=0.1~(0.04)$, $\alpha_{LNB}=0.1~(0.04)$, $\alpha_{2}^{(5)}=0.1~(0.04)$, $\alpha_{dN}=\alpha_{QN}=\alpha_{uN}=\alpha_{LNQd}= \alpha_{LdQN}=0.5~(0.04)$.

\begin{table}[h!]
	\centering
	\begin{tabular}{|c||c|c|c|}
		\hline
		& $\Lambda=4$ TeV & $\Lambda=1.5$ TeV &$\Lambda=500$ GeV  \\
		\hline
		$\mathcal{B}(Z\to \nu N\to2\nu + \gamma)$ & $\alpha_{LNB/W}\le1.88 \ (5.9)$  &$\alpha_{LNB/W}\le0.26 \ (0.84)$& $\alpha_{LNB/W}\le0.029 \ (0.09)$ \\
		\hline
		$\mathcal{B}(Z\to NN\to2\nu + 2\gamma)$ &$\alpha_{HN} \le1.04 \ (10.4)$ &$\alpha_{HN}\le0.14 \ (1.4)$ &$\alpha_{HN} \le0.01 \ (0.16)$ \\
		\hline
	\end{tabular}
	\caption{Constraints on $\alpha_{LNB/W}$ and $\alpha_{HN}$ from $Z-$width measurement~\cite{L3:1997exg,ParticleDataGroup:2018ovx} for BR($N\to\nu \gamma$)=1 (0.1).}\label{limit:decay}
\end{table}
\begin{table}[h!]
	\centering
	\begin{tabular}{|c||c|c|c|}
		\hline
		($\alpha_{2}^{(5)}$, $ \alpha_{LNH}$, $\alpha_{LNB/W}$)	& $\Lambda=4$ TeV&$\Lambda=1.5$ TeV &  $\Lambda=500$ GeV  \\
		\hline
		$(c,c,c)$ & $c\le0.115 $ &$c\le0.043$ & $c\le0.014$ \\
		\hline
		$(0,c,c)$ &$c\le1.93$ &$c\le0.5$ &$c \le0.242$ \\
		\hline
		$(c,0,c)$ &$c\le0.115$ & $c\le0.043$&$c \le0.0144$ \\	\hline
	\end{tabular}
	\caption{Constraints on $\alpha_2^5$, $ \alpha_{LNH}$ and $\alpha_{LNB/W}$ and  from invisible Higgs decay~\cite{ATLAS:2020cjb}.}\label{limit:higBr}
\end{table}
\section{Possible Production Mechanism}\label{prodxs}
In order to calculate the cross section, we build this dimension six $N_{R}$-EFT using FeynRules(v2.3) \cite{Alloul20142250} and generate corresponding UFO file. This UFO file can then use to evaluate the parton level cross section via Monte-Carlo simulator MadGraph5\_aMC@NLO(v2.6) \cite{Alwall:2014hca}.      

\subsection{Proton Proton Collider}

The LHC is the machine which has the capability to probe the physics that possibly lies at high energies. Currently the LHC is going through an upgradation and after that it will run at the centre of mass energy $\sqrt{s}$ = 14 TeV with a higher luminosity which will achieve the potential to collect 3000$~\text{fb}^{-1}$ data by the year 2030. The detailed plan for the high-luminosity LHC (HL-LHC) is presented in Ref.~\cite{Henderson:2021qrr,Nielsen:2020uvd,Gustavino:2019oxp}. In this paper, we will use 14 TeV LHC to propose various production processes for the RHNs.     

\begin{itemize}
	\item $\bf{gg \to h\to N~N/\nu}$
\end{itemize}
At LHC, SM Higgs boson is dominantly produced via gluon gluon fusion. The RHN field couples to the Higgs via Yukawa term at the renormalisable level. This coupling receives extra contribution via the operator $\mathcal{O}^{(5)}_{1}$, $\mathcal{O}^{(5)}_{2}$ and $\mathcal{O}_{LNH}$ at dimension five and dimension six respectively. These operators along with the $d = 4$ term allow the Higgs to decay into $\nu N$ and $NN$ modes if kinematically admissible.

\begin{figure}[t]
	\subfigure[]{\raisebox{20mm}{\includegraphics[width=0.45\textwidth]{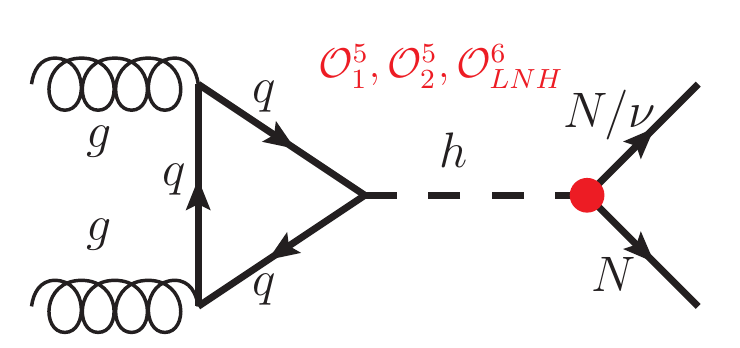}\label{Fig:ggFh}}	}
	\hspace{0.1cm}
	\subfigure[]{\includegraphics[width=0.45\textwidth,height=0.3\textheight]{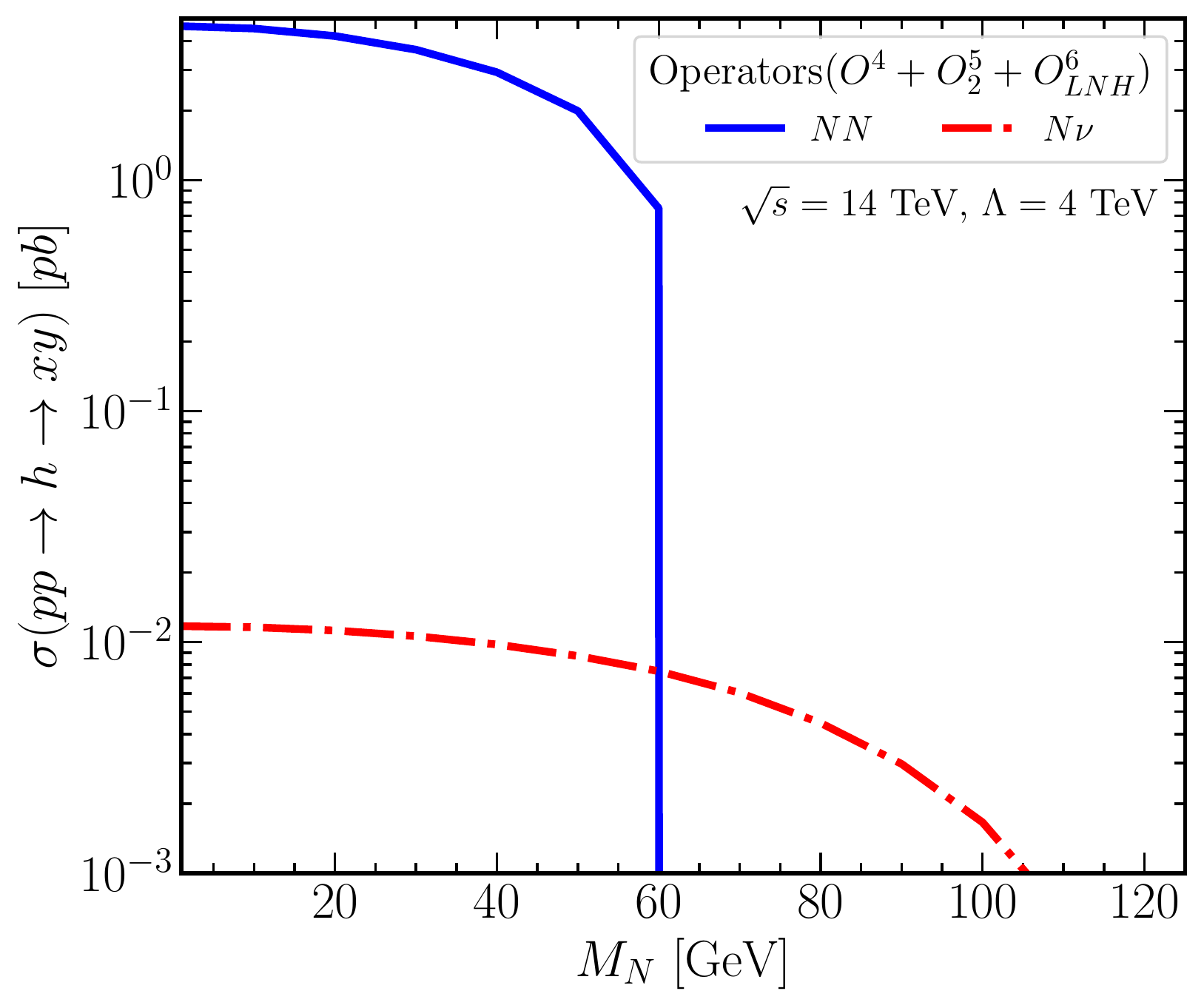}\label{fig:gghNN}}	
	\caption{Left panel: The Feynman diagram and the relevant operator for the process $pp\to NN/\nu$. Right panel: The variation of cross section for the process $\sigma(gg \to h\to NN/\nu)$  with $M_N$ for center of mass energy $\sqrt{s}=14$ TeV and cut-off scale $\Lambda=4$ TeV. $\sigma(gg \to h\to NN)$ is shown in blue solid line and $\sigma(gg \to h\to N\nu)$ is shown by the red dashed-dot line. See text for details.}
\end{figure}

In Fig[\ref{Fig:ggFh}] we present the Feynman diagram for this process. The vertices $\mathcal{C}^{6}_{h(\overline{\nu}N + \overline{N}\nu)}$ and $\mathcal{C}^{6}_{h\overline{N}N}$  are governed by the EFT parameters $\frac{\alpha^{(5)}_{1}}{\Lambda}$, $\frac{\alpha^{(5)}_{2}}{\Lambda}$, $\frac{\alpha_{LNH}}{\Lambda^{2}}$ along with the mixing angle. The smallness of the neutrino mass forces us to fix the $\alpha^{(5)}_{1}$ at zero. As mentioned in Section.~\ref{Sec:bound} the coupling $\alpha^{(5)}_{2}$ and $\alpha_{LNH}$ are considered to be 0.1 for $\Lambda=4$ TeV while respecting the current experimental bounds. To calculate the production cross section we have considered  $\tilde{\theta}=10^{-3}$ which is allowed by the recent electroweak precision data \cite{delAguila:2008pw}.
In Fig.~[\ref{fig:gghNN}], we present the production cross section for the process $gg \to h\to NN$ by the blue-solid line. As expected the cross section steeply falls at the mass around $\frac{m_{h}}{2}$ beyond which the Higgs decaying to a pair of on-shell $N$ fields is kinematically disallowed. At renormalizable level~($\mathcal{O}^{4}$) the coupling under consideration is controlled by the mixing angle $\tilde{\theta}$ and the smallness of this parameter leads to negligible cross-section. 
The result significantly alters once we include the operator $\mathcal{O}^{(5)}_{2}$ that enhanses the total cross-section upto $\mathcal{O}(10\,\text{pb})$ order.
\noindent
The cross-section of this process does not substantial change once we include the contribution coming from dimension six operator as this contributions depends on both the mixing angle and $\frac{\alpha_{LNH}}{\Lambda^{2}}$. We represent the cross section for $gg \to h \to N\nu$ in Fig.~\ref{fig:gghNN} by the red dashed-dotted line. Interestingly, the cross section due to the $\mathcal{O}^{4}$ term is comparable with the effect coming from $\mathcal{O}^{(5)}_{2}$. On the other hand the contribution coming from dimension six operator is independent of the mixing angle and enhances the total cross section by an order of magnitude. We would like to point out that the difference in the cross section between the $h \to NN$ and $h \to \nu N$ also arises due to the difference in associated kinematic factors.    

\begin{figure}
	\centering
	\subfigure[]{\raisebox{15mm}{\includegraphics[width=0.3\textwidth]{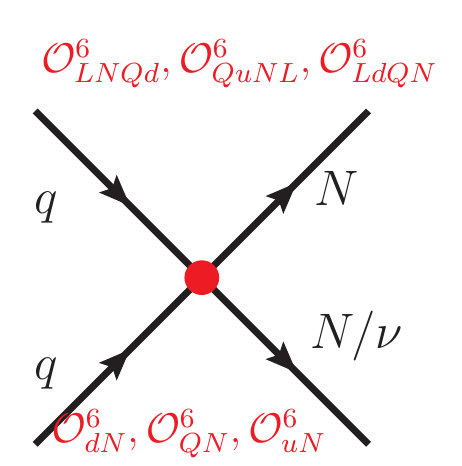}\label{fig:Feyn4fermipp}}}	
	\subfigure[]{\includegraphics[height=0.3\textheight]{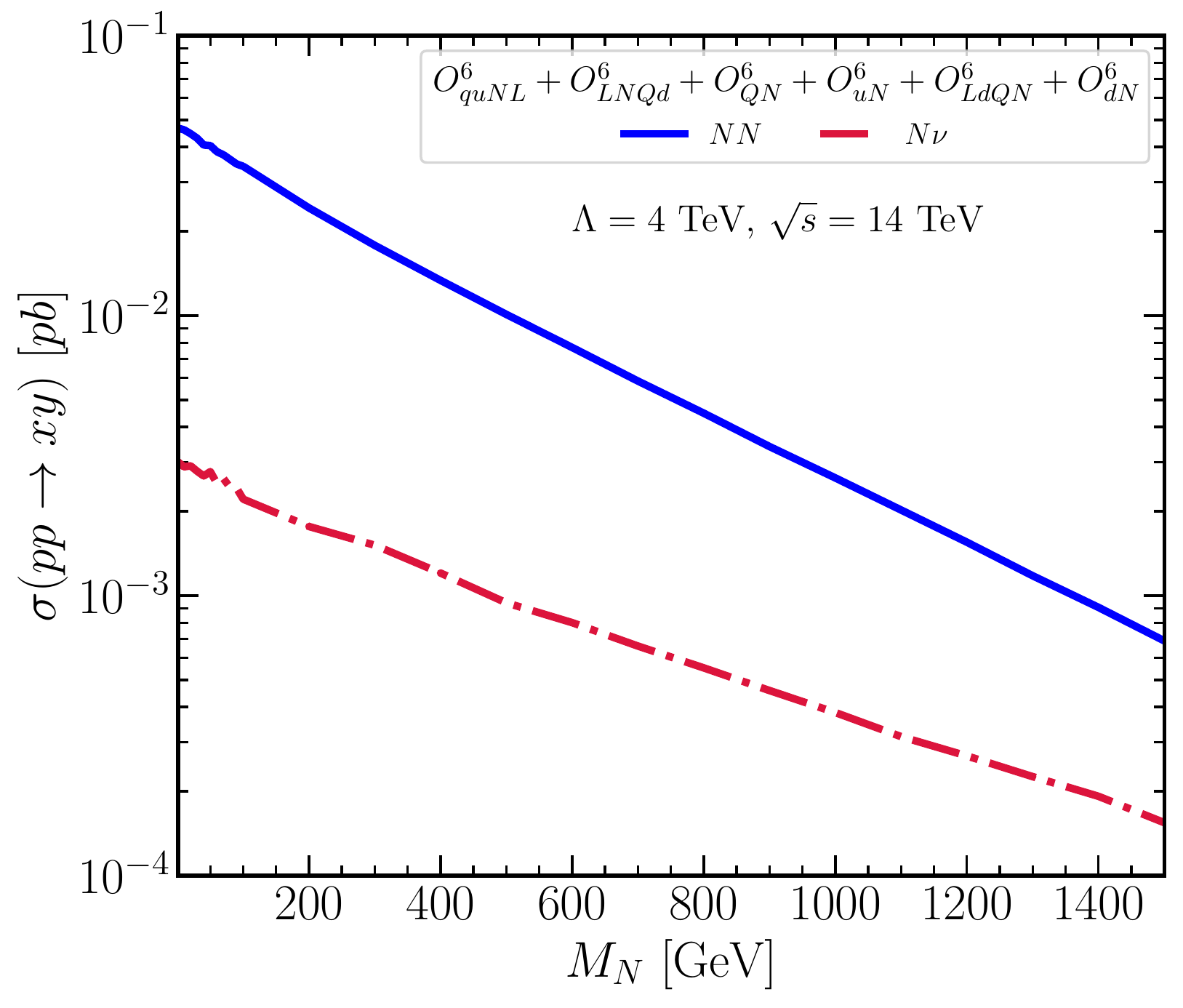}\label{fig:4fermipp}}
	\caption{Left panel: The Feynman diagram and the relevant operators for the process $pp\to NN/\nu$. Right panel: The variation of cross-section with $M_N$ for center of mass energy $\sqrt{s}=14$ TeV and cut-off scale $\Lambda=4$ TeV. $\sigma(pp \to NN)$ is shown in blue solid line and $\sigma(pp \to N\nu)$ is shown by the red dashed-dot line.}
\end{figure}

\begin{itemize}
	\item$\bf{pp \to NN}$ \textbf{via Four Fermi Operators}
\end{itemize}
We have discussed the pair production of the heavy neutrinos via Higgs decay. The problem with this production mode is one can only probe a certain range of $M_{N}$ owing to phase space suppression. In dimension six one can construct various four fermi operators which can produce these heavy neutral leptons with an appreciable cross section for a wider mass range of $M_{N}$. In proton proton collider these operators are $\mathcal{O}_{QN}$, $\mathcal{O}_{uN}$, $\mathcal{O}_{dN}$ and $\mathcal{O}_{QuNL}$ which can produce single as well as pair of RHN fields \cite{Cottin:2021lzz}. In Fig.~[\ref{fig:Feyn4fermipp}], we present the corresponding Feynman diagram of this process. To explicitly see how different operators participate in these process, we refer the readers to look into Appendix.~[\ref{App:Dim5Couplings}]. The contributions from $\mathcal{O}_{QN}, \mathcal{O}_{uN}~ \text{and} ~\mathcal{O}_{dN}$ are independent of the mixing angle. Contrary to that, the cross-section that is generated from $\mathcal{O}_{QuNL}$ depends on $\tilde{\theta}$ and hence will be suppressed.    

In Fig.~[\ref{fig:4fermipp}], blue line  represents the pair production cross section generated via four fermi operators at 14 TeV LHC. Lower parton density at high energy leads to decrease in cross-section as $M_N$ increases. The contribution of the $\mathcal{O}_{uN}$ is larger than the $\mathcal{O}_{dN}$ due to the difference in their corresponding pdf. Moreover the total cross-section of this process is primarily governed by the $\mathcal{O}_{QN}$ as it involves both $u$ and $d$ quarks in the initial state. The operators~($\mathcal{O}_{QuNL},~\mathcal{O}_{LNQd},\mathcal{O}_{LdQN}$) involving one RHN do not contribute substantially due to mixing suppression. 
We also present the cross section associated with the single $N$ by the red dashed-dotted line, where the operators~($\mathcal{O}_{QuNL},~\mathcal{O}_{LNQd},\mathcal{O}_{LdQN}$) involving one RHN are dominant. 
\begin{itemize}
	\item \textbf{RHN Production via Vector Boson Fusion Process}
\end{itemize}

The Vector Boson Fusion (VBF) process remains one of the intriguing channels which one can study at the LHC. Here we specify few VBF signals which are relevant for the $N$ production.  Apart from being one of the \emph{golden} channel for the Higgs discovery, VBF provides an excellent window to look for the unitarity of the underlying EWSB mechanism. Along with the Higgs, the $W$ and $Z$ can also be produced via this process. The large pseudo-rapidity ($\Delta\eta$) between the two leading jets helps to devise suitable cuts that can provide a relatively cleaner environment to search for the heavy neutrinos. 
\begin{itemize}
	\item[] $\bf{p~p \to h~j~j \to N~N/\nu~j~j}$
\end{itemize}

\begin{figure}
	\subfigure[]{\raisebox{20mm}{\includegraphics[width=0.45\textwidth]{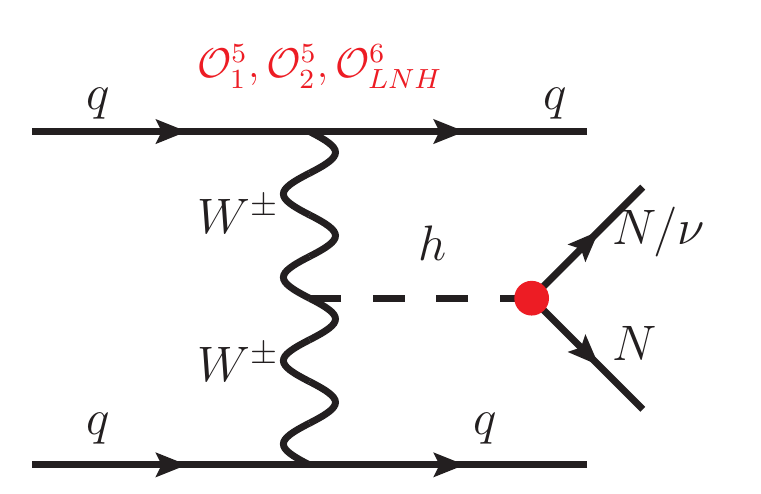}\label{fig:FeynVbfH}}}
	\subfigure[]{\includegraphics[height=0.3\textheight]{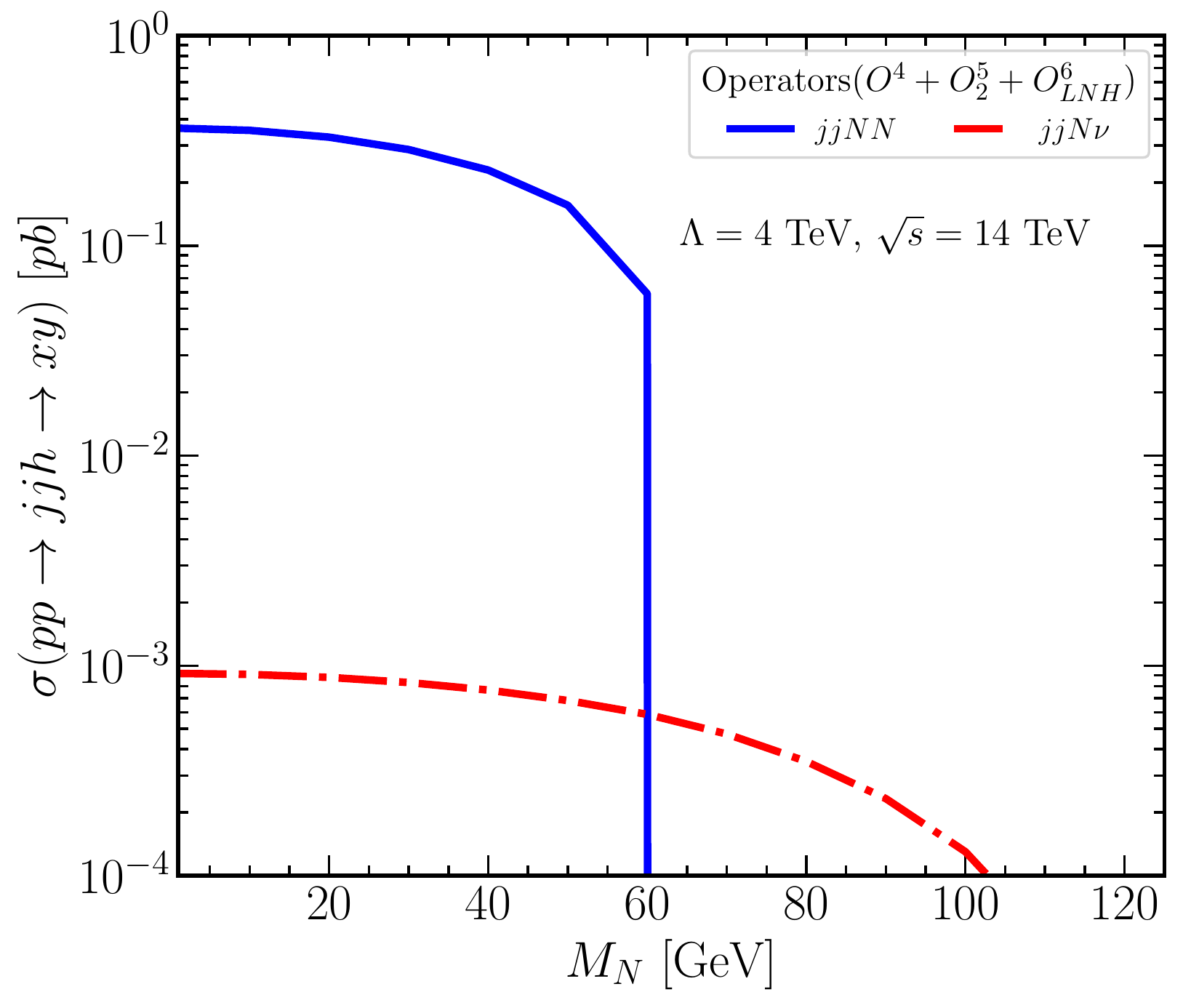}\label{fig:ppvbfhNN}}
	\caption{Left panel: The Feynman diagram with the corresponding relevant EFT operators, that contribute to the processes $pp \to NNjj$ and $pp \to N\nu j j$. Right panel: the variation of cross-section with $M_N$.  $\sigma(pp \to jj NN)$ is shown in blue solid line and $\sigma(pp \to jj N\nu)$ is shown by the red dashed-dot line.}
\end{figure}

In Fig.~[\ref{fig:FeynVbfH}], we show the Feynman digram for single and pair production of the sterile neutrinos via VBF channel. The operator dependence remain same as the gluon gluon fusion scenario. In Fig.~[\ref{fig:ppvbfhNN}] we display the cross section of $pp \to NNjj$ and $pp \to N\nu j j$ by the blue solid line and red dashed-dotted line, respectively. Although the operator dependence remains same as of Fig.~[\ref{fig:gghNN}], the overall yield of the VBF Higgs production is lower than the ggF Higgs production. This is reflected in the total cross section for this processes. The combine measurements in both these channels facilitate us to put suitable bounds on the EFT parameters.          

\begin{figure}[h!]
	\subfigure[]{\raisebox{20mm}{\includegraphics[width=0.45\textwidth]{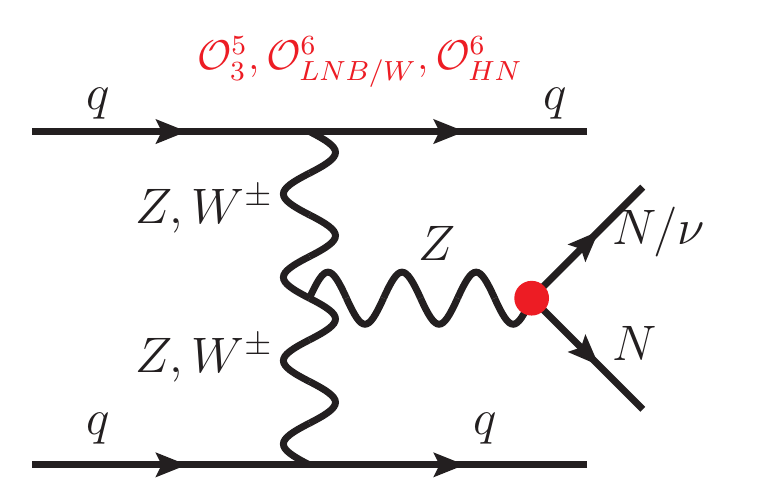}	\label{fig:vbfZpp}}}
	\subfigure[]{\includegraphics[width=0.5\textwidth,height=0.3\textheight]{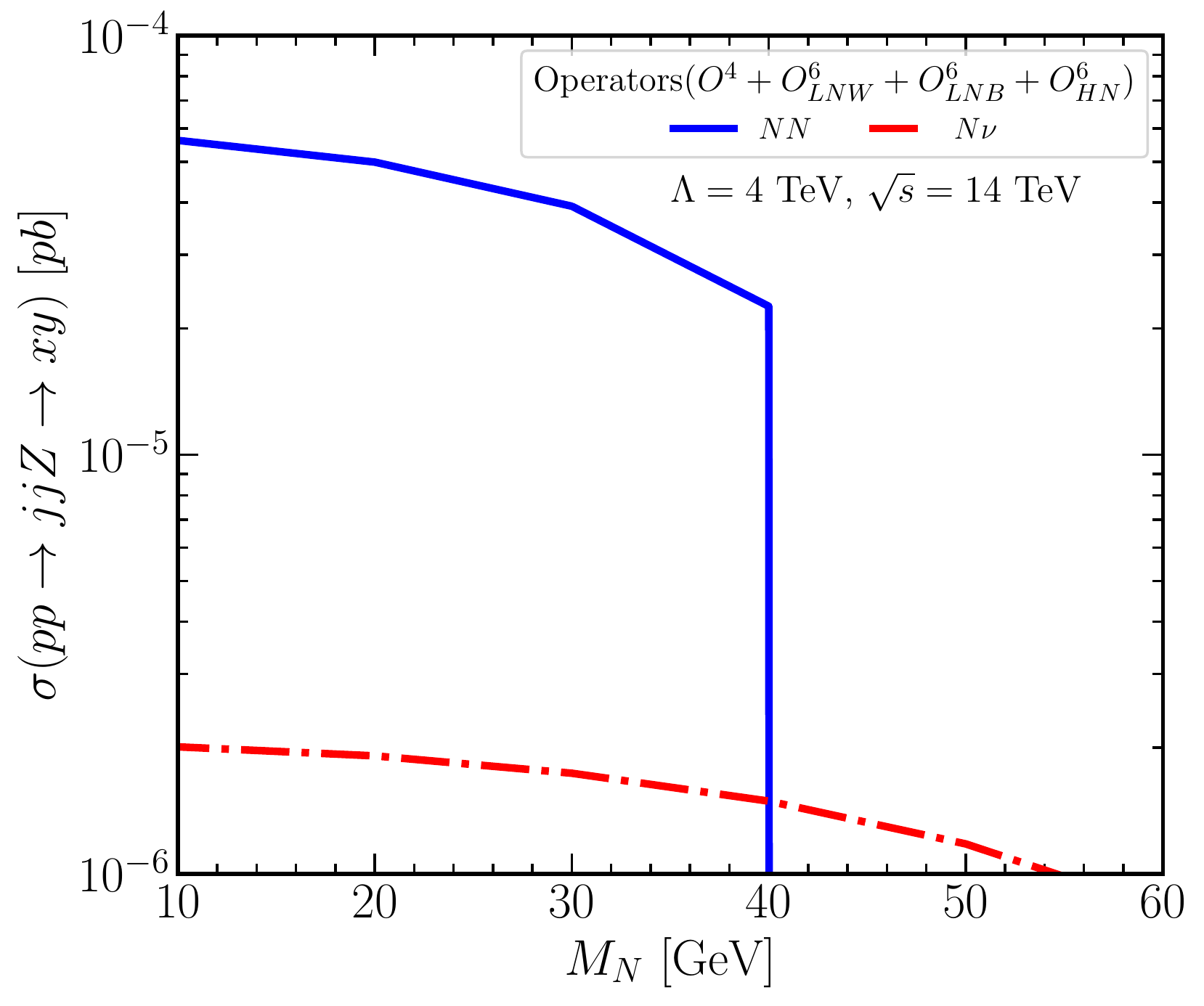}\label{fig:vbfzpp}}
	\caption{Left panel shows the Feynman diagram with the relevant contributing EFT operators for the VBF process $pp \to jj N\nu/ jj NN$. Right panel shows the variation of cross-section with heavy neutrino mass $M_N$.  The blue and red line represent $\sigma(pp \to jj NN)$ and $\sigma(pp \to jj N\nu)$ respctively.}
\end{figure}

\begin{itemize}
	\item[] $\bf{p~p \to Z~j~j \to N~N/\nu~j~j}$
\end{itemize}
Apart from the Higgs boson, the $Z$ boson can also be produced in VBF mode and it's decay can give rise to identical final states \emph{i.e} $jjNN$ and $jjN\nu$. The Feynman diagram corresponding to these processes are shown in Fig.~[\ref{fig:vbfZpp}] where we mark the EFT operators that contribute to these processes. Along with the renormalisable neutral current, dimension five operator $\mathcal{O}^{(5)}_{3}$, dimension six operators such as $\mathcal{O}_{NH}$, $\mathcal{O}_{LNW}$ and $\mathcal{O}_{LNB}$ participate in these processes. However, the operator $\mathcal{O}^{(5)}_{3}$ is in general antisymmetric in the flavour space which vanishes if we consider $Z$ boson decay into the same flavour leptons. \\ 

In Fig.~[\ref{fig:vbfzpp}], we demonstrate the corresponding cross sections. In case of $Z \to N N$, shown by the blue line, the cross section sharply falls down at the mass $\frac{M_{Z}}{2}$ after which the channel becomes kinematically forbidden. At dimension four the coupling $\mathcal{C}^{Z}_{\overline{N}N}$ is primarily controlled by the quadratic power of mixing angle which is chosen to be $10^{-3}$.
Hence, with the renormalizable dimension-4 coupling, the cross-section is highly suppressed $\mathcal{O}(10^{-18})$ pb. For $\mathcal{O}_{LNW}$ and $\mathcal{O}_{LNB}$, the $Z$ boson couples to heavy neutrino pair via the gauge state $W^{3}_{\mu}$ and $B_{\mu}$, respectively (see Eq.~\ref{Eq:ONB} and Eq.~\ref{Eq:ONW0} for details). From Table.~\ref{tab:vertex6} one can observe a relative minus sign between the \emph{Wilson} co-efficient correspond to these operators. This leads to destructive interference if we combine their effects in the cross section calculation.  
We like to point out, the cross section is larger if we consider  only $\mathcal{O}_{LNB}$ or $\mathcal{O}_{LNW}$ instead of both $\mathcal{O}_{LNW}$ and $\mathcal{O}_{LNB}$ together. 
The $\mathcal{O}_{HN}$ operator significantly enhance the number of signal production as the vertex dependency coming from the $\mathcal{O}_{HN}$ operator is $\tilde{\theta}$ independent. The red dashed-dotted line of Fig.~[\ref{fig:vbfzpp}] highlights the cross section associated with the $\nu N j j$ final state production via $Z$ decay.
Here also, the interference between the $\mathcal{O}_{LNB}$ and $\mathcal{O}_{LNW}$ exist similar to the case of $NNjj$. Finally the inclusion of $\mathcal{O}_{HN}$ increases the total cross section, but not much as the coupling coming from $\mathcal{O}_{HN}$ also has a mixing angle dependence.

\begin{figure}
	\subfigure[]{\raisebox{20mm}{\includegraphics[width=0.45\textwidth]{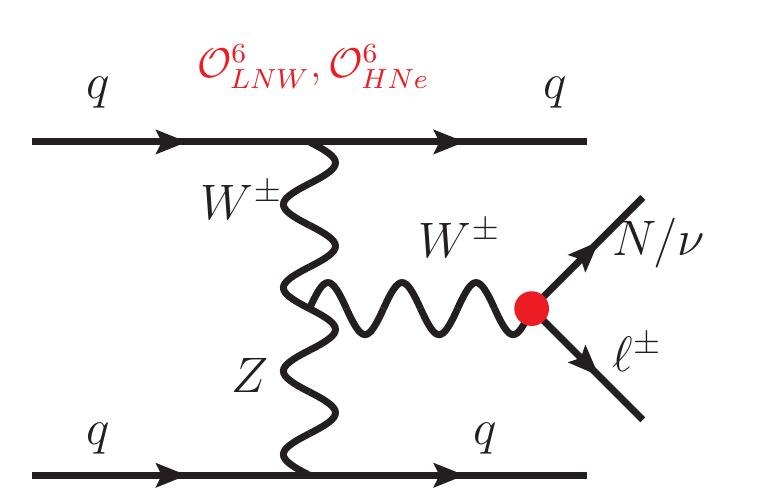}\label{Fig:feynVBFW}}}
	\subfigure[]{\includegraphics[width=0.5\textwidth,height=0.3\textheight]{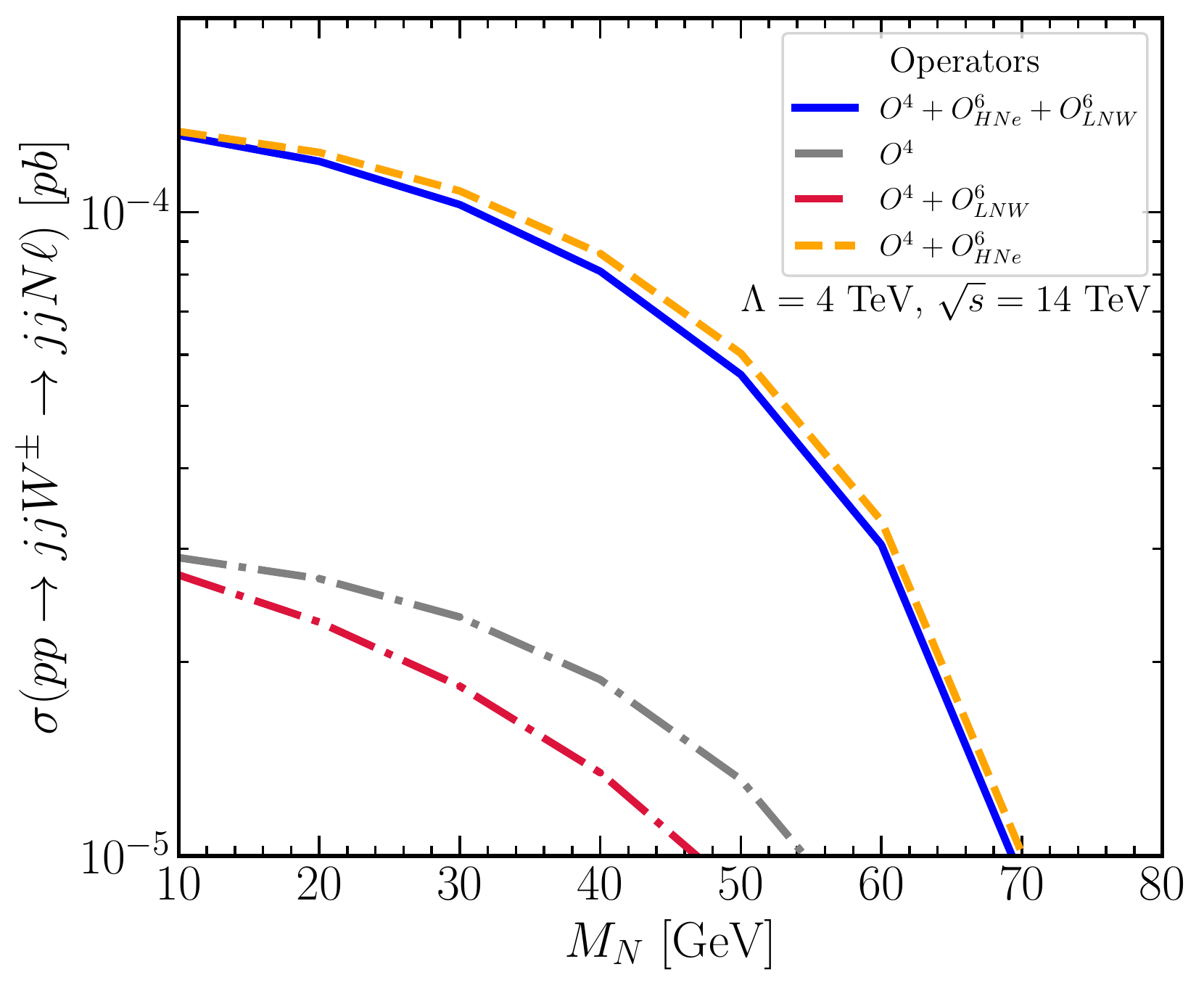}\label{Fig:ppVBFW}}
	\caption{Left panel shows the Feynman diagram with the relevant contributing operators for the VBF process $pp \to \ell^\pm N jj$. The right panel stands for variation of $\sigma(pp \to W^\pm jj\to \ell^\pm N jj)$ with $M_N$. The grey dot dashed, red dot dashed, orange dashed and thick blue line stands for the contribution to this cross section coming from mixing only, mixing+$\mathcal{O}_{LNW}$, mixing+$\mathcal{O}_{HNe}$ and combining all the operators.}	
\end{figure}

\begin{itemize}
	\item[] $\bf{pp \to W j j \to \ell N j j}$
\end{itemize}
The $W$ boson is another SM field that can be produced via VBF process and leads to a single $N$ production. The advantage of this process is that the RHN is produced along with a lepton which can be used to demarcate the signal from pure QCD events which in general appear during hadron collision. In Fig.~[\ref{Fig:feynVBFW}] and Fig.~[\ref{Fig:ppVBFW}], we present the Feynman digram and the associated cross section for this process respectively. At dimension four, the RHN production is only controlled by the mixing angle and for $\tilde{\theta}$ = $10^{-3}$ one cannot achieve any appreciable cross section as represented by the gray dot dashed line. As we include $\mathcal{O}_{LNW}$, the realtive sign difference between $\mathcal{O}^{4}$ and $\mathcal{O}_{LNW}$ coupling leads to a destructive interference, which is shown by the red dot dashed line.
This situation significantly improves once we consider $\mathcal{O}_{HNe}$ instead of $\mathcal{O}_{LNW}$ as shown by the orange dashed line in Fig.~[\ref{Fig:ppVBFW}]. 
The contribution of $\mathcal{O}_{LNW}$ is suppressed by a factor $1/16\pi^2$ mentioned in Subsection.~\ref{Sec:dim6}, which leads to lower contribution compared to that for the $\mathcal{O}_{HNe}$. Finaly, the total contribution from all these three operators has been indicated by the blue line. The total contribution is slightly lower than that for $\mathcal{O}_{HNe}+\mathcal{O}^{4}$ due to the destructive interference as discussed above.
\begin{itemize}
	\item \textbf{Drell Yan Production Mechanism}
\end{itemize}
The Drell-Yan process can serve as a viable production mode for $N$ fields. Both the $W$ and $Z$ can be created in s-channel via parton parton collision. These s-channel heavy states will further decay and generate RHN fields. Coupling of $N$ fields to $Z$ boson is primarily regulated by the choice of mixing angle except for the $\mathcal{O}_{HN}$. Further there exists experimental limit on $\text{BR}(Z\to NN\to2\nu + 2\gamma)$ as discussed in section.~\ref{Sec:bound}, which restricts the choice of $\alpha_{HN}$. Hence the expected cross section from this mode is significantly low and cannot be used to do meaningful phenomenological analysis. On the other hand the situation is relatively better in case of $W$ boson decay.

\begin{figure}
	\subfigure[]{\includegraphics[width=0.5\textwidth]{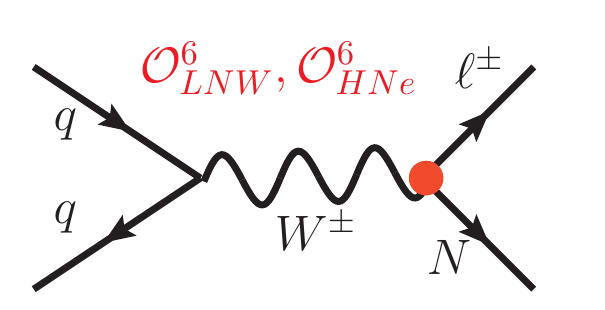}\label{fig:DYppW}}
	\hspace{1.5cm}
	\subfigure[]{\includegraphics[width=0.45\textwidth,height=0.3\textheight]{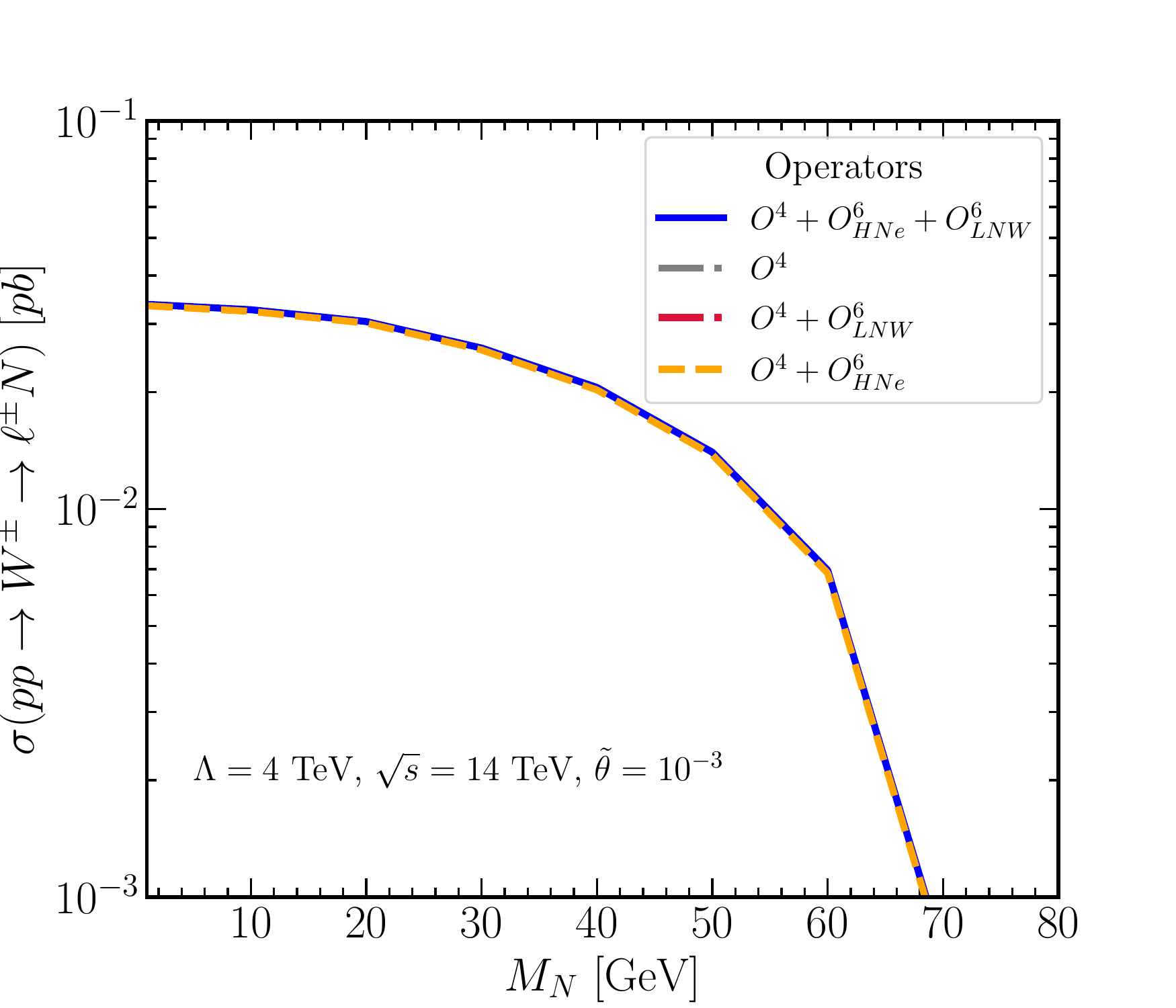}\label{fig:wlnpp}}
	\subfigure[]{\includegraphics[width=0.45\textwidth,height=0.3\textheight]{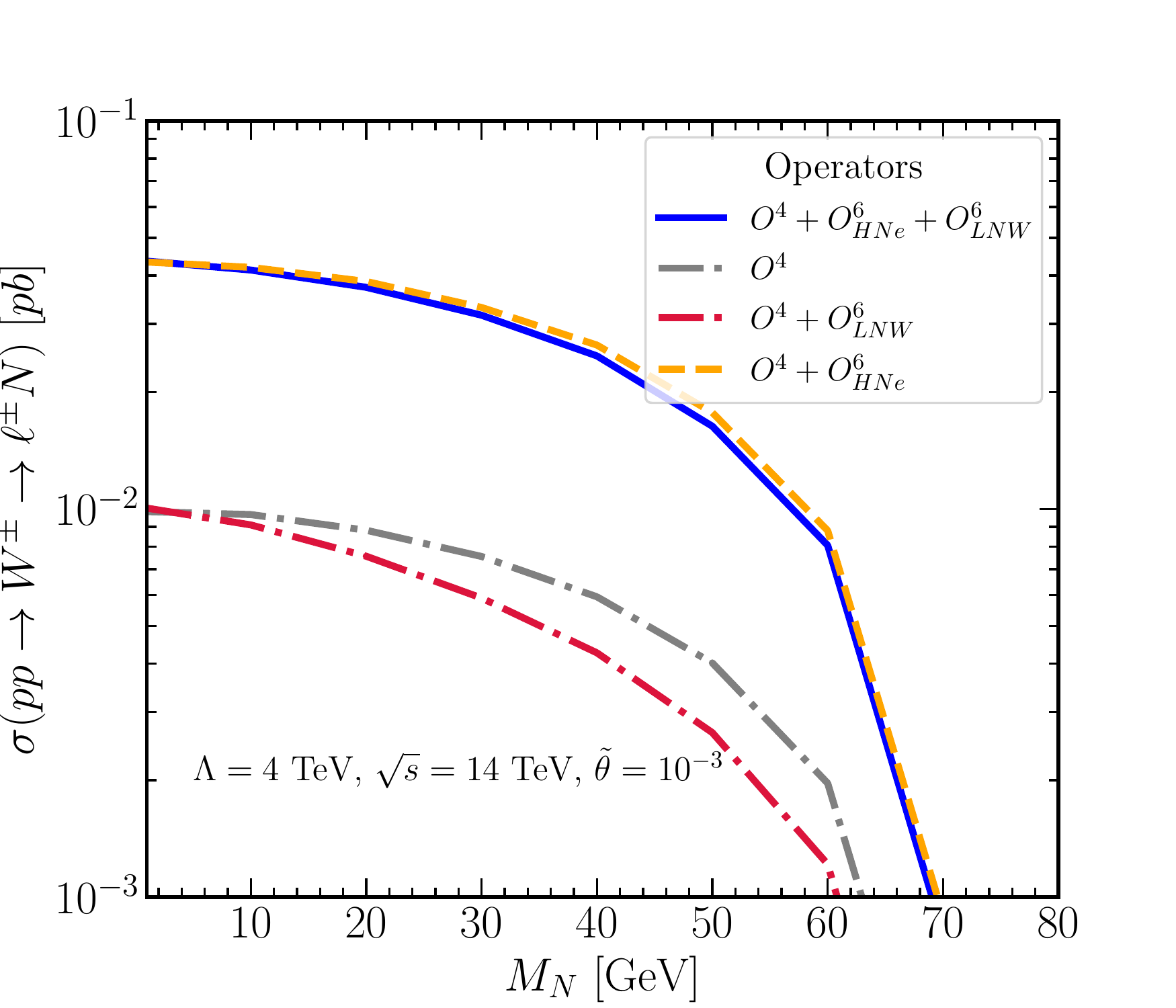}\label{fig:wlnpp3}}
	\caption{Fig.~\ref{fig:DYppW}: Feynman diagram and relevant operators for the process $pp \to W^\pm\to \ell^\pm N$. Fig.~\ref{fig:wlnpp} and Fig.~\ref{fig:wlnpp3} is variation of $\sigma(pp \to W^\pm\to \ell^\pm N)$ with $M_N$ for $\tilde{\theta}=10^{-6}$ and $10^{-3}$, respectively. In each panel the grey dot dashed, red dot dashed, orange dashed and thick blue line stands for the contribution to this cross section coming from mixing only, mixing+$\mathcal{O}_{LNW}$, mixing+$\mathcal{O}_{HNe}$ and combining all the operators.}
\end{figure}

In Fig,~[\ref{fig:DYppW}], we present the relevant Feynman diagram for $pp \to W^{*} \to \ell N$ that arises from the operators, $\mathcal{O}_{HNe}$, $\mathcal{O}_{LNW}$ as well as the $d=4$ charge current operator. The individual effects of these operators are illustrated in Fig.~[\ref{fig:wlnpp}] and Fig.~[\ref{fig:wlnpp3}]. We show the cross section for two values of $\tilde{\theta}=10^{-6}$~(Fig.~[\ref{fig:wlnpp}]), $10^{-3}$~(Fig.~[\ref{fig:wlnpp3}]).  For $\tilde{\theta}=10^{-6}$, the effect of the $d=4$ operator is suppressed. However, for $\tilde{\theta}=10^{-3}$ it has notable contribution, which enhances the cross-section as evident from Fig.~[\ref{fig:wlnpp3}]. 
As we include $\mathcal{O}_{LNW}$, the realtive sign difference between $\mathcal{O}^{4}$ and $\mathcal{O}_{LNW}$ coupling leads to a destructive interference, which is shown by the red dot dashed line. For $\tilde{\theta}=10^{-3}$ it is more prominant as the contribution of $\mathcal{O}^{4}$ become comparable to that of $\mathcal{O}_{LNW}$.
Finaly, the total contribution from all the three  operators has been indicated by the blue line. It is evident from Fig.~[\ref{fig:wlnpp}] and Fig.~[\ref{fig:wlnpp3}] that the operator $\mathcal{O}_{HNe}$ primarily controls the total cross-section. For $\tilde{\theta}=10^{-3}$, the total contribution is slightly lower than that for $\mathcal{O}_{HNe}+\mathcal{O}^{4}$ due to the relative sign difference in the effective coupling.

\subsection{Electron Proton Collider}
\begin{figure}[t]
		\centering
		\subfigure[]{\includegraphics[width=0.35\textwidth]{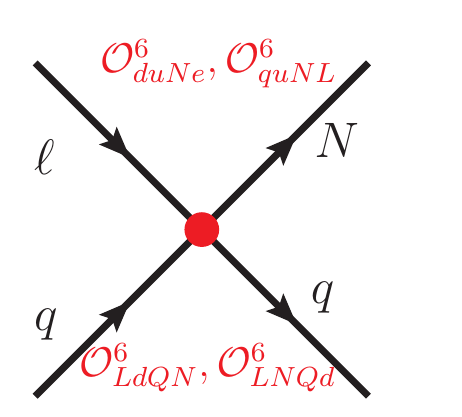}\label{fig:epjn4f}}
		\hspace{0.25cm}
		\subfigure[]{	\includegraphics[width=0.35\textwidth]{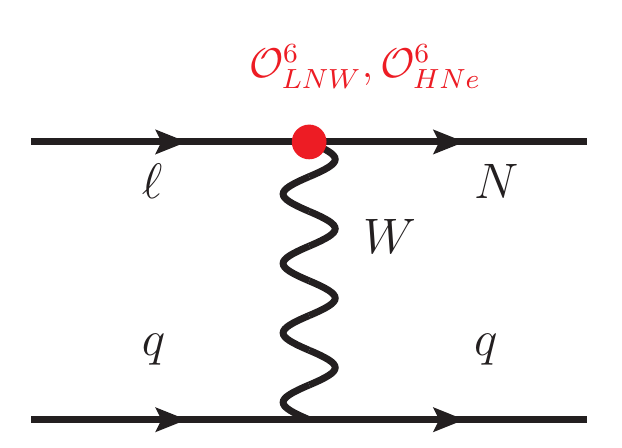}\label{fig:epjnw}}
		\subfigure[]{	\includegraphics[width=0.45\textwidth,height=0.3\textheight]{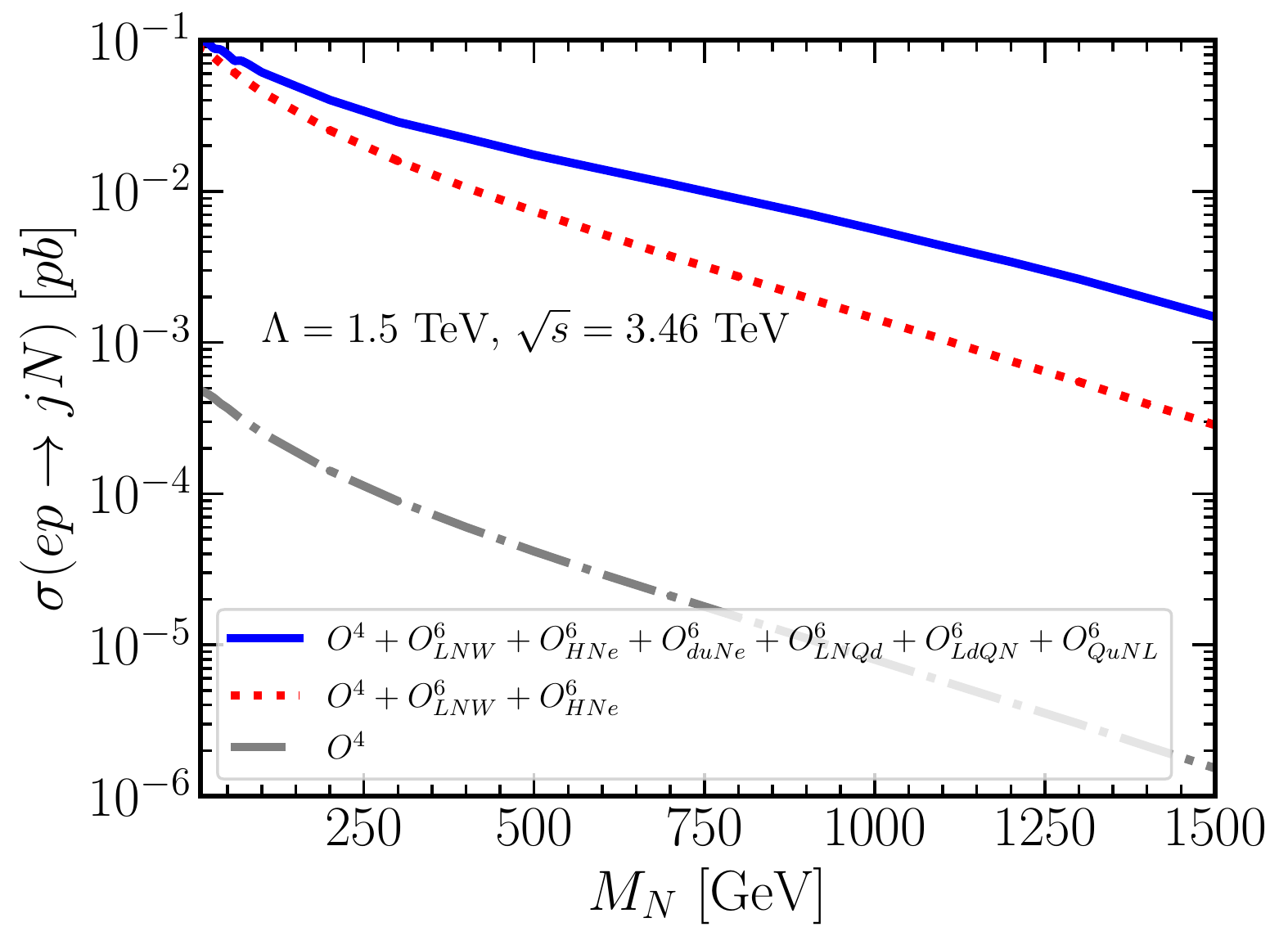}\label{fig:epjn}}
		\caption{Upper panel: The Feynman diagram  and the relevant operators for the process $e^-p\to j N$. Bottom panel: Variation of cross-section for the process $e^-p\to jN$ as function of sterile neutrino mass $M_N$. The grey dot dashed, red dotted and blue thick line stands for the contribution to the cross section coming from mixing only, mixing+$\mathcal{O}_{LNW}+\mathcal{O}_{HNe}$ and combination of all the operators, respectively.}
	\end{figure}

\begin{figure}[t]
		\subfigure[]{\includegraphics[width=0.45\textwidth]{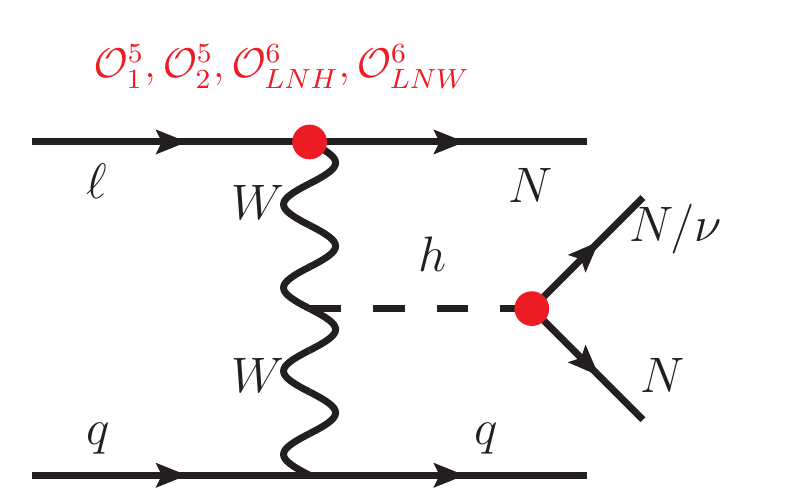}\label{fig:eph}}
		\hspace{0.75cm}
		\subfigure[]{\includegraphics[width=0.45\textwidth]{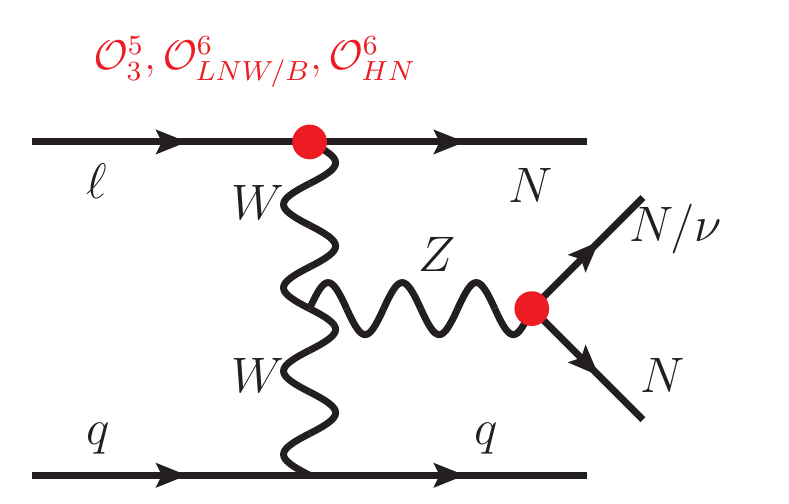}\label{fig:epz}}
		\hspace{1.5cm}	
		\subfigure[]{\includegraphics[width=0.45\textwidth,height=0.3\textheight]{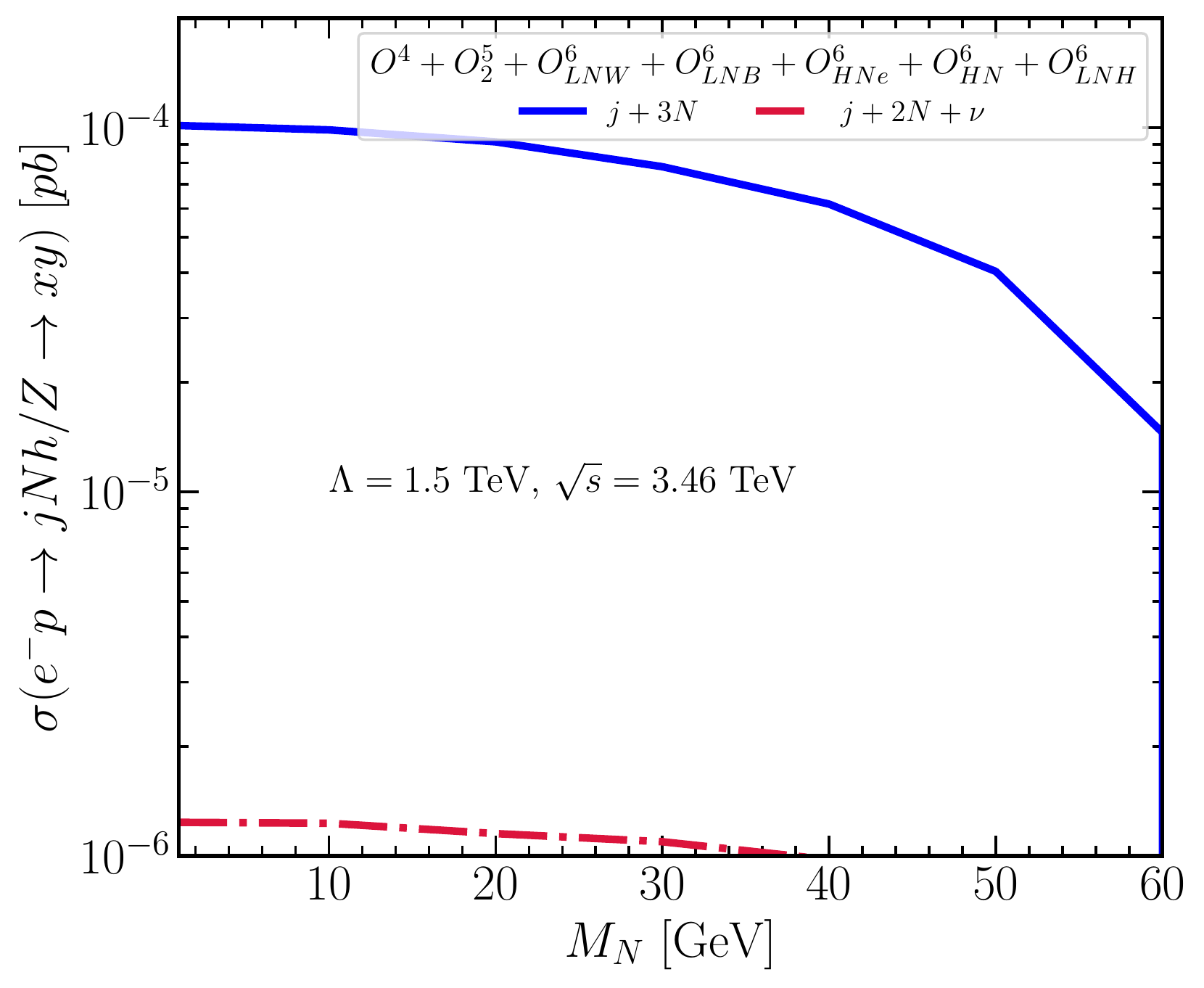} \label{fig:epj3n}}
		\caption{ Upper panel: The Feynman diagram  and the relevant operators for the process $e^{-} p \to (j + 3N) / (j + 2N + \nu)$.  the cross-section for the process $e^-p\to j+3N$. Bottom panel: Variation of cross section corresponding these processes as function of RHN mass $M_{N}$. The blue thick and red dot-dashed line signifies the cross section correspond to the process $e^{-}p \to j + 3N$ and $e^{-} p \to j + 2N + \nu$ respectively. The cross section is evaluated while taking into account all relevant operators.}
	\end{figure}

We consider the proposed $e^{-}p$ collider, FCC-eh \cite{Suarez:2022pcn} which will operate with a 60 GeV $e^{-}$ beam and 50 TeV proton beam providing a c.m. energy 3.46 TeV. Here we set the cut-off scale as $\Lambda=1.5$ TeV.
\begin{itemize}
	\item $\bf{e^-p\to jN}$:
\end{itemize}	
	 Production of RHN in association with a jet arises via two channels. One is the four-fermi interaction and the other is $W$ mediated process. Thus cross-section for this process is governed by the $d=4$ operator and as well as $d=6$ operators such as $\mathcal{O}_{duNe}^{6}$, $\mathcal{O}_{QuNL}^{6}$,$\mathcal{O}_{LNQd}^{6}$,$\mathcal{O}_{LdQN}^{6}$, $\mathcal{O}_{HNe}^6$ and $\mathcal{O}_{LNW}^6$ which is shown in Fig.~\ref{fig:epjn4f} and Fig.~\ref{fig:epjnw}. In Fig.~\ref{fig:epjn}, the gray dot dashed line indicates the $d=4$ contribution which is mixing suppresed ($\tilde\theta=10^{-3}$). The red dotted line shows the contribution from the $W$ mediated process including the effect of $\mathcal{O}_{HNe}^6$ and $\mathcal{O}_{LNW}^6$. The blue line represents the total contribution after including four-fermi interaction. As can be seen, large cross-section~$\sim \mathcal{O}(100\,\text{fb})$ is possible to obtain for $M_N \sim 50$ GeV.
	\begin{itemize}
	\item $\bf{e^-p\to j+3N/j+2N+\nu}$: 
	\end{itemize}
	Production of $h/Z$ with $j+N$ can take place via the two diagrams shown in Fig.~\ref{fig:eph} and Fig.~\ref{fig:epz}. These diagrams involve $d=6$ operator at two vertices, which can lead to $1/\Lambda^{x}$ with $x>4$ term in cross-section that can also arise at $d>6$ level. However, we restrict our calculation upto  $1/\Lambda^{4}$ term in cross-section and ignore higher power. The cross-section for $e^-p\to j+3N$  can be around 0.1 fb which yields $\sim100$ events with $1000~\text{fb}^{-1}$ luminosity. However, for $e^-p\to j+2N+\nu$ at most one event can be achieved with $1000~\text{fb}^{-1}$ luminosity. In this scenario, the Higgs mediated process is dominant over the $Z$ boson mediated process. Upon production the Higgs field further decays to  $N \nu$. The relevant coupling corresponding to $h \to \nu N$ is dependent on the mixing angle $\tilde{\theta}$, which suppress the total cross section of $e^{-} p \to j + 2N + \nu$ channel.
\subsection{Electron Positron Collider}
\begin{figure}[t]
	\subfigure[]{	\includegraphics[width=0.35\textwidth]{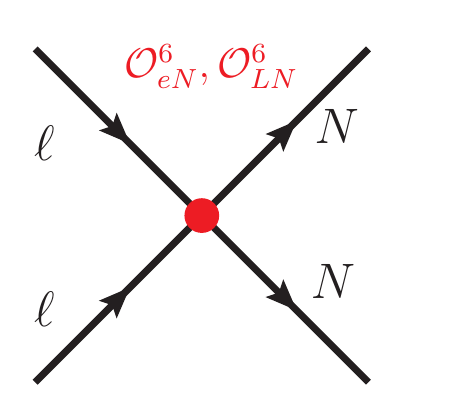}\label{fig:eeNN4f}}
			\hspace{0.75cm}
		\subfigure[]{ \includegraphics[width=0.5\textwidth]{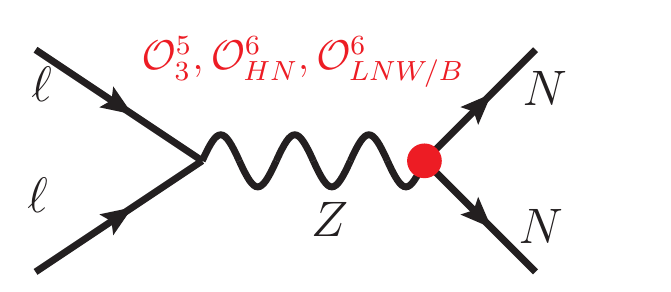}\label{fig:eeNNs}}
		\subfigure[]{	\includegraphics[width=0.45\textwidth,height=0.3\textheight]{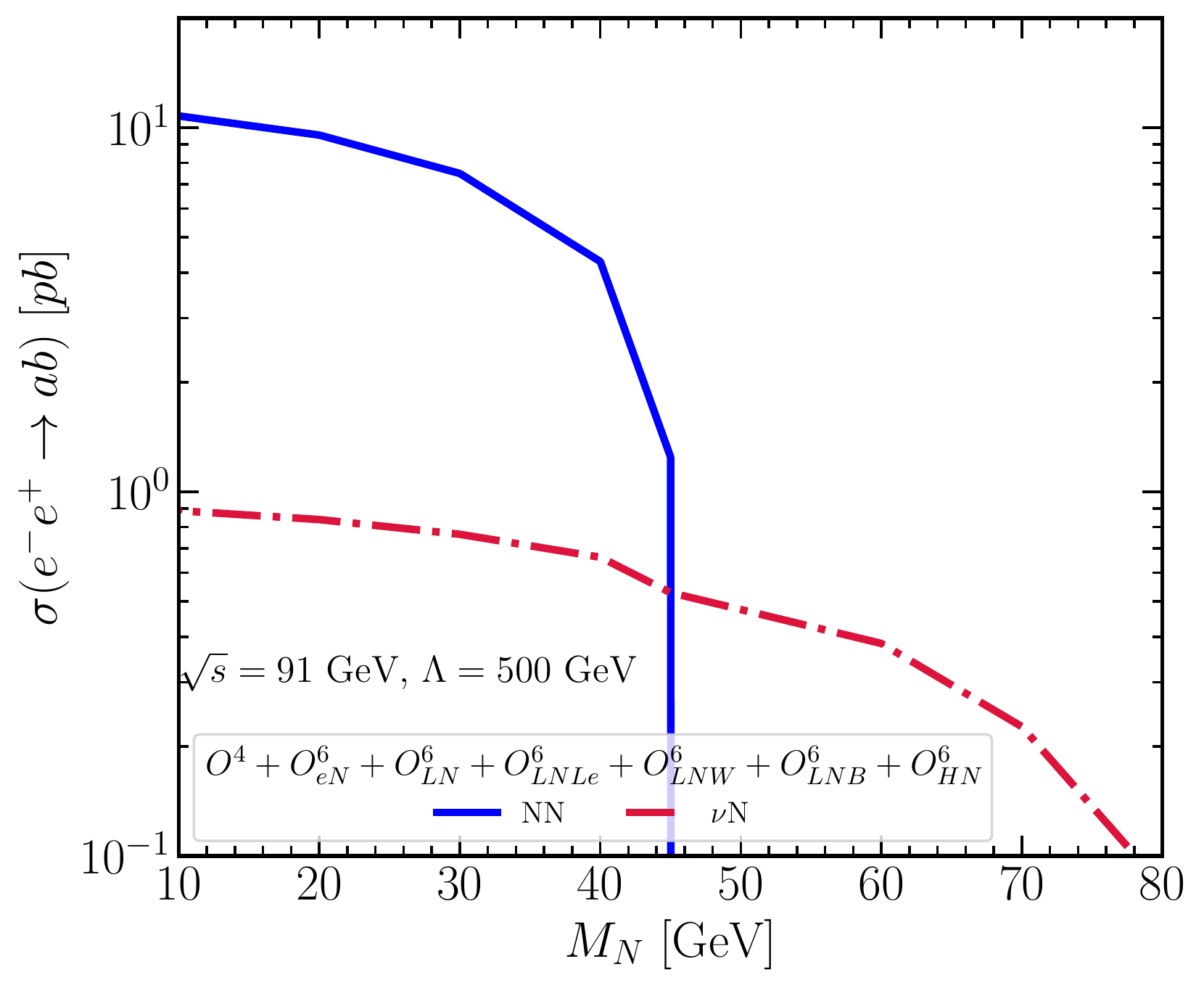}\label{fig:eeNN}}
		\subfigure[]{	\includegraphics[width=0.45\textwidth,height=0.3\textheight]{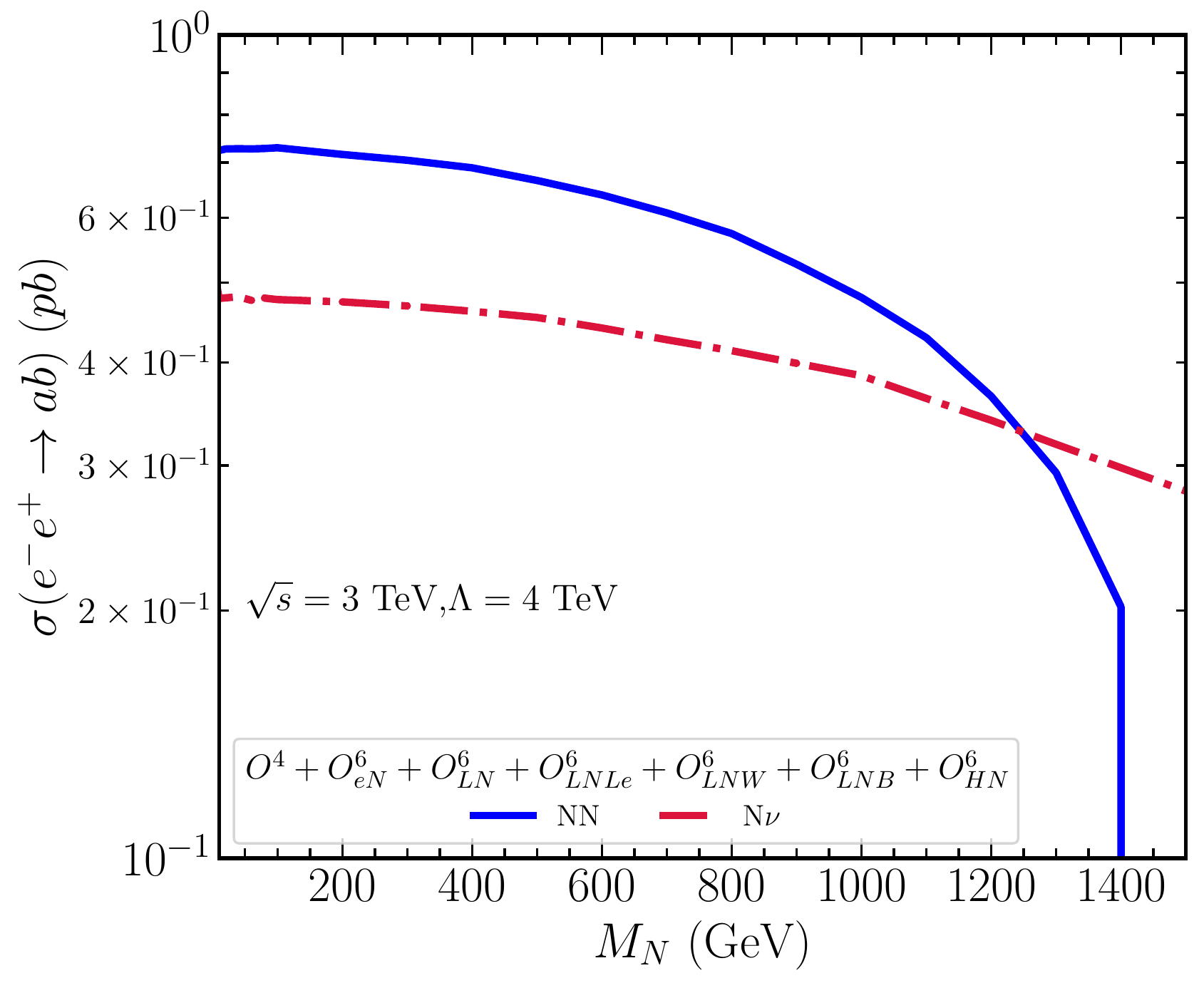}\label{fig:eeNN3}}
		\caption{In the upper panel we present the Feynman diagram and the relevant operators for the process $e^{+} e^{-} \to NN/ N\nu$. In the lower panel we show the cross-section for the process $e^+e^-\to NN/N\nu$ with $\sqrt{s}=91$ GeV~(\ref{fig:eeNN}) and $\sqrt{s}=3$ TeV~(\ref{fig:eeNN3}). The blue thick line and red dot-dashed line represents the process $e^{+}e^{-} \to N N$ and $e^{+} e^{-} \to N \nu$ respectively.}
	\end{figure}

We now discuss different production mechanism of the $N$ fields for the future electron positron collider. We choose two different c.m energy $\sqrt{s} = 91$ GeV and 3 TeV, where the corresponding cut-off scale $\Lambda$ are set to be 500 GeV and 4 TeV respectively.
\begin{itemize}
	\item $\bf{e^+e^-\to NN}$: 
\end{itemize}	
The four-fermi interaction, s-channel $Z/h/\gamma$ mediated diagram and t-channel $W^\pm$ mediated diagram lead to pair production of RHN at $e^{+}e^{-}$ collider.
	The $W^\pm$ mediated diagram receives extra $1/\Lambda^2$ suppression as it has two EFT vertex and hence it is ignored. On the other hand, the contribution coming from s-channel Higgs mediated process can also be ignored due to the smallness of $\mathcal{C}^{h}_{e^{+}e^{-}}$ coupling. Production cross-section for this process is governed by the operators $\mathcal{O}_{eN}^{6}$, $\mathcal{O}_{LN}^{6}$, $\mathcal{O}_{LNW/B}^6$, $\mathcal{O}_{HNe}^6$ and $\mathcal{O}_{HN}^6$ which is shown in Fig.~\ref{fig:eeNN4f} and Fig.~\ref{fig:eeNNs}. In Fig.~\ref{fig:eeNN} and \ref{fig:eeNN3} we illustrate the cross-section for two different c.m. energy  $\sqrt{s}=91$ GeV and $\sqrt{s}=3$ TeV, the corresponding cut-off scales are $\Lambda=500$ GeV and $\Lambda=4$ TeV.  Active sterile mixing is set to be $\tilde{\theta} =10^{-3}$, which makes the contributions of the renormalizable interaction terms (denoted by $\mathcal{O}^{4}$) suppressed.
	
For $\sqrt{s}=91$ GeV, $Z$-mediated process is dominant where the contribution from the operator $\mathcal{O}_{HN}^6$ plays major role for $e^-e^+\to NN$, as this interaction is not $\tilde{\theta}$ suppressed. Whereas, $\mathcal{O}_{LNW}^6$  and $\mathcal{O}_{LNB}^6$ involve $\tilde{\theta}$ dependancy. Contrary to that for $e^-e^+\to N\nu$, contribution from $\mathcal{O}_{HN}^6$ is $\theta$ suppressed and from the other two $d=6$ operators are independent of $\tilde{\theta}$. The contribution from 4F-operators are $\simeq 1$ pb. For $\sqrt{s}=3$ TeV, Contribution from 4F-operators are dominant.   
	
\begin{figure}[t]
		\subfigure[]{\raisebox{10mm}{\includegraphics[width=0.47\textwidth]{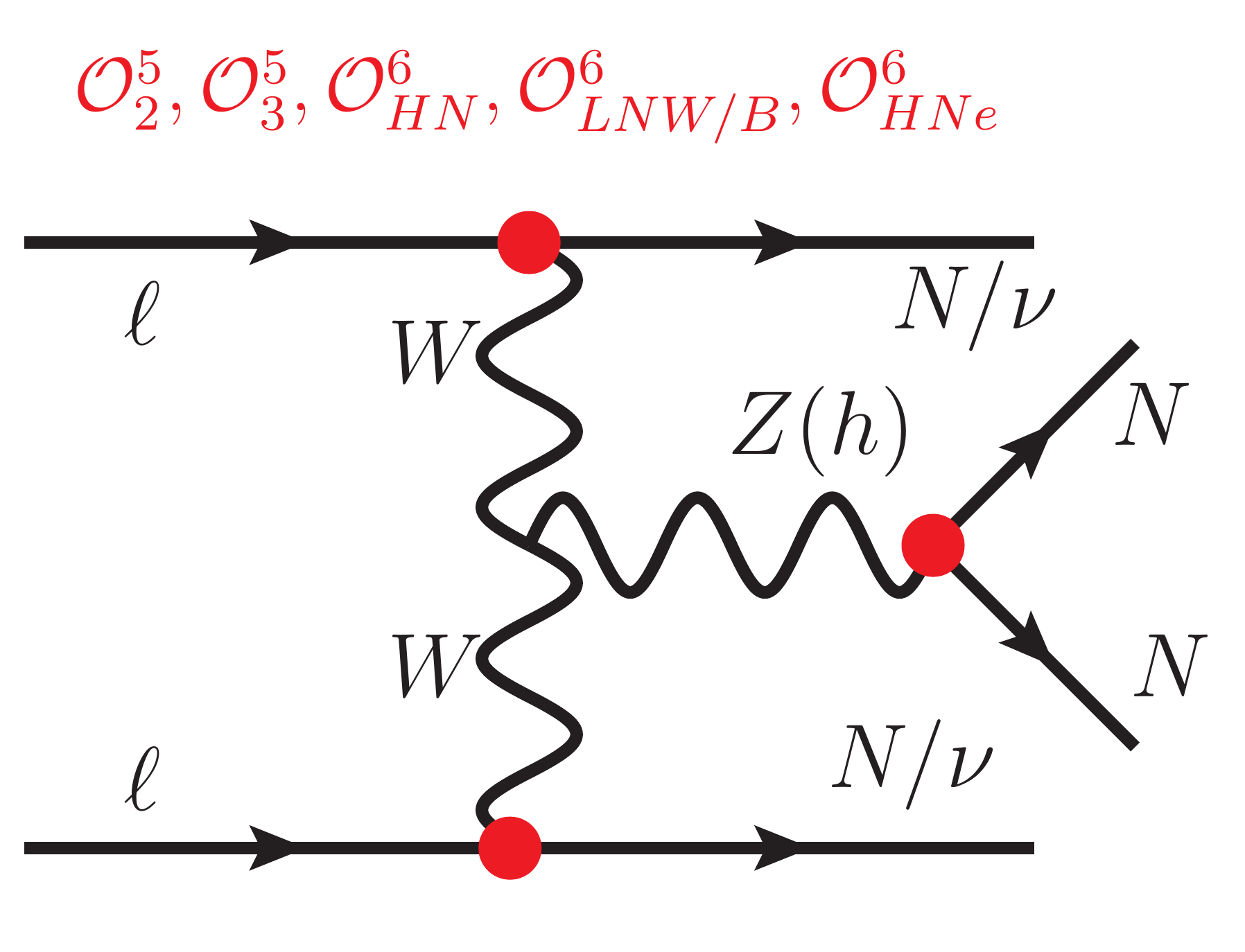}\label{fig:ee3Nfeyn}}}
		\subfigure[]{\includegraphics[width=0.5\textwidth,height=0.3\textheight]{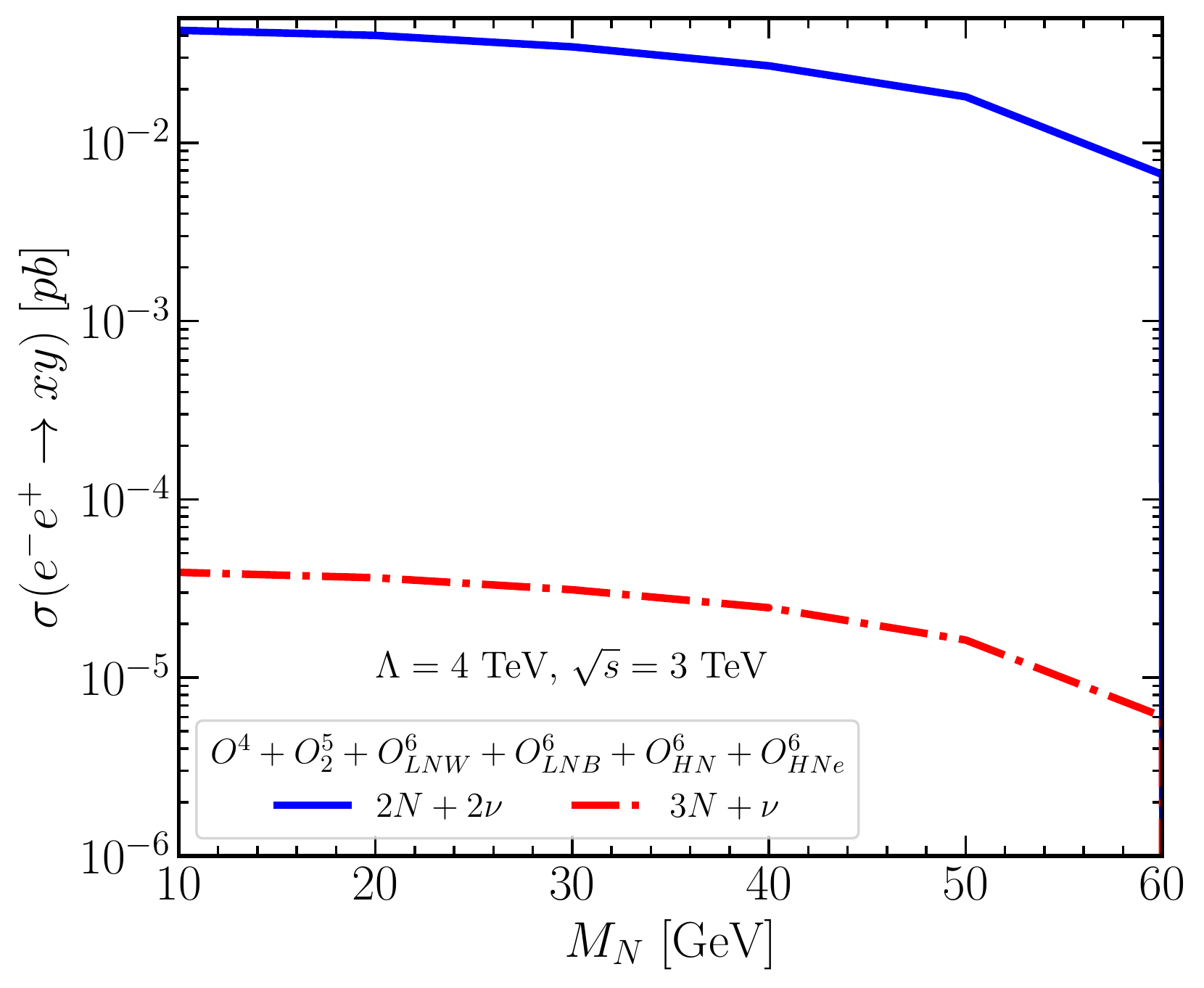}\label{fig:ee3N}}
		\caption{In Fig.~\ref{fig:ee3Nfeyn} we show the Feynman diagram and in Fig.~\ref{fig:ee3N} the cross-section for the process $e^+e^-\to 2N+2\nu$ and $e^+e^-\to 3N+\nu$. The blue thick line and red dot dashed line represents $e^+e^-\to 2N+2\nu$ and $e^+e^-\to 3N+\nu$ processes respectively.}
	\end{figure}

\begin{itemize}
\item $\bf{e^+e^-\to 2N+2\nu/ 3N +\nu}$: 
\end{itemize}
Fig.~\ref{fig:ee3Nfeyn} shows the Feynman diagram and corresponding operators for this process. As this diagram involves more than one $d=6$ vertex, in the amplitude level there exist terms with $1/\Lambda^n$, $n\ge3$. However, for simplicity we assume such contribution to be zero \cite{Brivio:2022pyi}. Fig.~\ref{fig:ee3N} shows the production rate for the processes $e^+e^-\to 2N+2\nu$ and $e^+e^-\to 3N+\nu$ by the blue line and red-dashed line, respectively. For $2N+2\nu$,  $\mathcal{C}^{W}_{\ell\nu}$ coupling is involved in two vertices and for $3N+\nu$, one $\mathcal{C}^{W}_{\ell\nu}$ and one $\mathcal{C}^{W}_{\ell N}$ couplings are involved. For$\mathcal{C}^{W}_{\ell\nu}$ coupling, the $\mathcal{O}^4$ contribution which is independent of $\tilde{\theta}$ is dominant compared to $\mathcal{O}^6$ contribution. The presence of extra $\mathcal{C}^{W}_{\ell\nu}$ coupling at one vertex makes the cross-section for the $2N+2\nu$ final state larger compared to that of $3N+\nu$.
\section{The Right Handed Neutrino Decay Modes}
\label{Sec:NRdecay}
\begin{figure}[h!]
\centering
\includegraphics[width=0.45\textwidth,height=0.25\textheight]{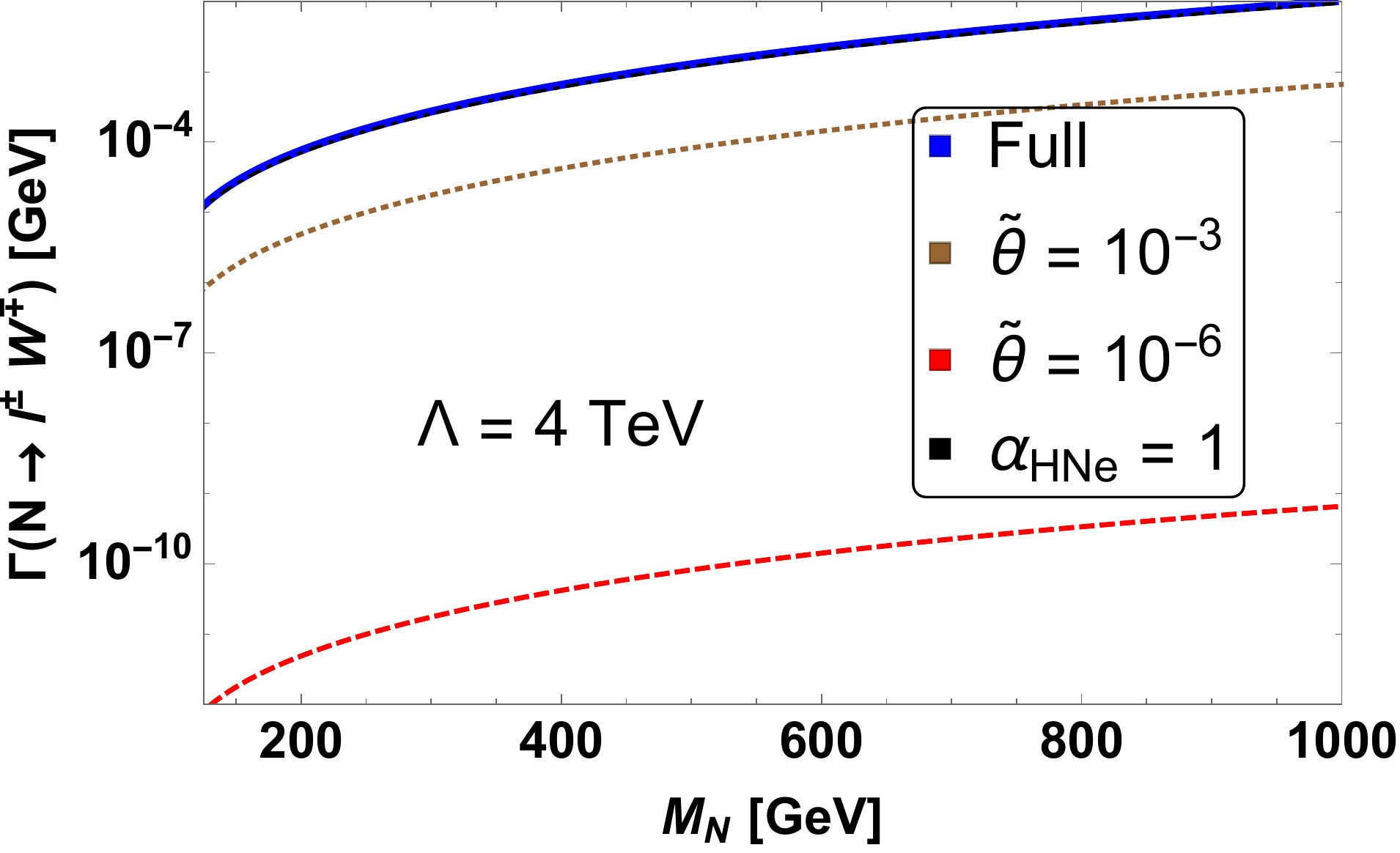}\label{Fig:2bodynlW}
\hspace{0.25cm}
\includegraphics[width=0.45\textwidth,height=0.25\textheight]{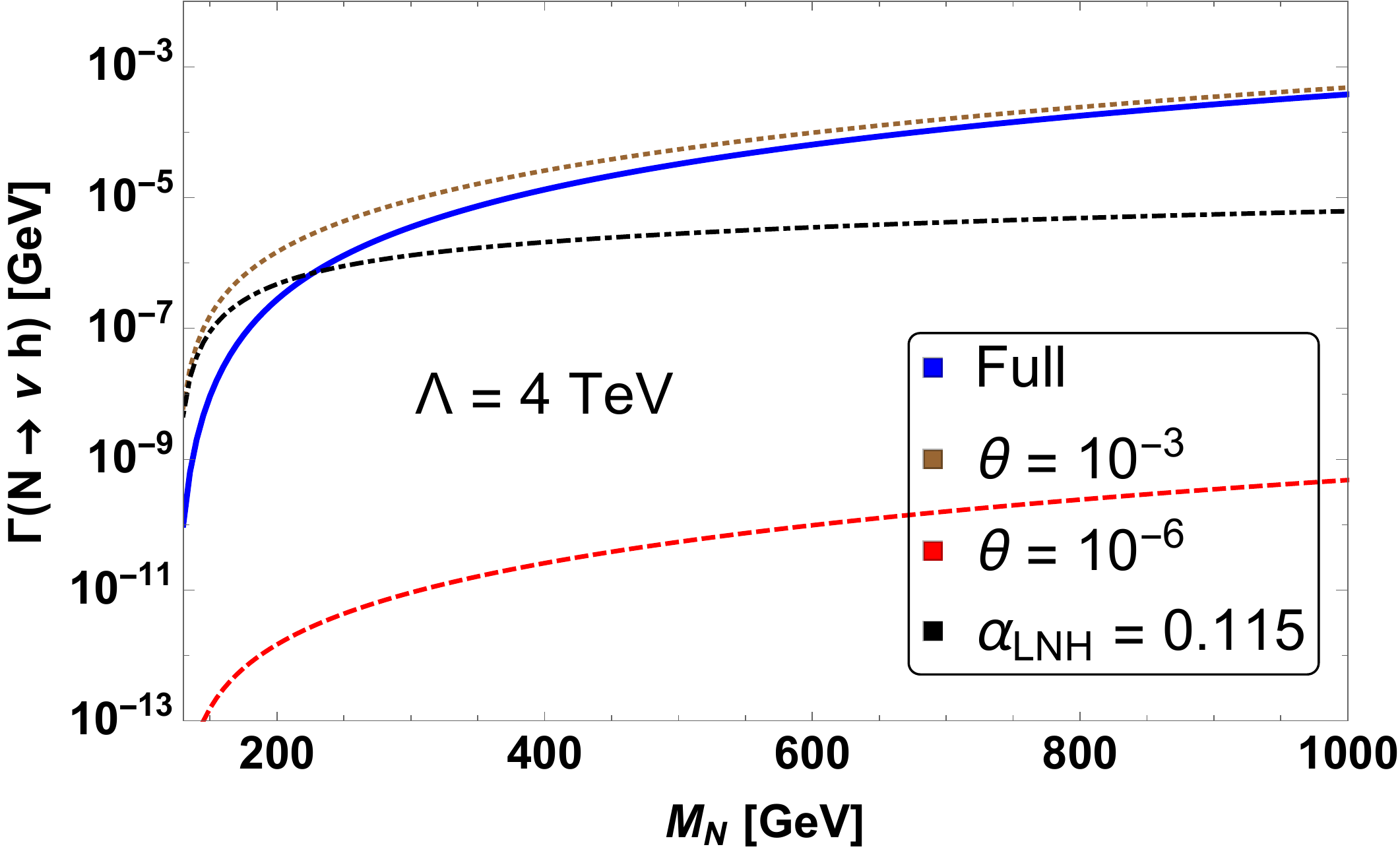}\label{Fig:2bodynuh}
\caption{The partial decay width correspond to $\Gamma(N \to \ell^{\pm} W^{\mp})$ (left) and $\Gamma(N \to \nu h)$ (right) respectively. For both these cases the cut-off scale is set at 4 TeV. See text for details.}
\label{Fig:NrlWnuh}
\end{figure}

Depending on the mass, the sterile neutrino can decay either via two body or three body modes. If we only consider the renormalisable part of the effective Lagrangian heavy neutrinos decays to $W\ell$, $\nu Z$ and $\nu h$ modes through the active-sterile mixing if it has sufficient mass. These decay modes receive additional contributions if we take into account the dimension five and dimension six operators. In Table.~\ref{tab:Decay5}, we present various two body decay modes as well as the operators that would contributes to each of these modes. In principle, $N_{i} \to N_{j} h/N_{j}Z/N_{j}\gamma$ (where $i \neq j$) mode is also possible. But for simplicity we have chosen degenerate mass values for all the RHNs. Hence, those decay modes are kinematically disallowed. For the decay modes such as $N_{i} \to \nu_j Z/\nu_j h/\nu_j \gamma$ both $i = j$ and $i \neq j$ scenarios are possible and the operators which would contribute to these modes are mentioned in the Table.~\ref{tab:Decay5}. However, it is important to point out that the $\mathcal{O}^{(3)}_{5}$ operator is anti symmetric in flavour space. Hence it would not contribute to $N_{i} \to \nu_{i} Z/\nu_{i}\gamma$ modes. All the other operators are symmetric in the flavour space and participate in both the $i = j$ and $i \neq j$ scenarios. There exist a great volume of work that have studied possible phenomenological aspects of these decay modes and interested readers can consult these in Ref.~\cite{Duarte:2016miz} for more comprehensive discussions. For completeness, here we briefly discuss the two body decay width and what role the different operators play in the corresponding decay width calculation.          
\begin{table}[h!]
\centering
\begin{tabular}{|p{3.5cm}||p{8cm}|}
\hline
~~~~Decay &~~~~~ Contributing Operators \\
\hline 
~~~~$\Gamma\left(N_{i} \to \ell_{j} W\right)$ := &~~~~~  $\mathcal{L}_{CC}, \mathcal{O}_{HNe}, \mathcal{O}_{LNW}$ \\
\hline
~~~~$\Gamma\left(N_{i} \to \nu_{j} Z\right)$ := &~~~~~ $\mathcal{L}_{NC}, \mathcal{O}^{(5)}_{3}, \mathcal{O}_{HN}, \mathcal{O}_{LNB}, \mathcal{O}_{LNW}$ \\
\hline
~~~~$\Gamma\left(N_{i} \to \nu_{j} h\right)$ := &~~~~~$\mathcal{L}_{yuk}, \mathcal{O}^{(5)}_{1}, \mathcal{O}^{(5)}_{2}, \mathcal{O}_{LNH}$ \\
\hline
~~~~$\Gamma\left(N_{i} \to \nu_{j} \gamma\right)$ := &~~~~~$\mathcal{O}^{(5)}_{3}, \mathcal{O}_{LNW}, \mathcal{O}_{LNB}$ \\
\hline
\end{tabular}
\caption{Different possible two body decay modes along with the operators that can contribute to these decays.}
\label{tab:Decay5}
\end{table}
\noindent
\begin{itemize}
\item$\bf{\Gamma\left(N_{i} \to \ell_{j} W\right)}$
\end{itemize}
We begin our discussion with $N_{i} \to \ell_{j} W$ channel which gets contributions from the standard renormalisable charged current interaction, which depends on active-sterile mixing as well as dimension six operator $\mathcal{O}_{HNe}$ and $\mathcal{O}_{LNW}$. From our discussion on Subsection~.\ref{Sec:dim6}, one can see that the coupling $\mathcal{C}^{W_{\mu}}_{\ell N}$ does not receive any modification from dimension five operators. On the other hand in dimension six this coupling does alter and one can find the explicit form of this coupling in Table~.\ref{tab:vertex6}. However, in general the operator $\mathcal{O}_{LNW}$ is loop mediated operator and the Wilson coefficient corresponding to this is suppressed by $\frac{1}{16\pi^{2}}$. Hence the contribution coming from this is very minimal \emph{w.r.t} both the $\mathcal{O}_{HNe}$ and $\tilde{\theta}$. Hence we can safely ignore its effect. With this assumption, the partial decay width of the above mentioned process is
\begin{align}
\Gamma\left(N_{i} \to \ell_{j} W\right) & = \frac{g^{2}}{64\pi M_{N}M^{2}_{W}}\{(|A|^{2} + |B|^{2})\left(M^{2}_{W}\left(M^{2}_{\ell_{j}} + M^{2}_{N}\right) + \left(M^{2}_{\ell_{j}} - M^{2}_{N}\right)^{2} - 2M^{4}_{W} \right)  \nonumber\\
							& - 12\text{Re}[A^{*}B]M_{\ell_{j}}M_{N}M^{2}_{W}\}\times \lambda^{\frac{1}{2}}\left(1, \frac{M^{2}_{\ell_{j}}}{M^{2}_{N}}, \frac{M^{2}_{W}}{M^{2}_{N}}\right),   			
\label{Eq:2bodylNW}
\end{align}
where parameter $A$ and $B$ is defined in the following fashion 
\[ ~~~ A = \tilde{\theta},~~~~~~~B = \frac{v^{2}\alpha_{HNe}}{\Lambda^{2}}.\]
\noindent
The left panel of Fig.~\ref{Fig:NrlWnuh} shows the decay width as a function of sterile neutrino mass $M_N$. For plotting purpose we fixed the cut-off scale at 4 TeV and chose the range of $M_{N}$ from 200 GeV to 1 TeV. The red dashed line and the brown dotted line represents sole contributions from renormalisable charged current interaction with $\tilde{\theta} = 10^{-6}$ and $\tilde{\theta} = 10^{-3}$ respectively. On the other hand the black dashed dotted curve highlight the effect of $\mathcal{O}_{HNe}$. The blue thick line shows the decay width when one accounts both the renormalisable part~($\tilde{\theta}=10^{-3}$) and dimension six operator. From the left panel of Fig.~\ref{Fig:NrlWnuh}, one can see the dominant contribution comes from the dimension six operator.  

\begin{itemize}
\item$\bf{\Gamma\left(N_{i} \to \nu_{j} h\right)}$
\end{itemize}
In case of $N$ decaying to $\nu h$ mode, the partial width is dependent on the mixing angle between active and sterile neutrinos and also on dimension six operator $\mathcal{O}_{LNH}$. In principle the operators $\mathcal{O}^{(5)}_{1}$ and  $\mathcal{O}^{(5)}_{2}$ can also contribute to this decay mode. However, from Table.~\ref{Tab:vertex5} one can notice that the relevant coupling coming from these operators are $\tilde{\theta}$ suppressed. Hence we can ignore their effects. The definition of the mixing angle varies with the mass dimension of underlying EFT. For a better understanding, remember upto dimension five, the mixing angle is defined as $M^{-1}_{N}\frac{Y_{\nu}v}{\sqrt{2}}$ which is independent of EFT parameters. In dimension six the operator $\mathcal{O}_{LNH}$ modifies the definition of the mixing angle as discussed in Eq.~\ref{Eq:Vdim6} and the mixing angle $\theta$ is replaced with $\tilde{\theta}$. To make a suitable comparison between the two different dimensions, here we write down the explicit part of $\tilde{\theta}$, where $\theta$ to be $M^{-1}_{N}\frac{Y_{\nu}v}{\sqrt{2}}$ and the additional part arises from $\mathcal{O}_{LNH}$. The partial decay width for the process can be written in the following fashion- 
\begin{align}
\Gamma\left(N_{i} \to \nu_{j} h\right)& = \frac{|A|^{2}}{32\pi M^{3}_{N}}\left(M^{2}_{N} - M^{2}_{h}\right)^{2},
\end{align}
where the co-efficient $A$ is expressed as 
\begin{align}
A = \left(\frac{3v^{2}}{2\sqrt{2}\Lambda^{2}}\alpha_{LNH} - \frac{\theta M_{N}}{v}\right) .
\end{align}
To realise the effect of individual parameters we refer to the right panel of Fig.~\ref{Fig:NrlWnuh}. The brown dotted and the black dash-dotted line represents the effects of $\theta$ and $\alpha_{LNH}$ respectively. The red dashed line shows the partial decay width if the mixing angle is $10^{-6}$. The relative minus sign in the definition of $A$ can be understood as the apparent destructive interference which appears in the full decay width calculation and is represented as the thick blue curve in Fig.~\ref{Fig:NrlWnuh}~(right hand side) which accounts both the renormalisable part~($\theta=10^{-3}$) and dimension six operator~$\alpha_{LNH}$. 

\begin{figure}[h!]
\centering
\includegraphics[width=0.45\textwidth,height=0.25\textheight]{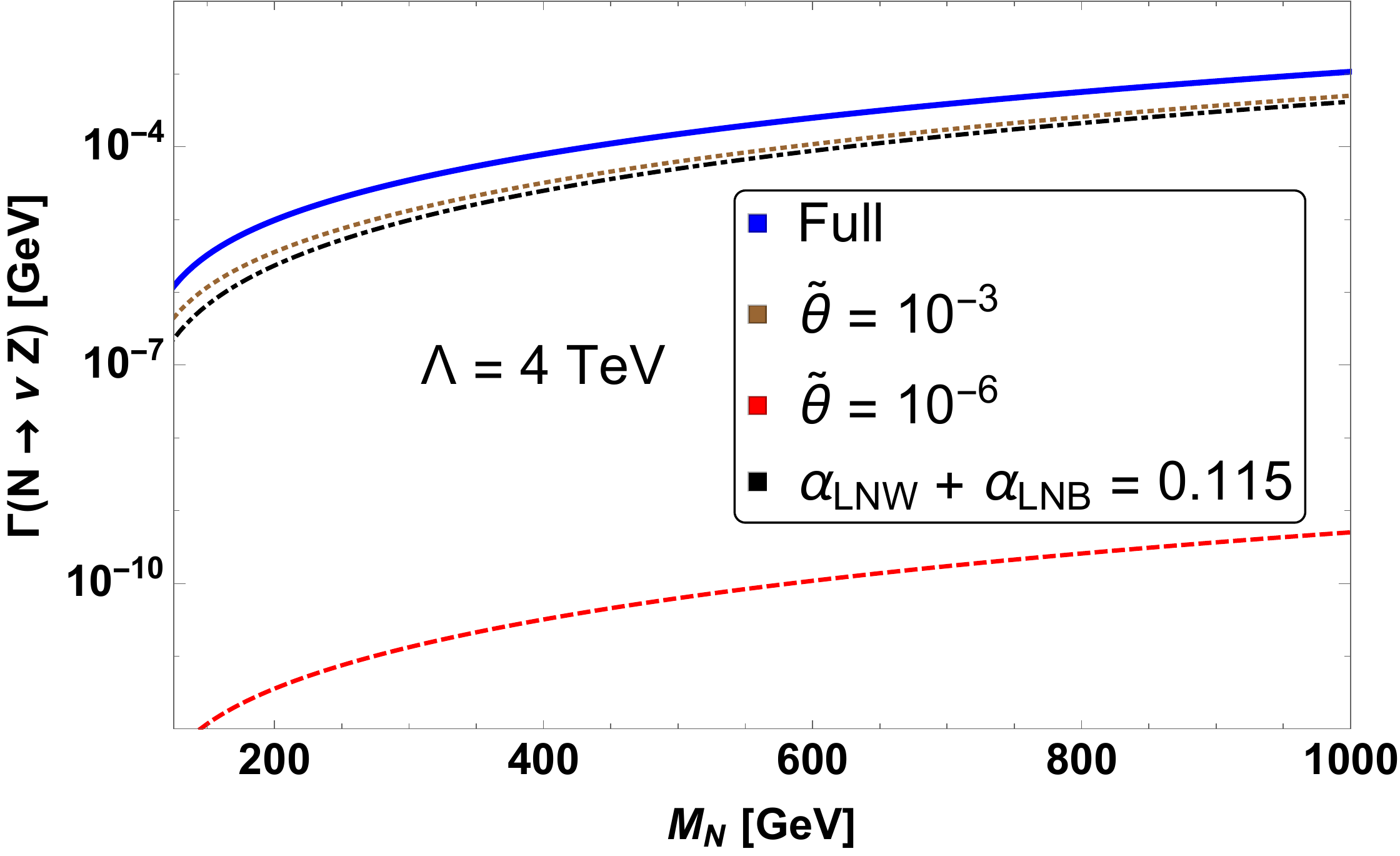}\label{Fig:2bodynuZ}
\hspace{0.25cm}
\includegraphics[width=0.45\textwidth,height=0.25\textheight]{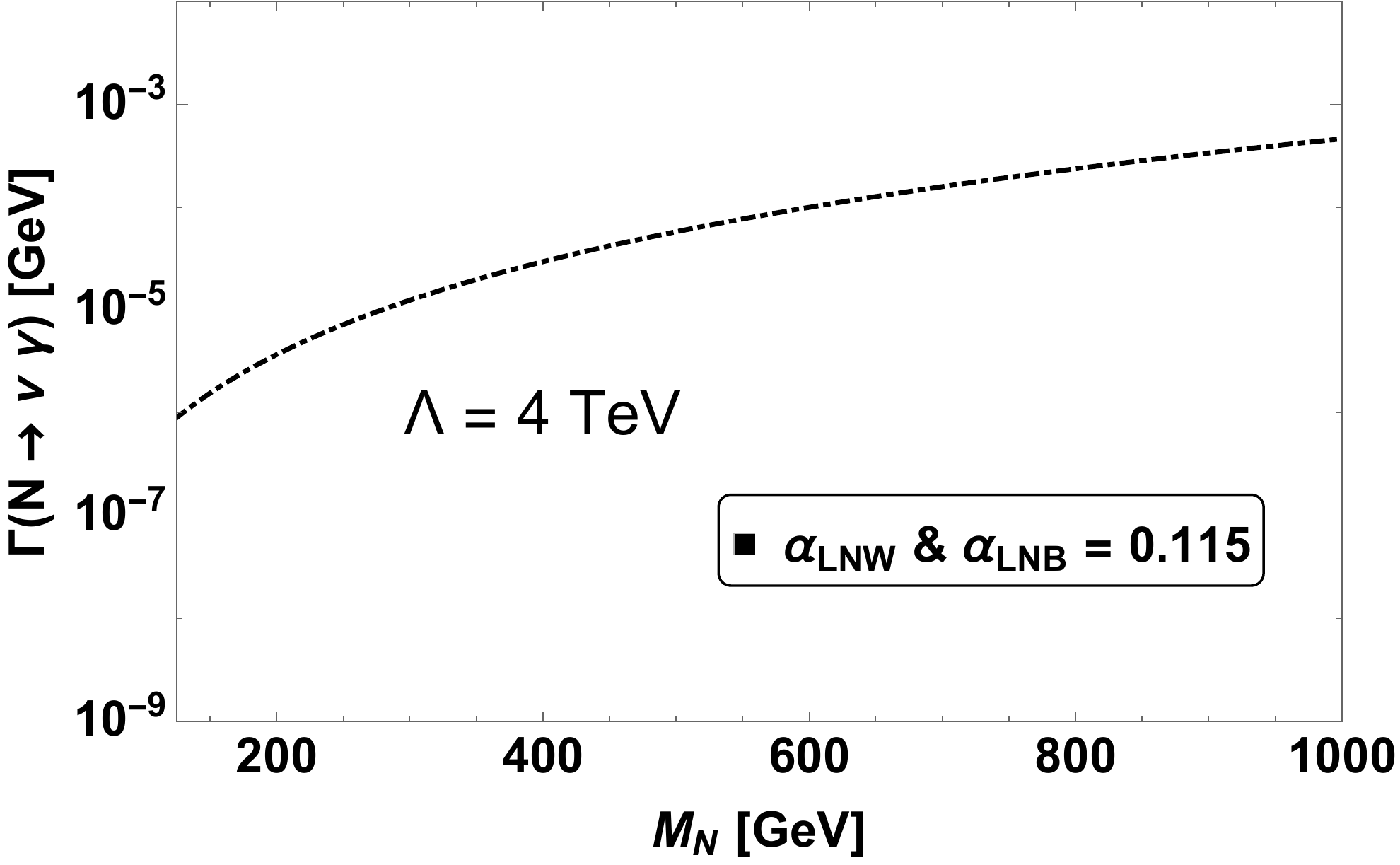}\label{Fig:2bodynuA}
\caption{The partial decay width correspond to $\Gamma(N\to \nu Z)$ (left) and $\Gamma(N\to \nu \gamma)$ (right) respectively. For both these cases the cut-off scale is set at 4 TeV. The value for $\alpha_{LNW}$ and $\alpha_{LNB}$ is consistent with the current experimental limits.}
\label{Fig:NrnuZnuA}
\end{figure}

\begin{itemize}
\item$\bf{\Gamma\left(N_{i} \to \nu_{i} Z\right)}$ $\&$ $\bf{\Gamma\left(N_{i} \to \nu_{i} \gamma\right)}$
\end{itemize}
We now turn our attention to RHN decays to $\nu Z$ and $\nu \gamma$ modes. In the absence of the EFT operators it can only decay to $\nu Z$ channel and the partial width correspond to this is regulated by the mixing angle. In the dimension five scenario the operator $\mathcal{O}^{(5)}_{3}$ invokes a new decay mode involving photon besides modifying the coupling $\mathcal{C}^{Z_{\mu}}_{(\overline{\nu}N + \overline{\nu}N)}$. However, this operator is antisymmetric in the flavour space and does not participate in the $N_{R}$ decay into the same flavour SM-neutrino. For different flavour SM-neutrino, the operator $\mathcal{O}^{(5)}_{3}$ does contribute but the relevant coupling is mixing angle suppressed (see  Table.~\ref{Tab:vertex5} for details.) Hence we can ignore this operator for present calculation. At dimension six level  $\nu Z$  receives contribution from the operators $ \mathcal{O}_{HN}, \mathcal{O}_{LNB}, \mathcal{O}_{LNW}$. But one can safely ignore $\mathcal{O}_{HN}$ as its effect in $\mathcal{C}^{6}_{Z_{\mu}(\overline{\nu}N + \overline{N}\nu)}$ coupling is negligible due to small mixing angle and $\frac{1}{\Lambda^{2}}$ factor. The other two operators can provide appreciable contribution in the decay width and in the following we present the definite formula for this case        
\begin{align}
\Gamma\left(N_{i} \to \nu_{j} Z\right) & = \frac{\left(M^{2}_{N} - M^{2}_{Z}\right)^{2}}{128\pi c^{2}_{w}M^{2}_{Z}M^{3}_{N}}\{g^{2}|\tilde{\theta}|^{2}\left(M^{2}_{N} + 2M^{2}_{Z}\right) + 64c^{2}_{w}M^{2}_{Z}|A|^{2} + \left(M^{2}_{Z} + 2M^{2}_{N}\right) \nonumber\\
						   &     + 48g c_{w}M_{N}M^{2}_{Z}\text{Re}[\tilde{\theta}^{*}A]\}
\end{align}
The dimension six part is expressed by introducing the new parameter $A$ which is defined as 
\[ ~~~~ A = \frac{c_{w}\alpha_{LNW}v}{\sqrt{2}\Lambda^{2}} - \frac{s_{w}\alpha_{LNB}v}{\sqrt{2}\Lambda^{2}}\]
The relative minus sign in the above expression can be understood from the mass basis definition of the $Z$ boson and photon. In SM, the $Z_{\mu} = c_{w}W^{3}_{\mu} - s_{w}B_{\mu}$ and $A_{\mu} = s_{w}W^{3}_{\mu} + c_{w}B_{\mu}$ where $W^{3}_{\mu}$ and $B_{\mu}$ are fields correspond to $T_{3}$ and $Y$ generators. The relative minus sign in the mass relation is responsible for the minus sign in $A$. Similarly, for the $\nu\gamma$ mode we present the explicit dependence of dimension five and dimension six operators in Eq.~\ref{Eq:2bodyNnuA}. 

\begin{align}
\Gamma\left(N_{i} \to \nu_{j} \gamma\right) & = \frac{M^{3}_{N}}{\pi}\left(|A|^{2} + |B|^{2}\right),
\label{Eq:2bodyNnuA}
\end{align}
\[ \text{where}~~~~~~~ A = - \frac{c_{w}}{\Lambda}\left(\tilde{\theta}\alpha^{(5)}_{3}\right) + \frac{c_{w}v}{\sqrt{2}\Lambda^{2}}\alpha_{LNB} +  \frac{s_{w}v}{\sqrt{2}\Lambda^{2}}\alpha_{LNW}, ~~~~~~~ B = \frac{c_{w}}{\Lambda}\left(\tilde{\theta}^{*}\alpha^{(3)}_{5}\right).\]
Here also, we set $\alpha^{(3)}_{5}$ to be zero for the reason discussed above. In Fig.~\ref{Fig:NrnuZnuA}, we represent the upshot of each operators and mixing angle along with their combine role. In left panel of Fig.~\ref{Fig:NrnuZnuA} we plot the partial decay width of $N \to \nu Z$ mode. The red dashed line shows the contributions coming from $\tilde{\theta} = 10^{-6}$. For the mixing angle $10^{-3}$ which is shown as brown dotted line in this figure, has dominant contribution over entire range of chosen $M_{N}$. The effects of dimension six terms shown as dot-dashed black curve is subdominant but have same order of magnitude as with the mixing angle $10^{-3}$. The blue line shows the full width for $N \to \nu Z$ which is calculated while taking into account both dimension six and mixing angle~($\tilde{\theta} = 10^{-3}$) contributions. The plot in the right side of Fig.~\ref{Fig:NrnuZnuA} shows the decay width for $N \to \nu \gamma$ channel and it is entirely dependent on the operators $\mathcal{O}_{LNW}$ and $\mathcal{O}_{LNB}$. In both cases the cut-off scale is at 4 TeV.    

\begin{figure}[h!]
\centering
\includegraphics[width=0.45\textwidth,height=0.25\textheight]{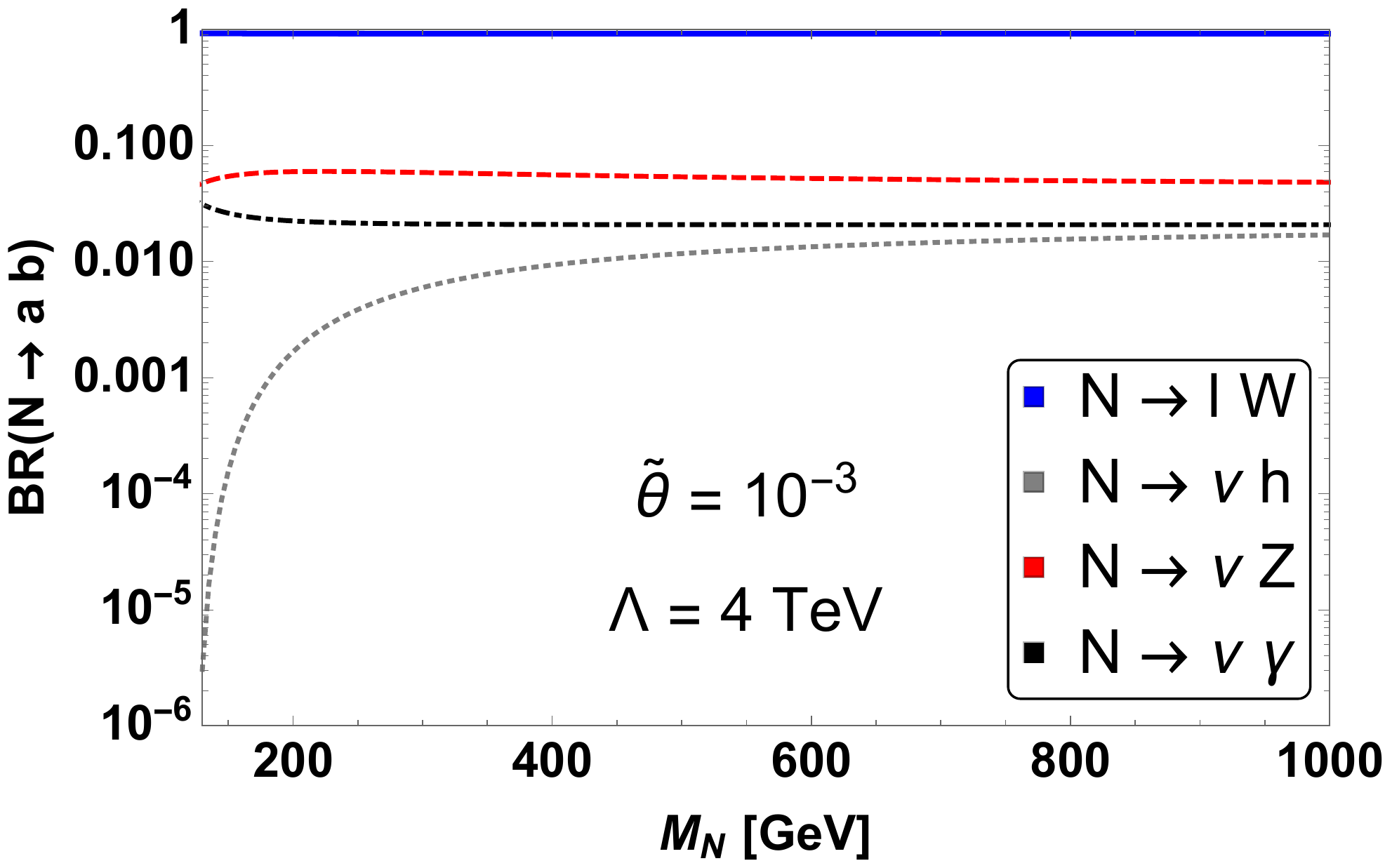}\label{Fig:2body1000}
\hspace{0.25cm}
\includegraphics[width=0.45\textwidth,height=0.25\textheight]{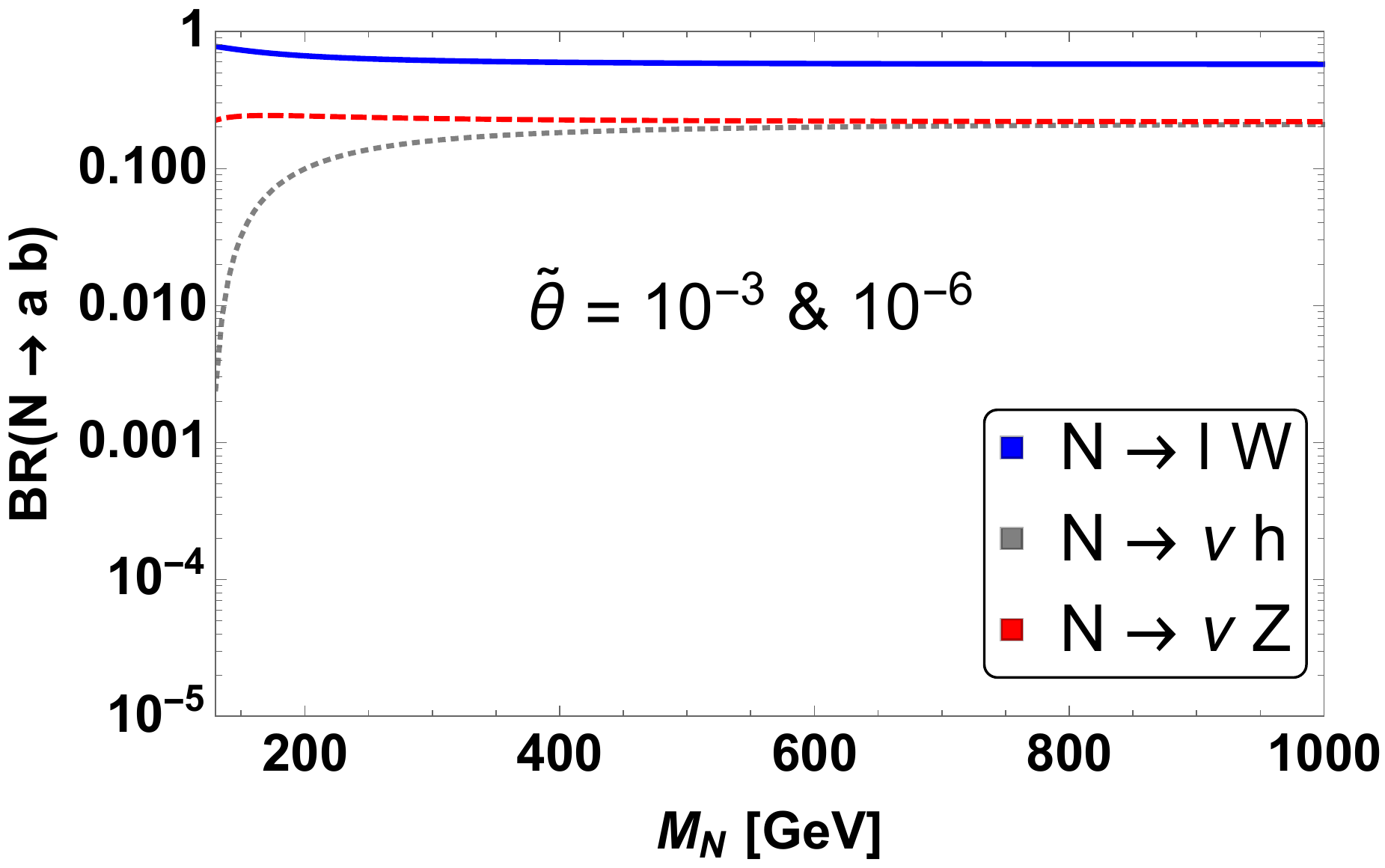}\label{Fig:2bodytheta}
\caption{The branching ratio correspond to different two body decay modes for the $M_{N}$ mass range from 200 GeV to 1 TeV. In the left panel, we show branching ratio correspond to $N_{R}$-EFT framework upto $d = 6$ where the cut-off scale $\Lambda$ is set to be 4 TeV and the mixing angle $\tilde{\theta} = 10^{-3}$. In the right panel we show the branching ratio of possible two body decay modes if we only consider the renormalisibale part of the Lagrangian. This plot correspond to both the mixing angle $\tilde{\theta} = 10^{-3}$ and $10^{-6}$ respectively. See text for details.}
\label{Fig:2bodyBR}
\end{figure}

We conclude our discussion on the two body decay modes while presenting the corresponding branching ratio in Fig.~\ref{Fig:2bodyBR} (left panel). As expected $\ell W$ mode which is represented by the thick blue line is dominated in the entire range of mass. The $\nu Z$ and $\nu \gamma$ channels are shown as red dashed and black dot-dashed line respectively and their corresponding BR is less than 10$\%$. Furthermore the BR of $\nu Z$ is always greater than $\nu \gamma$ as it receives additional contribution from mixing angle $\tilde{\theta}$. The BR of $\nu h$ which is shown as gray dotted line, remains the minimum for the entire mass range. 
\noindent
For comparison, in the right panel of Fig.~\ref{Fig:2bodyBR}  we present the branching ratio correspond to only renormalisable part of the Lagrangian. We have considered two values of mixing angle $\tilde{\theta} = 10^{-3}$ and $10^{-6}$. The pattern of the plot remain unaffected for the choice of mixing angle as the BR of each individual modes solely depend on their respective kinematic factor. Here also, $\ell W$ mode which is illustrated as blue thick line is dominated in the entire range of RHN mass. However \emph{w.r.t} the EFT counterpart the BR of the $N \to \ell W$ channel is relatively small. As a consequence the BR of $\nu h$ and $\nu Z$ mode can achieve around 20$\%$ value for $M_{N} \gtrsim$ 500 GeV. The $\nu\gamma$ curve is absent in this plot as the $N$ field can not decay into this mode as it only arise in the EFT framework.

\begin{itemize}
\item \textbf{Three Body Decay Modes}
\end{itemize}
Now we turn our attention to the mass range where $\tilde{M}_{N} \leq M_{W}$. Hence the only allowed two body decay mode is $\Gamma\left(N_{i} \to \nu_{j} \gamma\right)$. Having said that, the operators that control this process are loop suppressed. As a result the width corresponding to this channel is understandably low and we safely choose the corresponding BR value to be $ 5\%$ in the mass range $10 ~\text{GeV} \leq M_{N} \leq 80 ~\text{GeV}$. Instead, this particular mass range motivates us to consider different three body decay modes that can be perpetrated either via off shell decay of different SM states or via contact interactions. In the following we would discuss this in more detail.    

\begin{figure}[h!]
\centering
\includegraphics[width=0.45\textwidth,height=0.25\textheight]{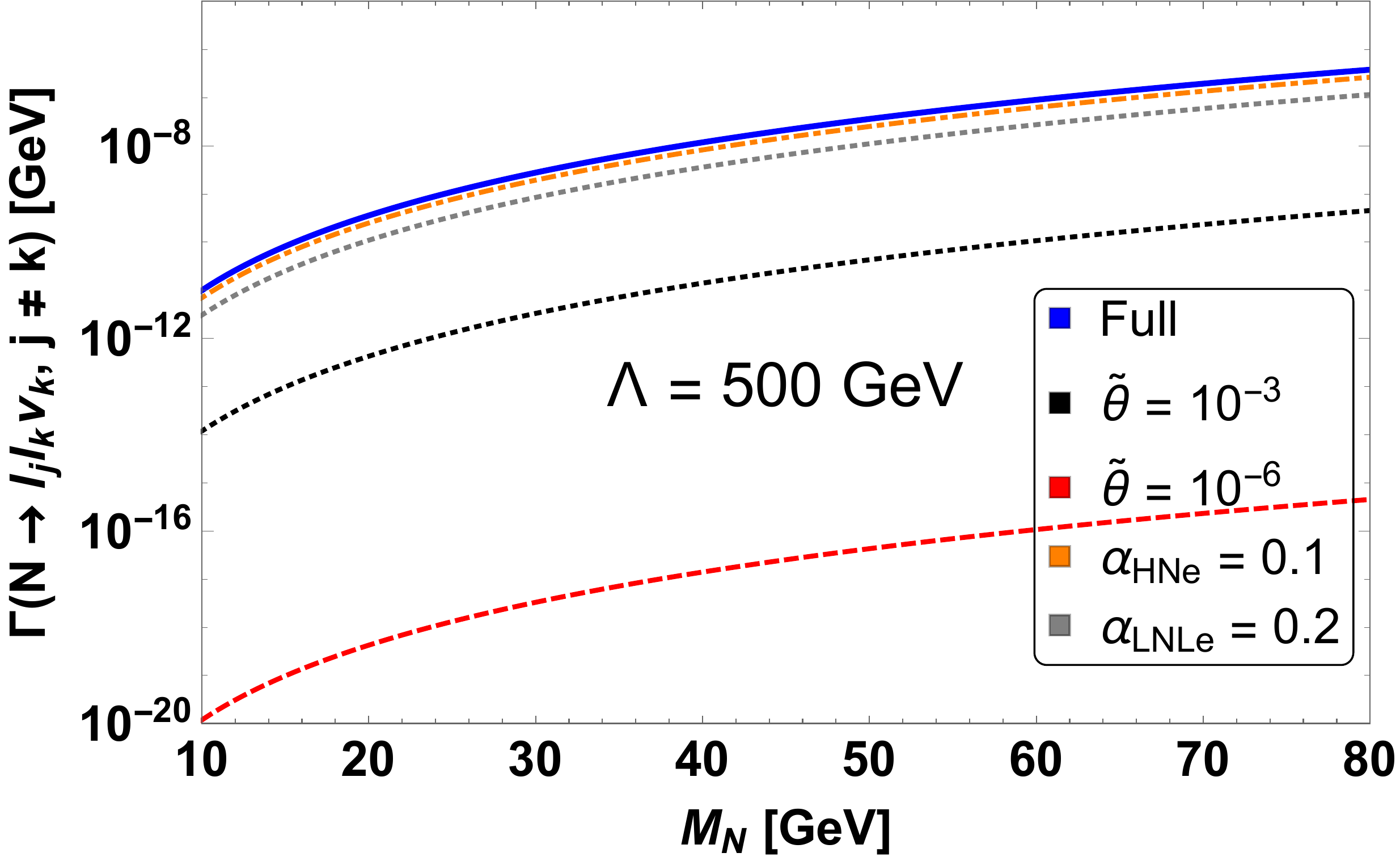}
\hspace{0.5cm}
\includegraphics[width=0.45\textwidth,height=0.25\textheight]{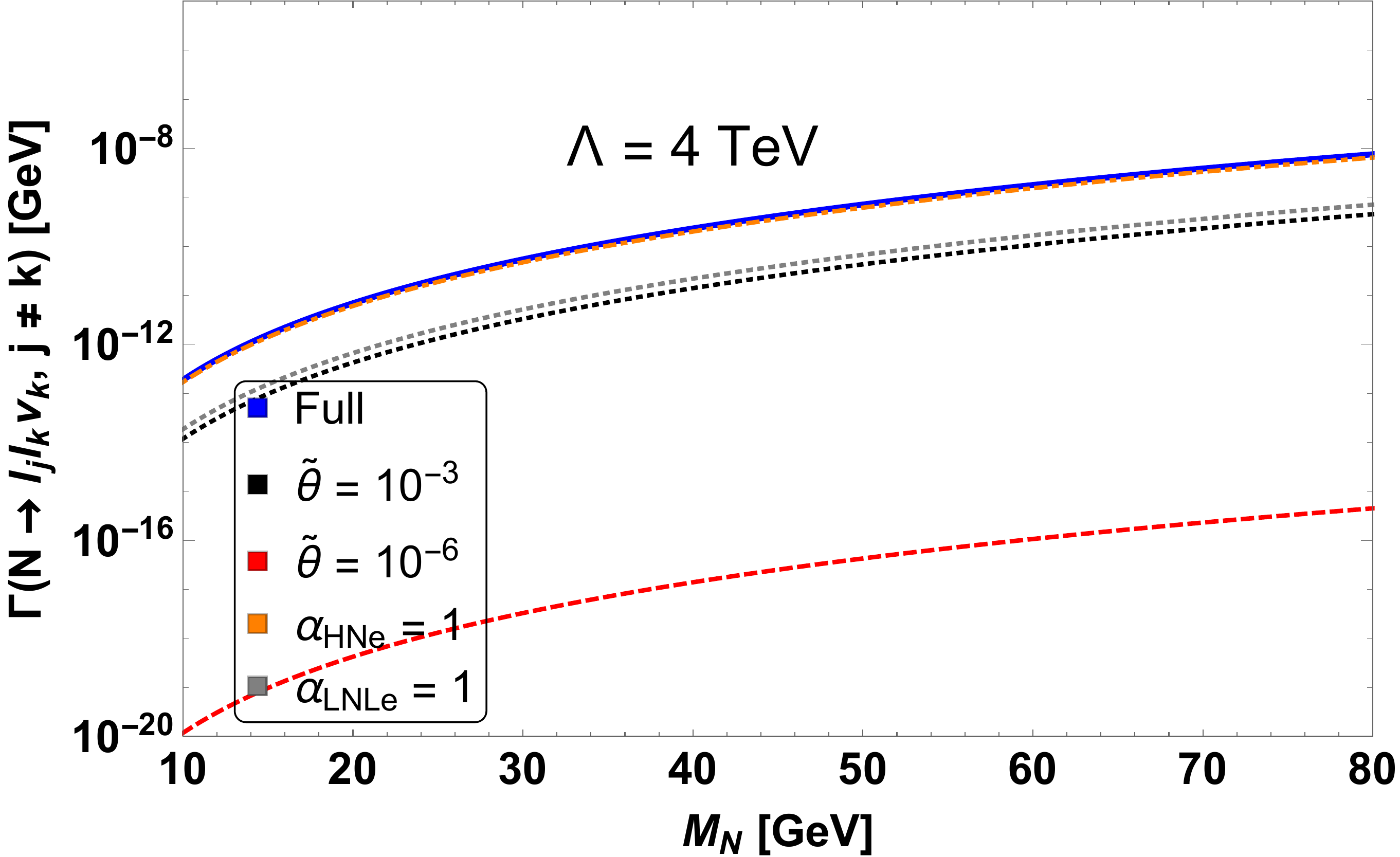}
\caption{Partial decay width corresponds to the decay mode $\Gamma\left(N_{i} \to \ell_{j}\,\ell_{k}\,\nu_{k}; \, j \neq k \right)$ for RHN mass ranging from 10~GeV to 80 GeV. The orange dot-dashed, black dotted~(red dashed) and grey dotted line stands for the individual contribution coming from $\mathcal{O}_{HNe}$, $\mathcal{L}_{CC}$ with $\tilde{\theta} = 10^{-3}~(\tilde{\theta} = 10^{-6})$ and $\mathcal{O}_{LNLe}$, respectively. The blue thick line denotes the total decay taking into account all the contributions. The left panel is for cut-off scale $\Lambda=500$ GeV and 4 TeV, respectively.}
\label{Fig:llnu}
\end{figure}

Considering operators upto $d = 5$ the three body decay modes are primarily mediated via the off shell decay of SM electroweak gauge bosons. In principle the RHN field can also decay into three body modes via off shell Higgs boson decay $N_{i} \to \nu_{j} h^{*}$. Upon production, this off shell Higgs boson $h^{*}$ can decay into SM light fermions. However the couplings between the Higgs boson and SM fermions are dependent on the light fermion masses which have negligible values. As a result, the partial width corresponds to $N$ field decay via off shell Higgs boson is minuscule and can be neglected in practical calculation. In case of dimension six one can write different four fermi operators. The contact interaction coming from these operators would significantly modify various three body decays. In Table.~\ref{tab:Decay6}, we illustrate these decay modes along with the Lagrangian term that will contribute to those processes. The latin indices $i, j$ and $k$ denote the flavour of the daughter leptons and the greek indices $\alpha, \beta$ and $\gamma$ are restricted for quark labels. The term $\mathcal{L}_{CC}$ stands for the $W$ boson mediated decay and $\mathcal{L}_{NC}$ denotes the decay via $Z$ boson. In contrast to Table.~\ref{Tab:vertex5}, Table.~\ref{tab:vertex6} and Table.~\ref{tab:vertexfermi} one can notice that we have only shown certain operators responsible for these decay modes. The rationale behind this choice will be argued for individual scenarios in the subsequent discussion.   
\begin{table}[h!]
\centering
\begin{tabular}{|p{5.5cm}||p{8cm}|}
\hline
~~~~Decay &~~~~~ Contributing Operators \\
\hline 
~~~~$\Gamma\left(N_{i} \to \ell_{j}\,\ell_{k}\,\nu_{k}; \, j \neq k \right)$ :=&~~~~~ $\mathcal{L}_{CC}, \, \mathcal{O}_{HNe}, \, \mathcal{O}_{LNLe}$ \\
\hline
~~~~$\Gamma\left(N_{i} \to \nu_{j}\,\ell_{k}\,\ell_{k}; \, j = k \right)$ :=&~~~~~ $\mathcal{L}_{CC}, \, \mathcal{L}_{NC}, \, \mathcal{O}_{HNe}, \, \mathcal{O}_{LNLe}$ \\
\hline
~~~~$\Gamma\left(N_{i} \to \nu_{j}\,\ell_{k}\,\ell_{k}; \, j \neq k \right)$ :=&~~~~~ $\mathcal{L}_{NC}, \, \mathcal{O}_{LNLe}$ \\
\hline
~~~~$\Gamma\left(N_{i} \to \ell_{j}\, u_{\alpha} \, \bar{d}_{\beta}; \, \alpha \neq \beta\right)$ := &~~~~~ $\mathcal{L}_{CC}, \, \mathcal{O}_{HNe}, \, \mathcal{O}_{QuNL}, \, \mathcal{O}_{Nedu}, \mathcal{O}_{LNq}, \mathcal{O}_{LdqN}$ \\ 
\hline
~~~~$\Gamma\left(N_{i} \to \nu_{j}\, u_{\alpha}\, \bar{u}_{\alpha}\right)$ := &~~~~~ $\mathcal{L}_{NC}, \, \mathcal{O}_{QuNL}$ \\
\hline
 ~~~~$\Gamma\left(N_{i} \to \nu_{j}\, d_{\alpha}\, \bar{d}_{\alpha}\right)$ := &~~~~~ $\mathcal{L}_{NC}, \, \mathcal{O}_{LdqN}$ \\
\hline
~~~~$\Gamma\left(N_{i} \to \nu_{j}\, \nu\, \bar{\nu}\right)$ := &~~~~~ $\mathcal{L}_{NC}$ \\
\hline
\end{tabular}
\caption{Various three body decay modes along with the operators that contribute to the process. Depending upon the final state flavour label of the particles, the operators might vary.}
\label{tab:Decay6}
\end{table}
\begin{itemize}
\item $\bf{\Gamma\left(N_{i} \to \ell_{j}\,\ell_{k}\,\nu_{k}; \, j \neq k \right)}$ 
\end{itemize}
This decay is mediated either via $W$ boson off-shell decay, dimension six operator $\mathcal{O}_{HNe}$ or via four fermi decay operator $\mathcal{O}_{LNLe}$. The diagram that involves $W$ boson can get contributions from $\mathcal{O}_{HNe}$ and $\mathcal{O}_{LNW}$ on top of tree level charged current $\mathcal{L}_{CC}$. However we have discussed before that the vertex correspond to $\mathcal{O}_{LNW}$ operator is loop suppressed. Hence the effect of this is not sizeable enough with respect to the other operators for our choice of parameters and will be ignored hereafter. The matrix element correspond to the $W$ mediated processes can be expressed in the following fashion 
\begin{equation}
\mathcal{M}_{W} = \mathcal{M}_{CC} + \mathcal{M}^{(6)}_{HNe} =  - \frac{g^{2}}{2M^{2}_{W}}\overline{u}(k_{1})\gamma^{\mu}\left(\tilde{\theta}^{ji}P_{L} + \frac{v^{2}}{\Lambda^{2}}\alpha^{ij}_{HNe}P_{R}\right)u(p)\overline{u}(k_{2})\gamma_{\mu}P_{L}v(k_{3}) ,
\label{Eq:AmpW}
\end{equation}    
\noindent
where $p$ is the momentum of decaying right handed neutrino and $k_{1}, k_{2}, k_{3}$ are the momentums of the outgoing leptons ($k_{1}$, $k_{3}$ correspond to singly charged leptons and $k_{2}$ corresponds to light neutrino) respectively. Along with that, the four fermi operator $\mathcal{O}_{LNLe}$ will participate in this process and the scattering matrix can be expressed as

\begin{equation}
\mathcal{M}^{(6)}_{LNLe} = - \frac{\alpha^{jikk}}{2\Lambda^{2}}\overline{u}(k_{1})P_{R}u(p)\overline{u}(k_{2})P_{R}v(k_{3}) + \frac{\alpha^{jikk}}{8\Lambda^{2}}\overline{u}(k_{1})\sigma_{\mu\nu}u(p)\overline{u}(k_{2})\sigma^{\mu\nu}v(k_{3}).
\label{Eq:AmpLNLe}
\end{equation}
\noindent
In a more general set up the dimension six operator $\mathcal{O}_{eN}$ and $\mathcal{O}_{LN}$ grant non-zero effects. However the coupling associated with these operators are proportionate with the mixing angle in addition to the quadratic cut-off scale suppression. As an outcome, one can safely ignore their effect while calculating the partial decay width. Adding Eq.~\ref{Eq:AmpW} and Eq.~\ref{Eq:AmpLNLe} one can find the total amplitude associated with the process $\Gamma\left(N_{i} \to \ell_{j}\,\ell_{k}\,\nu_{k}; \, j \neq k \right)$
\begin{equation}
\mathcal{M}_{\text{total}}(\Gamma\left(N_{i} \to \ell_{j}\,\ell_{k}\,\nu_{k}; \, j \neq k \right)) = \mathcal{M}_{W} + \mathcal{M}^{(6)}_{LNLe}. 
\end{equation}
\noindent 
Taking square of this total amplitude and performing the full phase space integral one can obtain the partial decay width for this process. The final form of this can be explicitly expressed as 
 
 \begin{align}
\Gamma\left(N_{i} \to \ell_{j} \ell_{k} \nu_{k}; j \neq k\right) & = \frac{M^{5}_{N}}{512\pi^{3}}\{\left(\frac{g^{4}}{M^{4}_{W}}\left(|A|^{2} + |B|^{2}\right)+ \frac{7}{4}|C|^{2}\right)\mathcal{I}_{1}\left(x_{\nu_{k}}, x_{\ell_{k}}, x_{\ell_{j}}\right)\nonumber  \\
									&~ + 6|C|^{2}\mathcal{I}_{5}\left(x_{\nu_{k}}, x_{\ell_{k}}, x_{\ell_{j}}\right) + \frac{2g^{4}\text{Re}[A^{*}B]}{M^{4}_{W}}\mathcal{I}_{2}\left(x_{\nu_{k}}, x_{\ell_{k}}, x_{\ell_{j}}\right) \nonumber \\
								         &~ + \frac{g^{2}\text{Re}[A^{*}C]}{M^{2}_{W}}\mathcal{I}_{3}\left(x_{\nu_{k}}, x_{\ell_{k}}, x_{\ell_{j}}\right) + \frac{g^{2}\text{Re}[B^{*}C]}{M^{2}_{W}}\mathcal{I}_{4}\left(x_{\nu_{k}}, x_{\ell_{k}}, x_{\ell_{j}}\right)\}, \label{Eq:D2lnujneqk} 
\end{align}
\noindent
where the co-efficients A, B and C stand for
\[A = \frac{v^{2}\alpha_{HNe}}{\Lambda^{2}},~~~~B = \tilde{\theta},~~~~C = \frac{\alpha_{LNLe}}{\Lambda^{2}}.\]
\noindent
The explicit form of the integrals $\mathcal{I}_{i} (i = 1~\text{to}~5)$ are given in Appendix.~\ref{App:Int}. From Eq.~\ref{Eq:AmpLNLe} one can notice that the operator $\mathcal{O}_{LNLe}$ has two distinct Lorentz structure, one is independent of $\gamma^{\mu}$ matrices and the other is dependent of $\sigma^{\mu\nu}$. As a result, one can see the pure terms arise from this operator is dependent on two integrals $\mathcal{I}_{1}$ and $\mathcal{I}_{2}$ respectively. The other three integrals $\mathcal{I}_{i}$ ($i =$ 2 to 4) arise due respective interference terms between $A$, $B$ and $C$. To realise the impact of each of these terms one needs to digrammatically express this for different values of mixing angle and the cut-off scale $\Lambda$. In Fig.~\ref{Fig:llnu}, we have shown the partial decay width of this channel where we take the mixing angle and relevant Wilson coefficients consistent with the experimental bounds.
In each panel of Fig.~\ref{Fig:llnu} the orange dot-dashed, black dotted~(red dashed) and grey dotted line represents the individual effect of $\mathcal{O}_{HNe}$, $\mathcal{L}_{CC}$ with $\tilde{\theta} = 10^{-3}~(\tilde{\theta} = 10^{-6})$ and $\mathcal{O}_{LNLe}$ respectively. The blue thick line denotes the total decay width which consists of individual operator contribution as well as the corresponding interference term. Note that when we are showing the total decay width we choose the value of mixing angle as $\tilde{\theta}=10^{-3}$. The difference between the left and right panel of Fig.~\ref{Fig:llnu} is the cut-off scale and different Wilson coefficient consistent with the experimental constraint for that cut-off scale. From the figure one can see the dominant contributions are usually coming from the dimension six operators. One can obtain a rough estimate of the significance of these operator while calculating the ratio between $A$, $B$ and $C$.
\begin{itemize}
\item $\bf{\Gamma\left(N_{i} \to \nu_{j}\,\ell_{k}\,\ell_{k}; \, j = k \right)}$ 
\end{itemize}
In this case as the flavour labels $j$ and $k$ is same, in addition to the operators such as $\mathcal{L}_{CC}, \, \mathcal{O}_{HNe}, \, \mathcal{O}_{LNLe}$, one needs to add the neutral current contribution which mediates via $Z$ boson propagation. If we only consider the renormalisable part of the Lagrangian, the coupling $\mathcal{C}^{Z_{\mu}}_{\overline{\nu}N + \overline{N}\nu}$ is dependent on the active sterile mixing angle. Moreover in case of $N_{R}$-EFT this coupling gets modification both from dimension five as well as dimension six operators. In dimension five, the operator $\mathcal{O}^{(5)}_{3}$ and in dimension six, three operators $\mathcal{O}_{HN}$, $\mathcal{O}_{NB}$ and $\mathcal{O}_{LNW}$ give some contributions which might not be tangible for the present calculation. For example, the relevant coupling coming from $\mathcal{O}_{HN}$ operator is proportional to mixing angle as well as the $\frac{\alpha_{HN}}{\Lambda^{2}}$, which suggest that one can ignore this term. On the other hand, the rest of these three EFT operators are loop suppressed. So same like charged current, one can avoid its inclusion into the calculation. Bringing all these points together, one can write down the matrix element for the $Z$ boson mediated digram in the following manner\footnote{We define $g_{L} = 2\sin^{2}\theta_{w} - 1$ and $g_{R} = 2\sin^{2}\theta_{w}$ and $\sin^{2}\theta_{w}$ is the \emph{Weinberg} angle.}     
\begin{equation}
\mathcal{M}_{Z} = - \frac{g^{2}}{2M^{2}_{W}}\overline{u}(k_{1})\gamma^{\mu}\theta^{ki}P_{L}u(p)\overline{u}(k_{2})\gamma_{\mu}\left(g_{L}P_{L} + g_{R}P_{R}\right)v(k_{3}).
\label{Eq:AmpZ}
\end{equation}
\noindent
Along with the $\mathcal{M}_{Z}$ one should take into account of $\mathcal{M}_{W}$ and $\mathcal{M}_{LNLe}$ which are illustrated in Eq.~\ref{Eq:AmpW} and Eq.~\ref{Eq:AmpLNLe} respectively. The total scattering matrix for this process is 
\begin{equation}
\mathcal{M}_{\text{total}}(\left(N_{i} \to \nu_{j}\,\ell_{k}\,\ell_{k}; \, j = k \right)) = \mathcal{M}_{Z} + \mathcal{M}_{W} + \mathcal{M}_{LNLe}.
\end{equation}
\noindent
Using this one can compute the partial decay width of this process which takes the subsequent analytic structure - 
\begin{align}
\Gamma\left(N_{i} \to \ell_{k} \ell_{k} \nu_{k} \right) & = \frac{M^{5}_{N}}{512\pi^{3}}\{\left(\frac{g^{4}}{M^{4}_{W}}\left(|A|^{2} + |B|^{2}\left(g^{2}_{R} + (g_{L} - 1)^{2}\right)\right)+ \frac{7}{4}|C|^{2}\right)\mathcal{I}_{1}\left(x_{\ell_{k}}, x_{\ell_{k}}, x_{\nu_{k}}\right) \nonumber \\
									&~ + 6|C|^{2}\mathcal{I}_{5}\left(x_{\ell_{k}}, x_{\ell_{k}}, x_{\nu_{k}}\right) + \frac{3g^{2}\text{Re}[A^{*}C]}{M^{2}_{W}}\mathcal{I}_{2}\left(x_{\ell_{k}}, x_{\ell_{k}, x_{\nu_{k}}}\right) \nonumber \\
									&~+ \left(\frac{2g^{2}\text{Re}[A^{*}C]}{M^{2}_{W}} + \frac{2|B|^{2}g^{4}}{M^{4}_{W}}g_{R}(g_{L} - 1)\right)\mathcal{H}_{1}\left(x_{\nu_{k}}, x_{\ell_{k}}\right) \nonumber \\
									&~+ \left(\frac{g^{4}\text{Re}[A^{*}B]}{M^{4}_{W}}(g_{L} - g_{R} - 1) + \frac{g^{2}\text{Re}[B^{*}C]}{M^{2}_{W}}(2(g_{L} - 1)- g_{R})\right)\mathcal{H}_{2}\left(x_{\nu_{k}}, x_{\ell_{k}}\right) \nonumber\\
									&~+ \frac{3\text{Re}[B^{*}C]}{2M^{2}_{W}}g^{2}(g_{L} - 1)\mathcal{H}_{3}\left(x_{\nu_{k}}, x_{\ell_{k}}\right)\} \label{Eq:2lnujeqk}
\end{align}
where the co-efficient $A$, $B$ and $C$ signifies
\[~~~A = \frac{v^{2}\alpha_{HNe}}{\Lambda^{2}},~~~~B = \tilde{\theta},~~~~C = \frac{\alpha_{LNLe}}{\Lambda^{2}}.\]
The explicit form of the integrals $\mathcal{H}_{i}$ (where $i$ = 1 to 3) is given in Appendix.~\ref{App:Int}. The $Z$ boson couplings to SM fermions do vary depending upon the chirality. As a result in Eq.~\ref{Eq:2lnujeqk}, the term correspond to $|B|^{2}$ has both $g_{L}$ and $g_{R}$ dependence. The dependence on the active-sterile mixing angle also come from $W$ boson mediated process which justify the $(g_{L} - 1)$ factor in that term. It is important to highlight that, the existing SM couplings receive some alteration in the EFT setup - examples include $g_{Z\nu\overline{\nu}}$ or $g_{W\ell\nu}$. However, we have assumed all those correction to be negligible as they are both cut-off scale and mixing suppressed. 
 \begin{figure}[h!]
\centering
\includegraphics[width=0.45\textwidth,height=0.25\textheight]{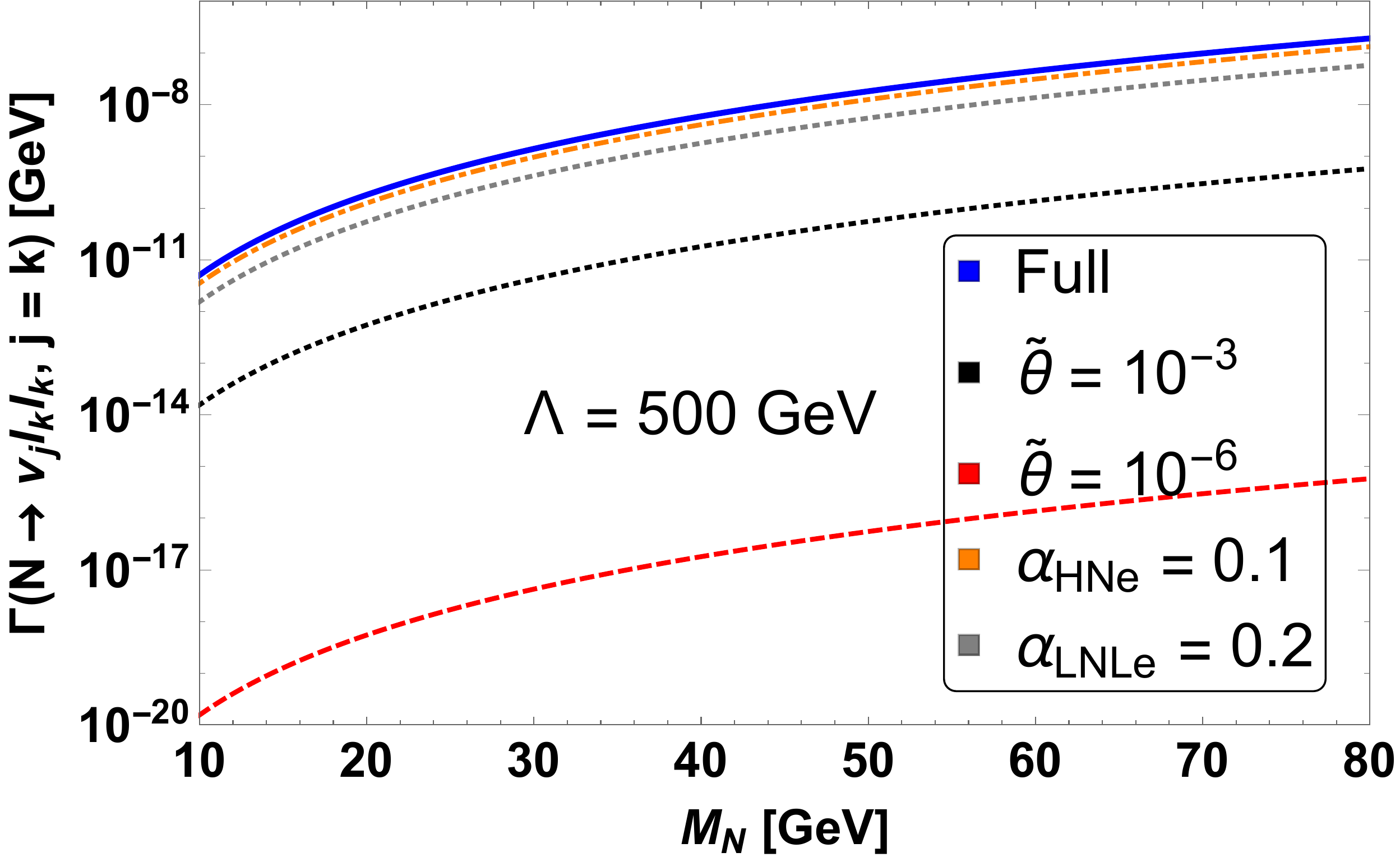}
\hspace{0.25cm}
\includegraphics[width=0.45\textwidth,height=0.25\textheight]{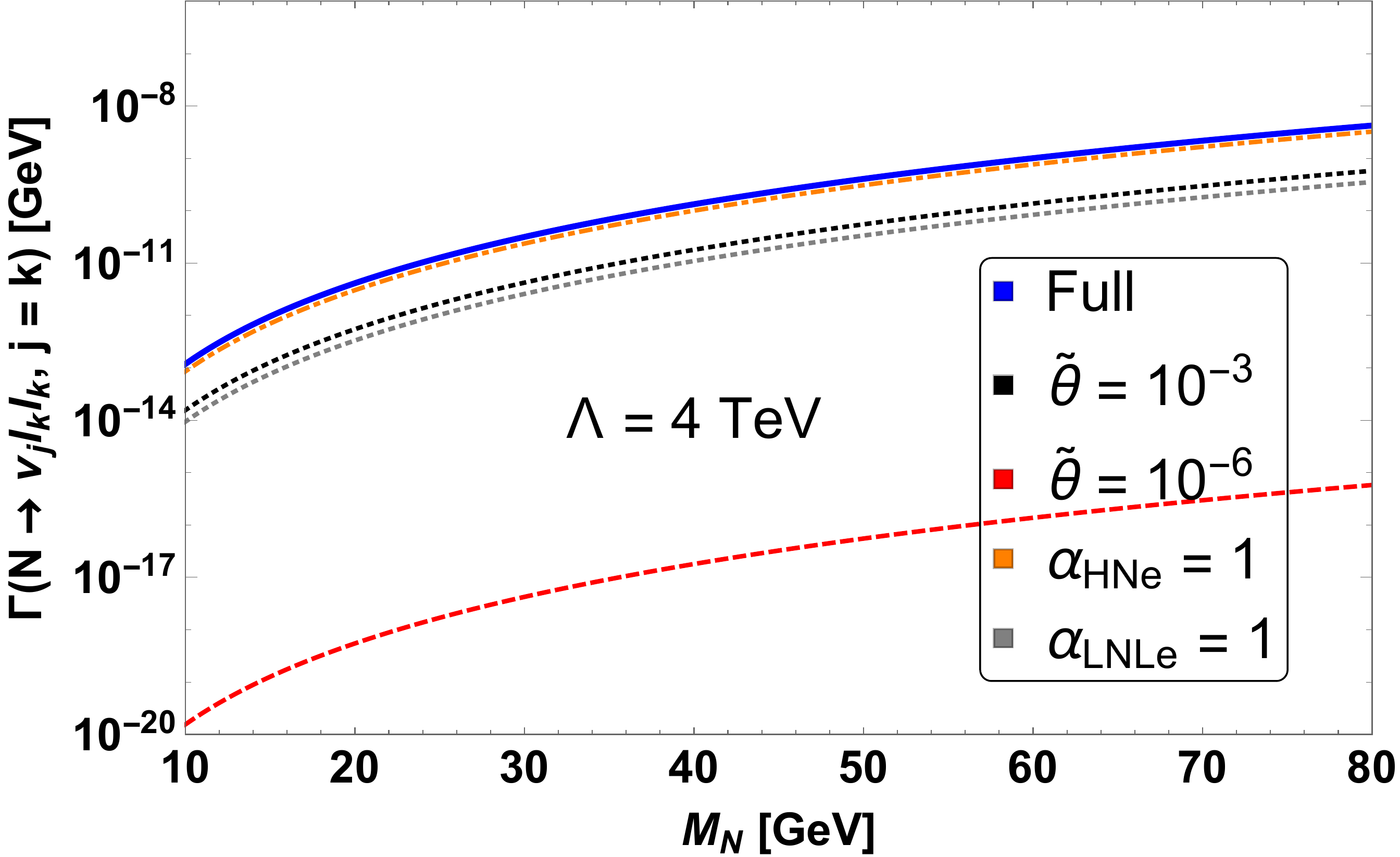}
\caption{Partial decay width corresponds to the decay mode $\Gamma\left(N_{i} \to \nu_{j}\,\ell_{k}\,\ell_{k}; \, j = k \right)$ for RHN mass ranging from 10~GeV to 80 GeV. The meaning of color code is same as in Fig.~\ref{Fig:llnu}. }
\label{Fig:nu2ljeqk}
\end{figure}
As expected, the pure term coming from $A$ and $C$ remains same as of Eq.~\ref{Eq:D2lnujneqk}, but the interference effect do vary due to the emergence of $Z$ propagating decay mode. In Fig.~\ref{Fig:nu2ljeqk}, we illustrate the effect of these operators to this decay mode and the color code is same as in Fig.~\ref{Fig:llnu}. 
 
\begin{figure}[h!]
\centering
\includegraphics[width=0.45\textwidth,height=0.25\textheight]{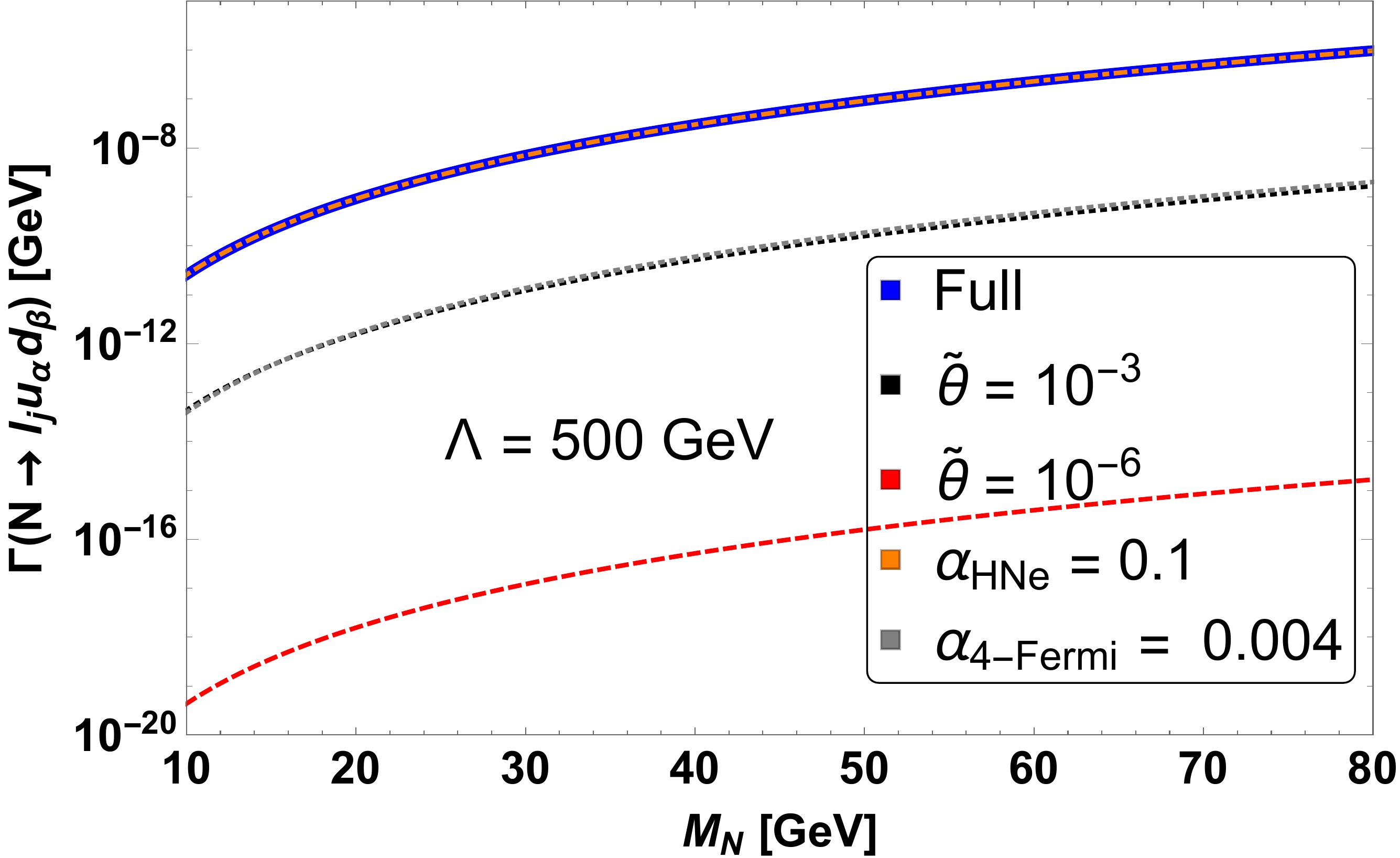}
\hspace{0.25cm}
\includegraphics[width=0.45\textwidth,height=0.25\textheight]{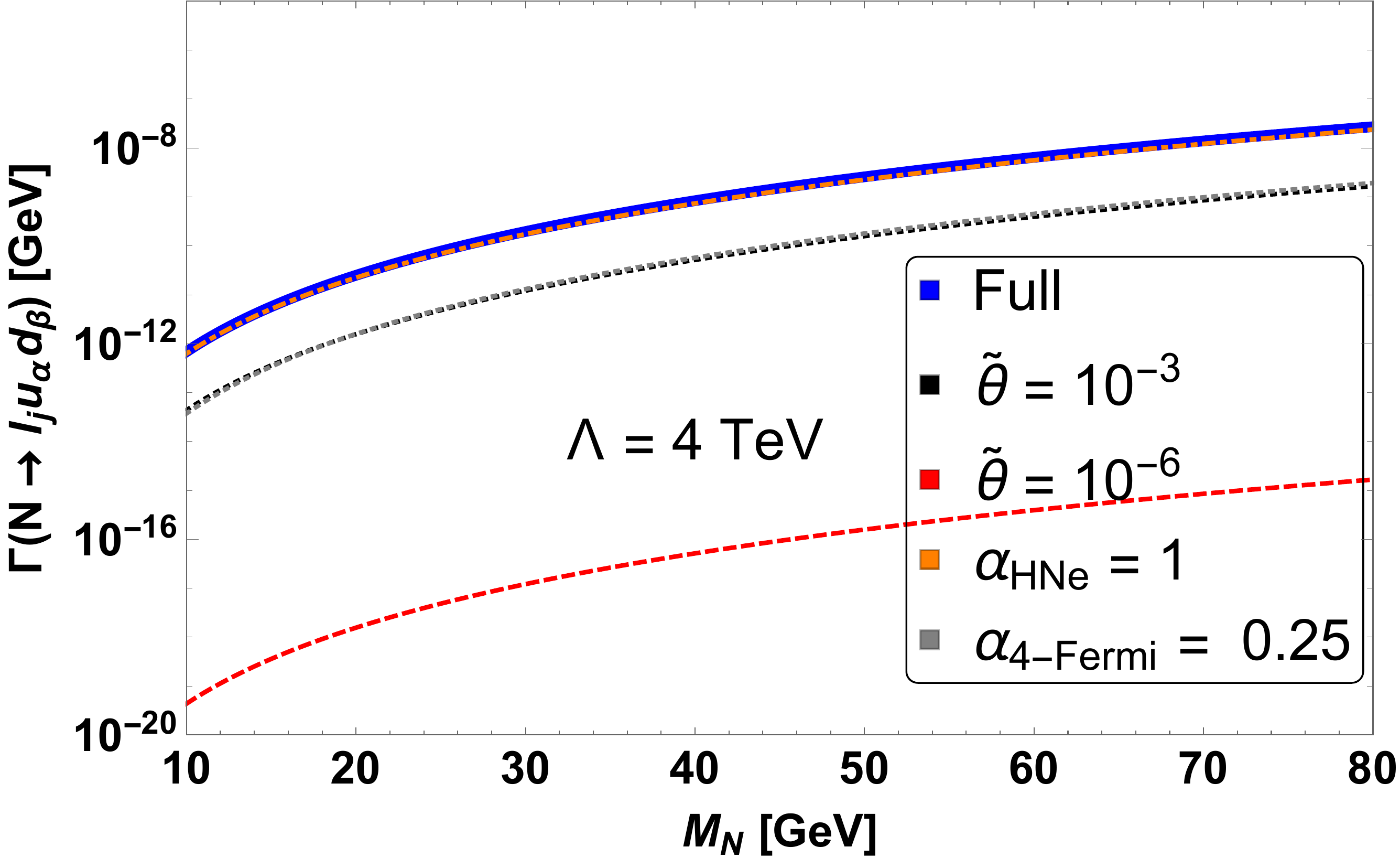}
\caption{Partial decay width corresponds to the decay mode $\Gamma\left(N_{i} \to \ell_{j}\, u_{\alpha} \, \bar{d}_{\beta}; \, \alpha \neq \beta\right)$ for RHN mass ranging from 10~GeV to 80 GeV. In each panel, the black dotted or red dashed, orange dot-dashed and grey line stands for the contribution coming from the mixing angle, $\mathcal{O}_{HNe}$ and from combination of four fermi operator. The blue thick line represents the total contribution with the assumption of $\tilde{\theta} = 10^{-3}$ and left, right panel is for two different cut-off scale.}
\label{Fig:Nrlud}
\end{figure}

\begin{itemize}
\item $\bf{\Gamma\left(N_{i} \to \nu_{j}\,\ell_{k}\,\ell_{k}; \, j \neq k \right)}$ 
\end{itemize}
\noindent
Apart from these modes the RHNs can potentially decay via another leptonic mode i.e. $N_{i} \to \nu_{j}\,\ell_{k}\,\ell_{k}; \, j \neq k $. The assigned flavour label suggests that, the charged leptons in this three body mode should have same same flavour as opposed to the active neutrino. Invoking the idea of electromagnetic charge conservation one can see that the process like this can only appear via $Z$ boson off-shell decay along with the four fermi operator $\mathcal{O}_{LNLe}$. The matrix element for this process should not contain the $\mathcal{M}_{W}$ term and it can be written as -
\begin{equation}
\mathcal{M}_{\text{total}}\left(N_{i} \to \nu_{j}\,\ell_{k}\,\ell_{k}; \, j \neq k \right) = \mathcal{M}_{Z} + \mathcal{M}_{LNLe} ,
\end{equation} 
 \noindent
 Adopting the explicit form $\mathcal{M}_{Z}$ and $\mathcal{M}_{LNLe}$ as described in Eq.~\ref{Eq:AmpZ} and Eq.~\ref{Eq:AmpLNLe}, one can determine the analytic from of this decay width which can be written as
 \begin{align}
\Gamma\left(N_{i} \to \nu_{j}\ell_{k}\ell_{k}\right) & =  \frac{M^{5}_{N}}{512\pi^{3}}\{\left(\frac{g^{4}|B|^{2}}{M^{4}_{W}}(g^{2}_{L} + g^{2}_{R}) + \frac{7}{4}|C|^{2}\right)\mathcal{I}_{1}\left(x_{\ell_{k}}, x_{\ell_{k}}, x_{\nu_{j}}\right) + \frac{2|B|^{2}g^{4}}{M^{4}_{W}}g_{L}g_{R}\mathcal{H}_{1}\left(x_{\nu_{j}}, x_{\ell_{k}}\right) \nonumber \\
									&~+ 6|C|^{2}\mathcal{I}_{5}\left(x_{\ell_{k}}, x_{\ell_{k}}, x_{\nu_{j}}\right) + \frac{\text{Re}(B^{*}C)}{M^{2}_{W}}g^{2}(2g_{L} + g_{R})\mathcal{H}_{2}\left(x_{\nu_{j}}, x_{\ell_{k}}\right) \nonumber \\
									&~+ \frac{3\text{Re}(B^{*}C)}{2M^{2}_{W}}g^{2}g_{L}\mathcal{H}_{3}\left(x_{\nu_{j}}, x_{\ell_{k}}\right)\}, \label{Eq:lklknuj}
\end{align}
where $B$, $C$ stand for
\[~~~~B = \tilde{\theta},~~~~C = \frac{\alpha_{LNLe}}{\Lambda^{2}}.\]
\begin{figure}[h!]
\centering
\includegraphics[width=0.45\textwidth,height=0.25\textheight]{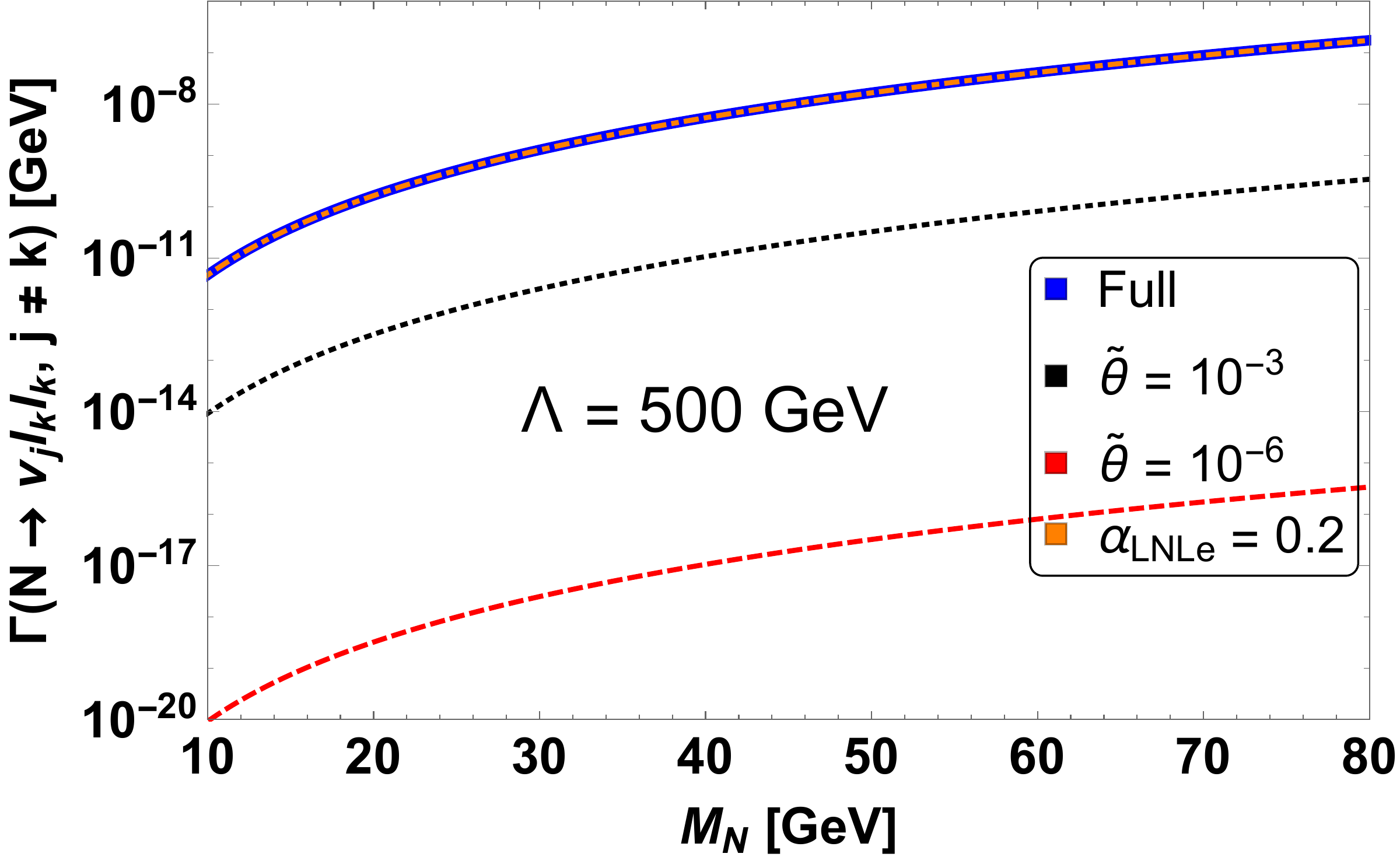}
\hspace{0.25cm}
\includegraphics[width=0.45\textwidth,height=0.25\textheight]{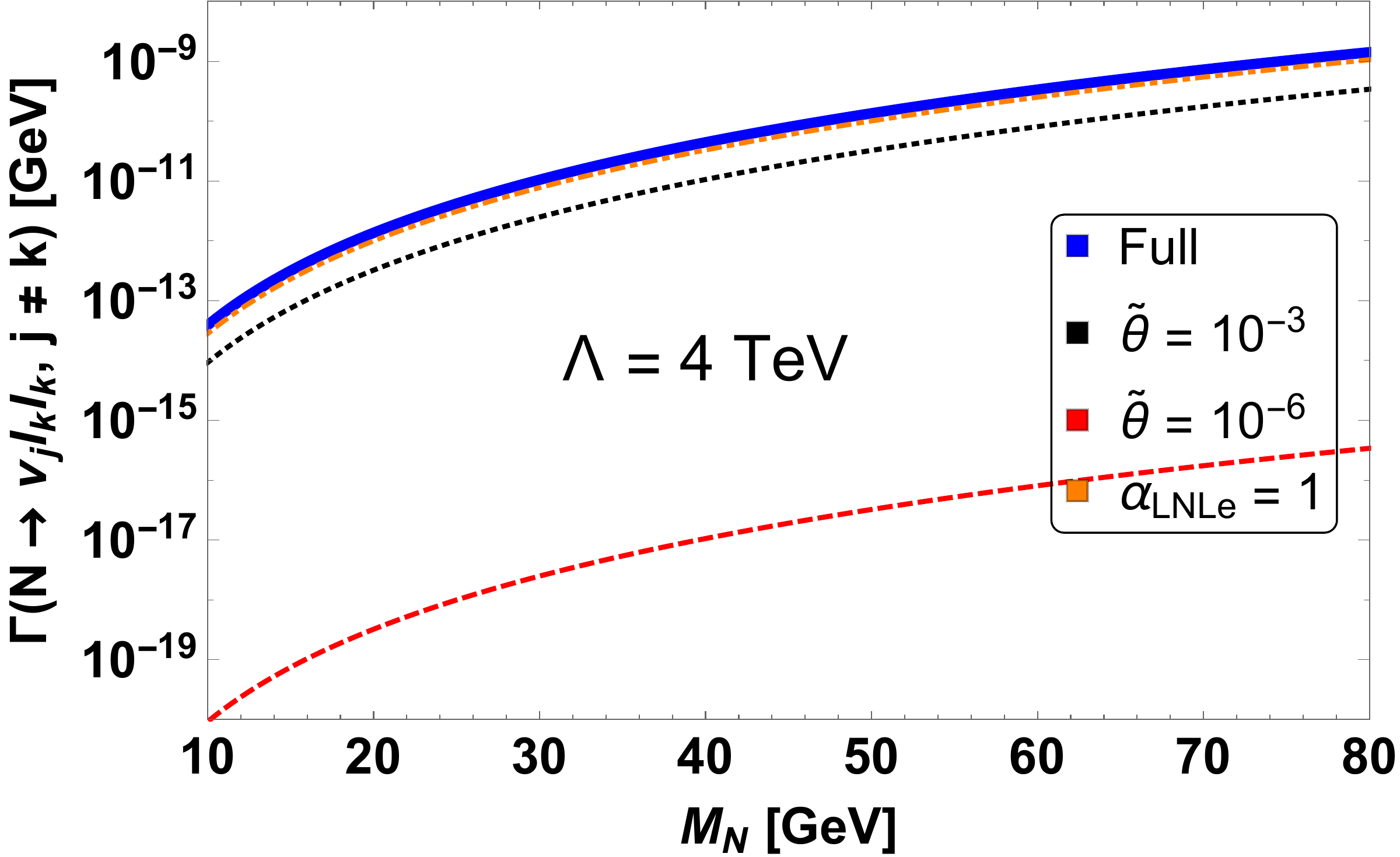}
\caption{Partial decay width corresponds to the decay mode $\Gamma\left(N_{i} \to \nu_{j}\,\ell_{k}\,\ell_{k}; \, j \neq k \right)$ for RHN mass ranging from 10~GeV to 80 GeV. The orange dot-dashed, black dotted~(red dashed) line stands for the contribution coming from $\mathcal{O}_{LNLe}$ and mixing angle $\tilde{\theta} = 10^{-3}$~($\tilde{\theta} = 10^{-6}$). The blue thick line represnts the total contribution with the assumption of $\tilde{\theta}=10^{-3}$. Left and right panel are for cut-off scale $\Lambda=500$ GeV and 4 TeV, respectively.}
\label{Eq:nujlklk}
\end{figure} 
 In Fig.~\ref{Eq:nujlklk}, we illustrate the effects of various operators to this decay mode. The orange dot-dashed line represents the impact of $\mathcal{O}_{LNLe}$ operator whereas the black dotted~(red dashed) line shows the effect of mixing angle $\tilde{\theta} = 10^{-3}$~($\tilde{\theta} = 10^{-6}$) that is coming from $Z$ boson mediated decay. We see that the dominating contribution comes from the operator $\mathcal{O}_{LNLe}$. The blue thick line stands for the total contribution taking into account both the mixing part~($\tilde{\theta} = 10^{-3}$) and dimension six operator. The left and right panel is for two different cut-off scale $\Lambda=500$ GeV and 4 TeV, respectively. 
\begin{itemize}
\item $\bf{\Gamma\left(N_{i} \to \ell_{j}\, u_{\alpha} \, \bar{d}_{\beta}; \, \alpha \neq \beta\right)}$ 
\end{itemize} 
So far we have discussed the RHN decays via leptonic modes. The RHN can also decay via hadronic modes along with either a charged leptons or a light neutrino. We begin our discussion about RHN decay to a semi-leptonic final state - $N_{i} \to \ell_{j}\, u_{\alpha} \, \bar{d}_{\beta}; \, \alpha \neq \beta$. The presence of both up and down type quark along with charged leptons indicates a charge current mediated process. Hence, in evaluating this contribution, we consider the standard renormalisable charge current interaction part. There will be also contribution from dimension six operators such as, $\alpha_{HNe}$ and $\alpha_{LNW}$. Similar to before we consider, $\alpha_{LNW}$ to be zero due to loop suppression.  In addition to that, a tower of four-fermi operators would give appreciable contribution in this process. From Table.~\ref{tab:Ope6}, one can see, these operators are $\mathcal{O}_{QuNL}$, $\mathcal{O}_{duNe}$, $\mathcal{O}_{LNqd}$ and $\mathcal{O}_{LdqN}$. There contribution to this decay width is given as
\begin{align}
\mathcal{M}^{(6)}_{QuNL} & = \frac{\alpha^{*\beta\alpha ij}_{QuNL}}{\Lambda^{2}}\overline{u}(k_{1})P_{R}u(p)\overline{u}(k_{2})P_{L}v(k_{3}), \nonumber\\
\mathcal{M}^{(6)}_{duNe} & = \frac{\alpha^{*ij\beta\alpha}_{duNe}}{\Lambda^{2}}\overline{u}(k_{1})\gamma^{\mu}P_{R}u(p)\overline{u}(k_{2})\gamma_{\mu}P_{R}v(k_{3}), \nonumber\\
\mathcal{M}^{(6)}_{LNqd} & = - \frac{\alpha^{ij\alpha\beta}_{LNqd}}{\Lambda^{2}}\overline{u}(k_{1})P_{R}u(p)\overline{u}(k_{2})P_{R}v(k_{3}), \nonumber\\
\mathcal{M}^{(6)}_{LdqN} & = - \frac{\alpha^{j\beta\alpha i}_{LdqN}}{2\Lambda^{2}}\overline{u}(k_{1})P_{R}u(p)\overline{u}(k_{2})P_{R}v(k_{3}) - \frac{\alpha^{j\beta\alpha i}_{LdqN}}{8\Lambda^{2}}\overline{u}(k_{1})\sigma_{\mu\nu}u(p)\overline{u}(k_{2})\sigma^{\mu\nu}v(k_{3}).\label{Eq:Mlud} 
\end{align}
Considering the chirality of the fermion fields as well as the intrinsic space-time transformation properties one can identify the distinction among each of these operator. For example, both the matrix elements $\mathcal{M}_{QuNL}$ and $\mathcal{M}_{LNqd}$ correspond to charged scalar mediated graph in high scale UV complete model. On the other hand, matrix element $\mathcal{M}_{duNe}$ hints upon a vector-like charged currents which one can realise in BSM theories with $SU(2)_{R}$ extensions. The operator $\mathcal{O}_{LdqN}$ contains two separate Lorentz structure similar to $\mathcal{O}_{LNLe}$. Using appropriate Fierz transformation one can separate the $\mathcal{M}^{(6)}_{LdqN}$ matrix into a scalar as well as tensor objects. Adding the matrix elements that are mentioned in Eq.~\ref{Eq:Mlud} with $\mathcal{M}_{W}$ and performing the full phase space integrals one can obtain the partial decay width for this channel. In Eq.~\ref{Eq:ljuadb}, we write down its explicit form        
\begin{align}
\Gamma(N_{i} \to \ell_{j}u_{\alpha}\bar{d}_{\beta}) & = \frac{M^{5}_{N}N_{c}}{512\pi^{3}}\{\left(\frac{g^{4}|V_{CKM}|^{2}}{M^{4}_{W}}\left(|A|^{2} + |B|^{2}\right) + 4|C_{2}|^{2} + |C_{1}|^{2} + |C_{3}|^{2} + \frac{3}{2}|C_{4}|^{2}\right)\mathcal{I}_{1}\left(x_{u}, x_{d}, x_{\ell_{j}}\right) \nonumber \\
									     &~ +\frac{2\text{Re}[A^{*}B]g^{4}}{M^{4}_{W}}\mathcal{I}_{2}\left(x_{u}, x_{d}, x_{\ell_{j}}\right) + \left(3|C_{4}|^{2} - \frac{8g^{2}\text{Re}[A^{*}C_{2}]}{M^{2}_{W}}\right)\mathcal{I}_{5}\left(x_{u}, x_{d}, x_{\ell_{j}}\right) \nonumber \\
									     &~ +\left(\frac{3g^{2}\text{Re}[B^{*}C_{4}]}{2M^{2}_{W}} - \frac{g^{2}\text{Re}[A^{*}C_{1}]}{M^{2}_{W}}\right)\mathcal{G}_{1}\left(x_{u}, x_{d}, x_{\ell_{j}}\right) \nonumber \\
									     &~ +\left(\frac{3g^{2}\text{Re}[A^{*}C_{4}]}{2M^{2}_{W}} - \frac{g^{2}\text{Re}[A^{*}C_{3}]}{M^{2}_{W}} - 3\text{Re}[C^{*}_{2}C_{4}]\right)\mathcal{G}_{1}\left(x_{d}, x_{u}, x_{\ell_{j}}\right) \nonumber  \\
									     &~ +\left(\frac{3g^{2}\text{Re}[A^{*}C_{4}]}{2M^{2}_{W}} + \frac{g^{2}\text{Re}[B^{*}C_{1}]}{M^{2}_{W}} + 2\text{Re}[C^{*}_{2}C_{3}] - 3\text{Re}[C^{*}_{2}C_{4}]\right)\mathcal{G}_{2}\left(x_{u}, x_{d}, x_{\ell_{j}}\right) \nonumber \\
									     &~ +\left(\frac{g^{2}\text{Re}[B^{*}C_{3}]}{M^{2}_{W}} + \frac{3g^{2}\text{Re}[B^{*}C_{4}]}{2M^{2}_{W}} + 2\text{Re}[C^{*}_{1}C_{2}]\right)\mathcal{G}_{2}\left(x_{d}, x_{u}, x_{\ell_{j}}\right) \nonumber \\
									     &~ +\left(\frac{4g^{2}\text{Re}[B^{*}C_{2}]}{M^{2}_{W}} - 4\text{Re}[C^{*}_{1}C_{3}]\right)\mathcal{G}_{3}\left(x_{u}, x_{d}, x_{\ell_{j}}\right)\} \label{Eq:ljuadb}
\end{align}
where different vertex factors are defined as 
\[~~~A = \tilde{\theta},~~B = \frac{v^{2}\alpha_{HNe}}{\Lambda^{2}},~~C_{1} = \frac{\alpha_{QuNL}}{\Lambda^{2}},~~C_{2} = \frac{\alpha_{Nedu}}{\Lambda^{2}},~~C_{3} = \left(\frac{\alpha_{LNqd}}{\Lambda^{2}} + \frac{\alpha_{LdqN}}{\Lambda^{2}}\right),~~C_{4} = \frac{\alpha_{LdqN}}{\Lambda^{2}}\]
and $N_{c} = 3$ is the colour factor. The integrals $\mathcal{G}_{i}$ (where $i$ = 1 to 3) are given in Appendix.~\ref{App:Int}. The scalar piece of operator $\mathcal{O}_{LNqd}$ contributes to coefficient $C_{3}$ along with $\mathcal{O}_{LdqN}$, whereas the tensorial part is treated as a separate coefficient $C_{4}$. Without loss of generality, one can choose the value of both these coefficients are in the same order while understanding their role in the decay width formula. In Fig.~\ref{Fig:Nrlud}, we display the decay width as a function of mass $M_N$.    
As before, the black dotted~(red dashed) line correspond to the mixing angle $\tilde{\theta} = 10^{-3}$~($\tilde{\theta} = 10^{-6}$) and the orange dot-dashed line correspond to $\mathcal{O}_{HNe}$, respectively. The grey dotted lines show the combine effects of all four-Fermi operators. The blue thick line stands for the total contribution taking into
account both the mixing part~($\tilde{\theta} = 10^{-3}$) and dimension six operator. The left and right panel is for two different cut-off scale $\Lambda = 500$ GeV and 4 TeV, respectively. For these two cut off scale we set the Wilson coefficient correspond to four-Fermi operators to be $\alpha_{4-\text{Fermi}}=0.004$ and 0.25 respectively which is consistent with the current experimental bounds.

\begin{figure}[h!]
\centering
\includegraphics[width=0.45\textwidth,height=0.25\textheight]{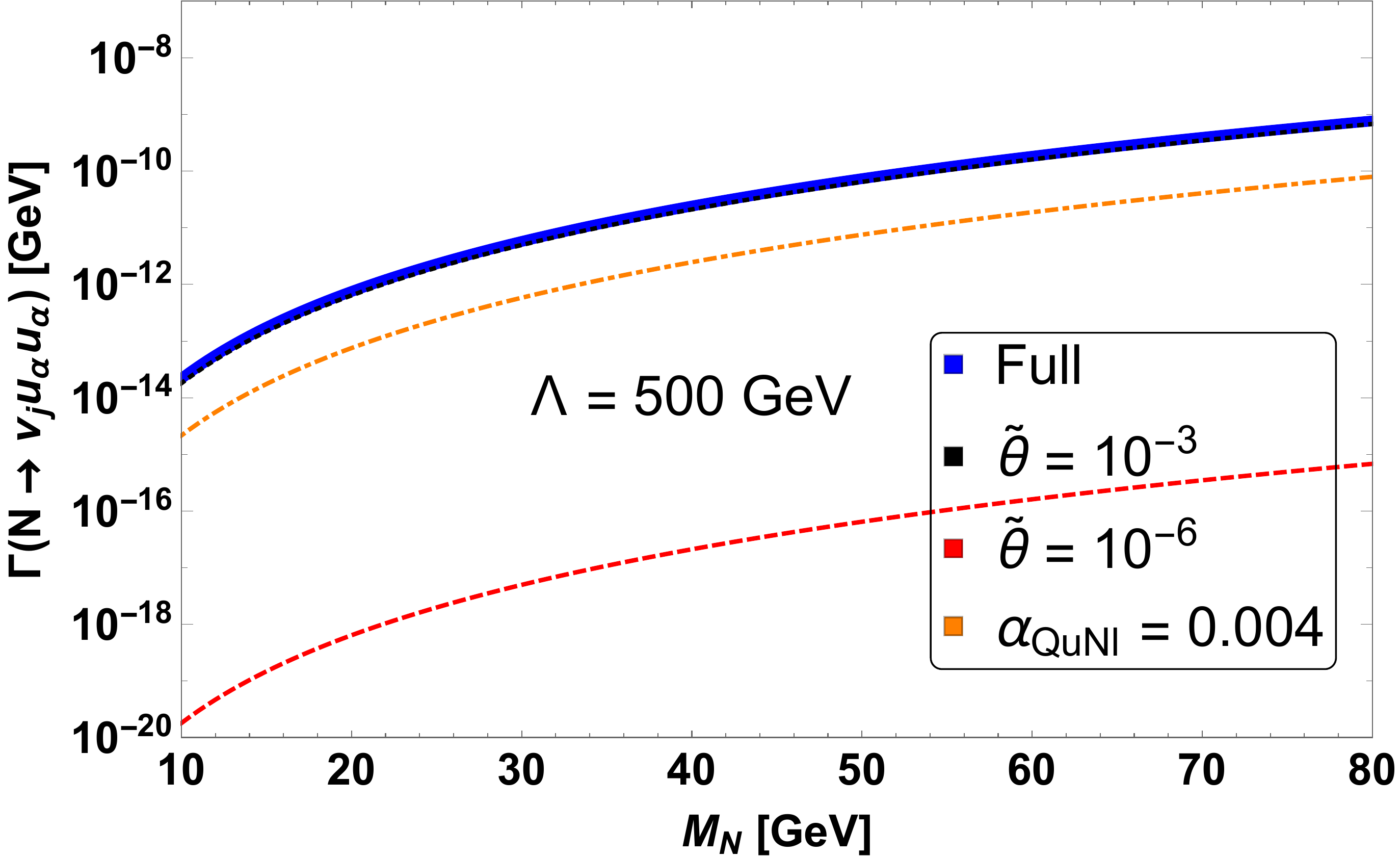}
\hspace{0.25cm}
\includegraphics[width=0.45\textwidth,height=0.25\textheight]{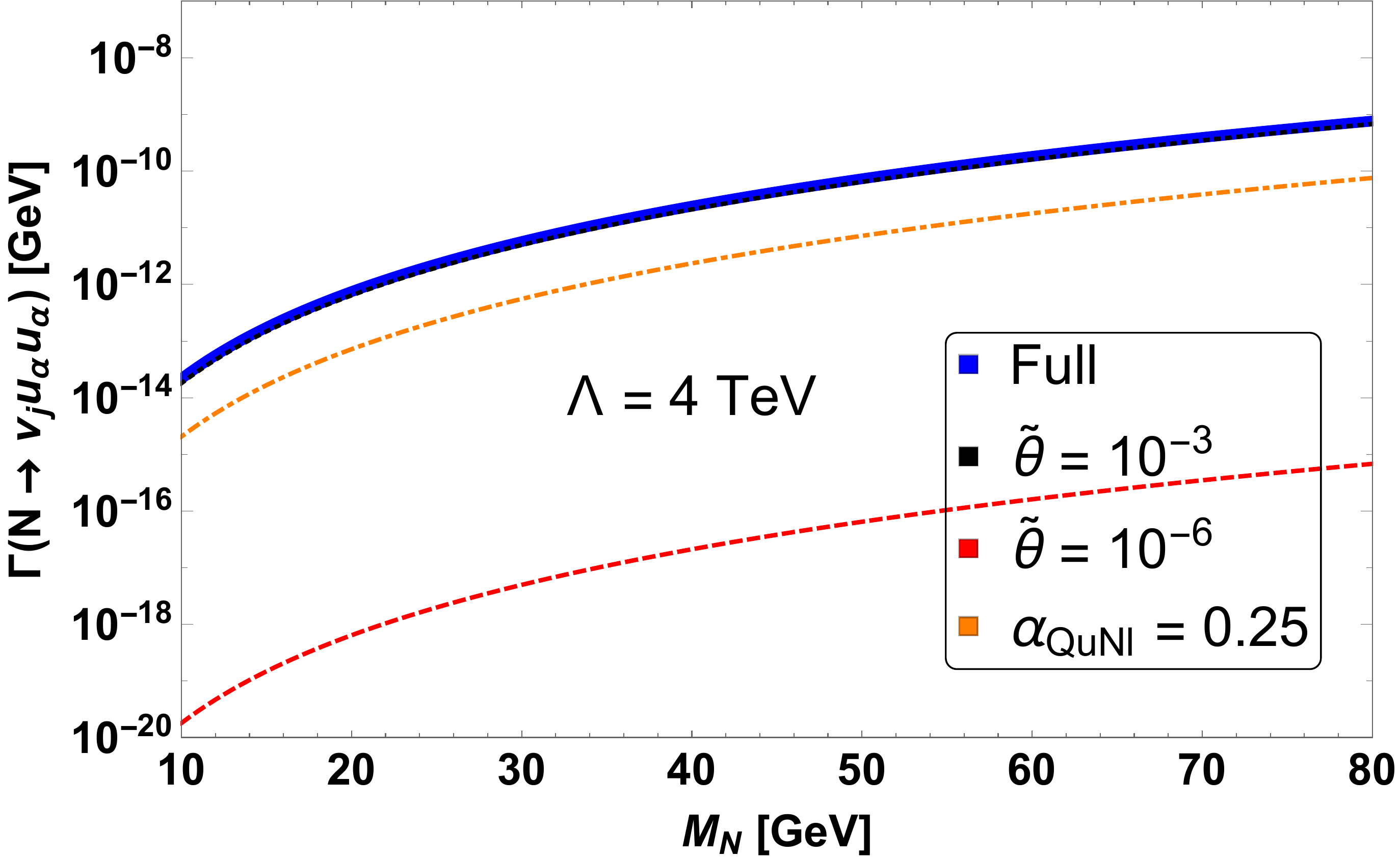}
\caption{Partial decay width corresponds to the decay mode $\Gamma\left(N_{i} \to \nu_{j}\, u_{\alpha}\, \bar{u}_{\alpha}\right)$ of RHN having mass in the range of 10~GeV to 80 GeV. In each panel, the black dotted or red dashed and orange dot-dashed line stands for the contribution coming from the mixing angle and $\mathcal{O}_{QuNl}$. The blue thick line represnts the total contribution with the assumption of $\tilde{\theta} = 10^{-3}$ and left, right panel is for two different cut-off scale.} 
\label{Fig:Nuuu}
\end{figure}

\begin{figure}[h!]
\centering
\includegraphics[width=0.45\textwidth,height=0.25\textheight]{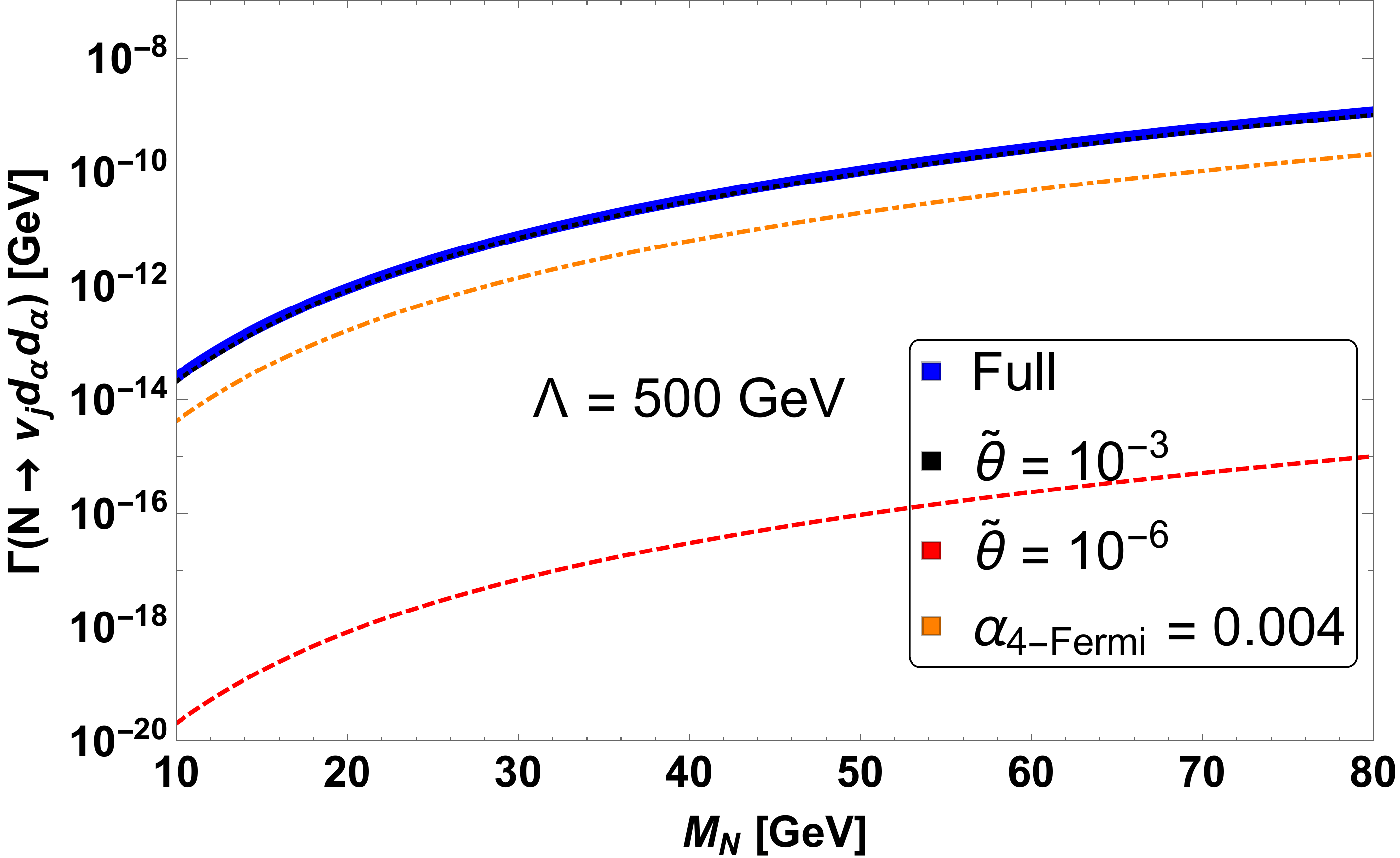}
\hspace{0.25cm}
\includegraphics[width=0.45\textwidth,height=0.25\textheight]{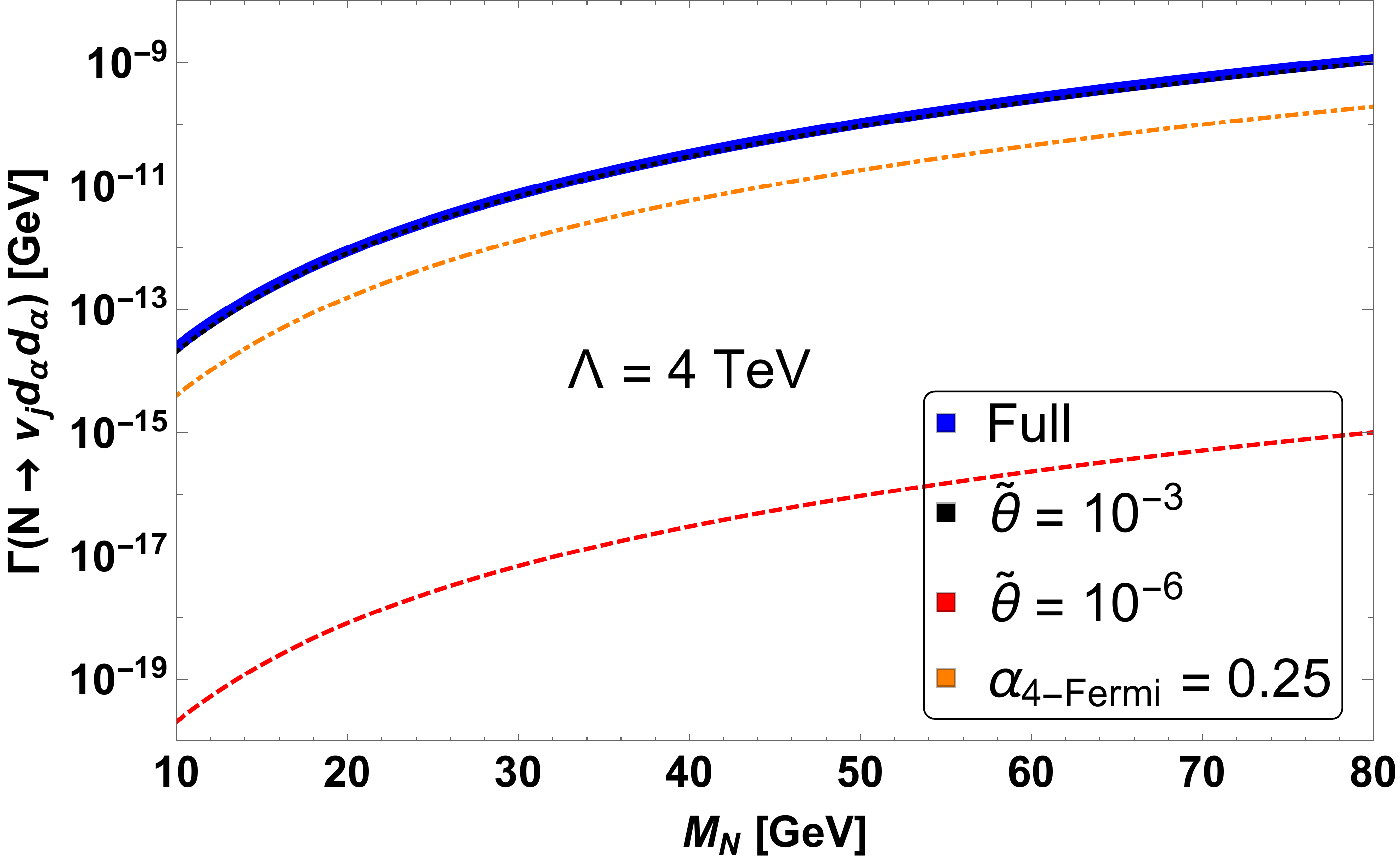}
\caption{Partial decay width corresponds to the decay mode $\Gamma\left(N_{i} \to \nu_{j}\, d_{\alpha}\, \bar{d}_{\alpha}\right)$ of RHN mass ranging from 10~GeV to 80~GeV. The color code is same as in Fig.~\ref{Fig:Nuuu} except the orange dot-dashed line now stand for four-Fermi operators.}
\label{Fig:Nrnudd}
\end{figure}

\begin{figure}[h!]
\centering
\includegraphics[width=0.45\textwidth,height=0.25\textheight]{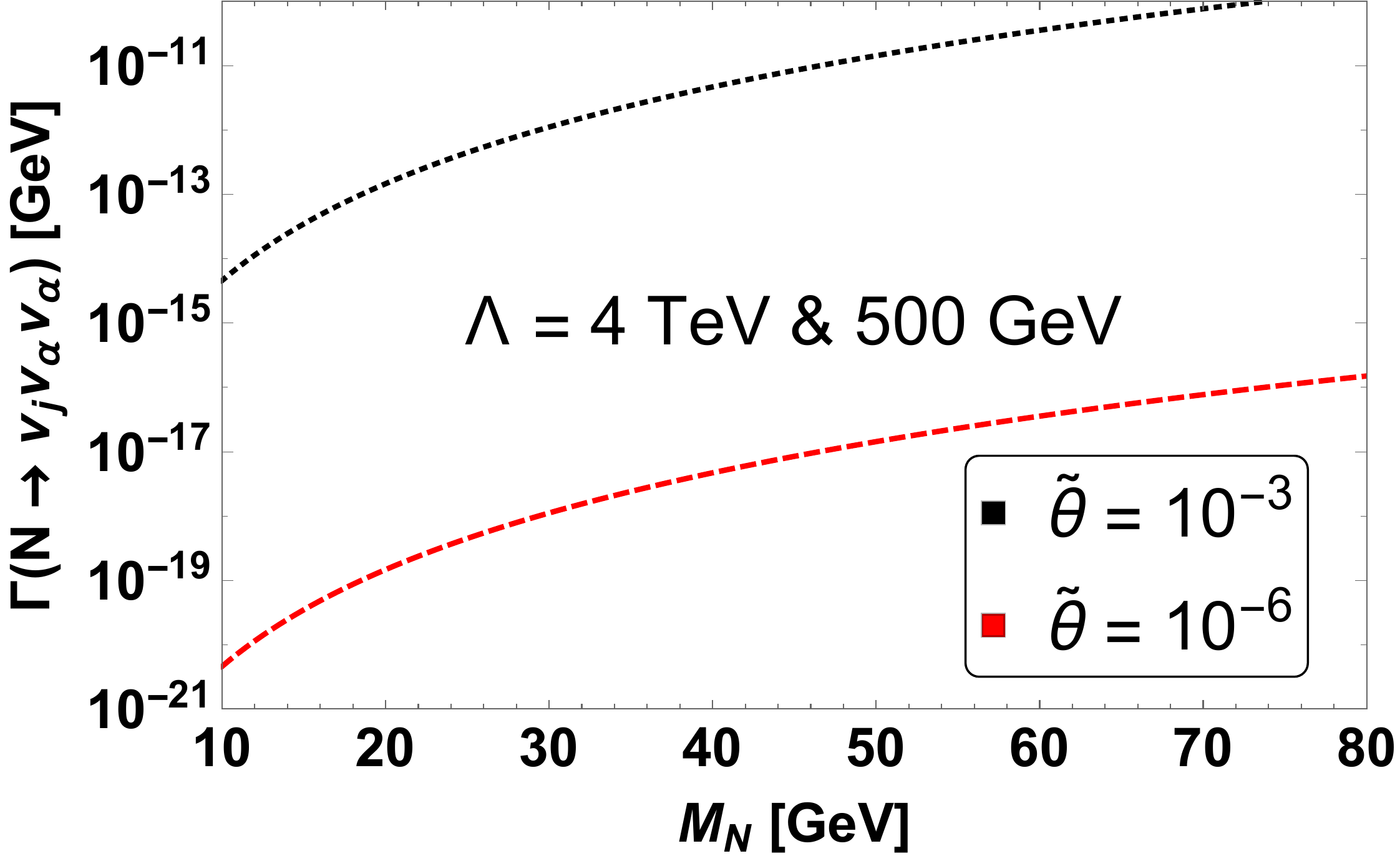}
\caption{Partial decay width corresponds to the decay mode $\Gamma\left(N_{i} \to \nu_{j}\, \nu\, \bar{\nu}\right)$ of RHN mass ranging from 10~GeV to 80~GeV. The black dotted and red dashed line stands for mixing angle $\tilde{\theta}=10^{-3}$ and $\tilde{\theta}=10^{-6}$, respectively.}
\label{Fig:Nr3nu}
\end{figure}

\begin{figure}[h!]
\centering
\includegraphics[width=0.45\textwidth,height=0.25\textheight]{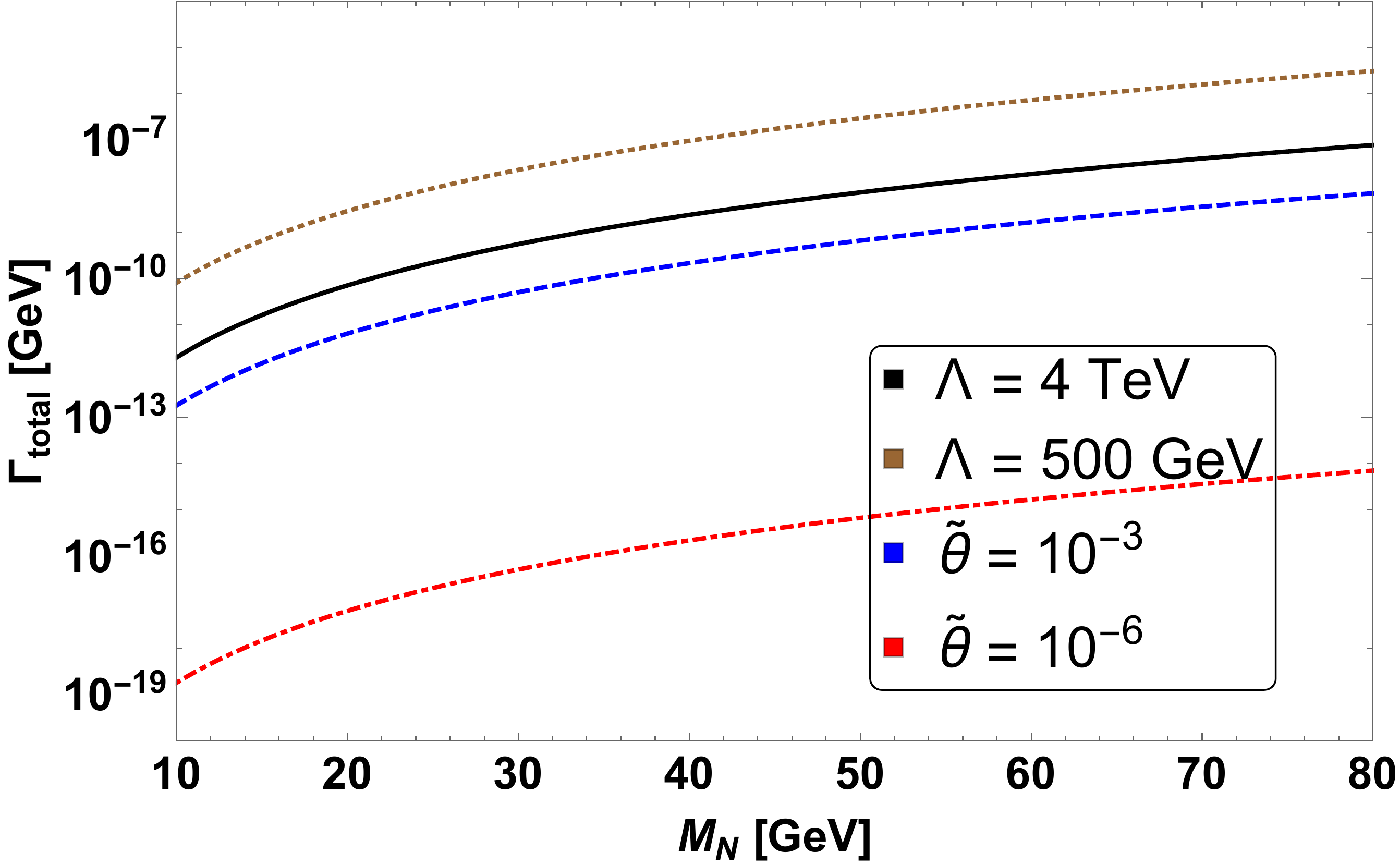}
\caption{The total decay width correspond to RHN field where the $\tilde{M}_{N}$ is ranging from 10~GeV to 80~GeV. The blue dashed and red dot-dashed line stands for the mixing angle  $\tilde{\theta}=10^{-3}$ and $\tilde{\theta}=10^{-6}$, respectively. The black solid line represents the total decay width of the $N$ field under full $d = 6$ $N_{R}-$EFT where the cut off scale $\Lambda$ is set to be 4 TeV. Similarly the brown dotted line stands for the total decay width of the $N$ field with cut off scale $\Lambda$ = 500~GeV.}
\label{Fig:Nr3total}
\end{figure}

\begin{itemize}
\item $\bf{\Gamma\left(N_{i} \to \nu_{j}\, u_{\alpha}\, \bar{u}_{\alpha}\right)}$ $\&$ $\bf{\Gamma\left(N_{i} \to \nu_{j}\, d_{\alpha}\, \bar{d}_{\alpha}\right)}$. 
\end{itemize}
The RHN can also decay via same flavour hadronic modes along with missing energy. The flavour neutrality in the final states suggest that the decay can occur due to the $Z$ boson mediation at renormalisable level Lagrangian. Furthermore, depending upon the quark flavour, different four Fermi operators would participate in it. In case of up-type quarks these operators are $\mathcal{O}_{uN}$, $\mathcal{O}_{QN}$ and $\mathcal{O}_{QuNL}$. However, out of these three operators the first two are mixing suppressed, hence do not play much role in the decay mode $\Gamma\left(N_{i} \to \nu_{j}\, u_{\alpha}\, \bar{u}_{\alpha}\right)$. For the current calculation we add the matrix elements $\mathcal{M}_{Z}$ and $\mathcal{M}_{QuNL}$ and the total matrix element of the process takes the following form 
\begin{equation}
\mathcal{M}_{\text{total}}\left(N_{i} \to \nu_{j}\, u_{\alpha}\, \bar{u}_{\alpha}\right) = \mathcal{M}_{Z} + \mathcal{M}_{QuNL}  .
\end{equation}      
\noindent
Taking the explicit form of $\mathcal{M}_{Z}$ and $\mathcal{M}_{QuNL}$ from Eq.~\ref{Eq:AmpZ} and Eq.~\ref{Eq:Mlud} one can obtain the corresponding partial decay width of this channel 
\begin{align}
\Gamma(N_{i} \to \nu_{j}u_{\alpha}\bar{u}_{\alpha}) & = \frac{M^{5}_{N}N_{c}}{512\pi^{3}}\{\left(|B|^{2} + \frac{g^{4}|A|^{2}}{M^{4}_{W}}\left(g^{2}_{L} + g^{2}_{R}\right)\right)\mathcal{I}_{1}\left(x_{u}, x_{u}, x_{\nu_{j}}\right) \nonumber  \\
										&~ - \frac{2|A|^{2}g^{2}}{M^{2}_{W}}g_{L}g_{R}\mathcal{G}_{3}\left(x_{u}, x_{u}, x_{\nu_{j}}\right) + \frac{\text{Re}[A^{*}B]g^{2}}{M^{2}_{W}}\left(g_{R} - g_{L}\right)\mathcal{G}_{1}\left(x_{u}, x_{u}, x_{\nu_{j}}\right)\} 
\end{align}
where $A$ and $B$ are
\[~~~A= \tilde{\theta},~~~B = \frac{\alpha_{QuNL}}{\Lambda^{2}}.\]
Similar to previous cases, we have only consider the renornmalisable neutral current interaction for the diagram correspond to $\mathcal{M}_{Z}$. In Fig.~\ref{Fig:Nuuu}, we illustrate the effects of these operators for this decay mode. The black dotted~(red dashed) line correspond to the mixing angle $\tilde{\theta} = 10^{-3}$~($\tilde{\theta} = 10^{-6}$) and the orange dot-dashed line correspond to $\mathcal{O}_{QuNL}$, respectively. The blue thick line stands for the total contribution taking into account both the mixing part~($\tilde{\theta} = 10^{-3}$) and dimension six operator $\mathcal{O}_{QuNL}$. We see that the contribution coming from renormalisable part~($Z$-mediated case) dominates over the dimension six contribution when mixing angle is $\tilde{\theta}=10^{-3}$. The blue thick line denotes the full decay width with the assumption on the mixing angle as $\tilde{\theta} = 10^{-3}$. 
\noindent
In case of down type quark the underlying calculation has an additional complication. As mentioned before, the decay receives substantial contribution from the $Z$ boson propagation. Along with that, the operators, $\mathcal{O}_{dN}$, $\mathcal{O}_{QN}$, $\mathcal{O}_{LNQd}$ and $\mathcal{O}_{LdqN}$ gives sufficient contributions. But the effects coming from the operators $\mathcal{O}_{dN}$ and $\mathcal{O}_{QN}$ are in general mixing dependent and one can safely ignore these for practical purpose. On the other hand the operators $\mathcal{O}_{LNQd}$ and $\mathcal{O}_{LdqN}$ does play a role here. Taking into account all of these operators, one can obtain the following form of partial decay width        
\begin{align}
\Gamma(N_{i} \to \nu_{j}d_{\alpha}\bar{d}_{\alpha}) & = \frac{M^{5}_{N}N_{c}}{512\pi^{3}}\{\frac{2|A|^{2}}{M^{4}_{W}}g^{4}g_{L}g_{R}\mathcal{H}_{1}\left(x_{\nu_{j}}, x_{d_{\alpha}}\right) \nonumber \\
							                     & + \left(\frac{|A|^{2}}{M^{4}_{W}}g^{4}\left(g^{2}_{L} + g^{2}_{R}\right) + \frac{|B|^{2}}{4} + \frac{3}{4}|C|^{2}\right)\mathcal{I}_{1}\left(x_{d_{\alpha}}, x_{d_{\alpha}}, x_{\nu_{j}}\right)  + 3|C|^{2}\mathcal{I}_{5}\left(x_{d_{\alpha}}, x_{d_{\alpha}}, x_{\nu_{j}}\right)  \nonumber \\
							                     & + \frac{g^{2}}{2M^{2}_{W}}\left[\left(g_{L} - g_{R}\right)\text{Re}[A^{*}B] - 3\left(g_{L} + g_{R}\right)\text{Re}[A^{*}C]\right]\mathcal{G}_{1}\left(x_{d_{\alpha}}, x_{d_{\alpha}}, x_{\nu_{j}}\right) \nonumber \\
							                    &  - \frac{3g^{2}\text{Re}[A^{*}C]}{2M^{2}_{W}}\left(g_{L} + g_{R}\right)\mathcal{H}_{3}\left(x_{\nu_{j}}, x_{d_{\alpha}}\right) \} ,
\end{align}
where the coefficients $A$, $B$ and $C$ are 
\[~~~A= \tilde{\theta},~~~B = \frac{\alpha_{LNQd}}{\Lambda^{2}} + \frac{\alpha_{LdqN}}{\Lambda^{2}},~~~ C =  \frac{\alpha_{LdqN}}{\Lambda^{2}}\]
\noindent
In Fig.~\ref{Fig:Nrnudd}, we illustrate the participation of these operators in the decay width for mass value 10~GeV to 80~GeV. The black dotted and red dashed line denotes the contribution from renormalisable neutral current process for $\tilde{\theta} = 10^{-3}$ and $\tilde{\theta} = 10^{-6}$, respectively. The orange dot-dashed line shows the subdominant contribution coming from the combination of four-Fermi operators. The blue thick line shows the overall effect coming from mixing~($\tilde{\theta}=10^{-3}$) and four-Fermi operator. We see that the contribution coming from renormalisable part ($Z$-mediated case) dominates over the dimension six contribution when mixing angle is $\tilde{\theta}=10^{-3}$ and because of this black dotted and thick blue line coincides in each panel.        
\begin{itemize}
\item $\bf{\Gamma\left(N_{i} \to \nu_{j}\, \nu\, \bar{\nu}\right)}$
\end{itemize}
The RHNs can also decay to pure active neutrino states. The decay can be mediated either via off-shell $Z$ decay or the Fermi operators $\mathcal{O}_{NN}$ and $\mathcal{O}_{NNNN}$ that arise in the dimension six set up. However the relevant terms coming from these dimension six operators are proportional to cubic power of the mixing angle along with the inverse of the quadratic $\Lambda$ suppression. Hence, once can ignore those terms for practical purposes. The partial decay is then can be written as   
\begin{align}
\Gamma\left(N_{i} \to \nu_{j} \nu \bar{\nu}\right) & = \frac{G^{2}_{F}M^{5}_{N}}{96\pi^{3}}|\tilde{\theta}|^{2}.
\label{Eq:Nr3nu}
\end{align}
In Fig.~\ref{Fig:Nr3nu}, we present the corresponding decay width for different choices of mixing angles. As the EFT operators does not participate in the process the change in the cut-off scale is irrelevant in this calculation.
In Fig.~\ref{Fig:Nr3total} we illustrate the the total decay width of the RHN fields while taking into account all possible decay modes in the mass range $10~\text{GeV} < M_{N} < 80~\text{GeV}$. The red dot dashed line and the blue dashed line correspond to mixing angle $\tilde{\theta} = 10^{-6}$ and $\tilde{\theta} = 10^{-3}$ respectively. On the other hand, the black thick line and the brown dotted line signifies the total decay width corresponds to full $d = 6$ $N_{R}$-EFT, where the cut-off scale $\Lambda$ = 4~TeV and 500~GeV respectively. In the EFT calculation we set the active-sterile mixing angle to be $\tilde{\theta} = 10^{-3}$. The value of the total decay width suggests that the RHN field would behave as a prompt particle in  both the EFT benchmark points as well as $\tilde{\theta} = 10^{-3}$ case for entire range of 10~GeV $< M_{N} <$ 80 GeV.   

\begin{figure}[h!]
\includegraphics[width=0.44\textwidth,height=0.25\textheight]{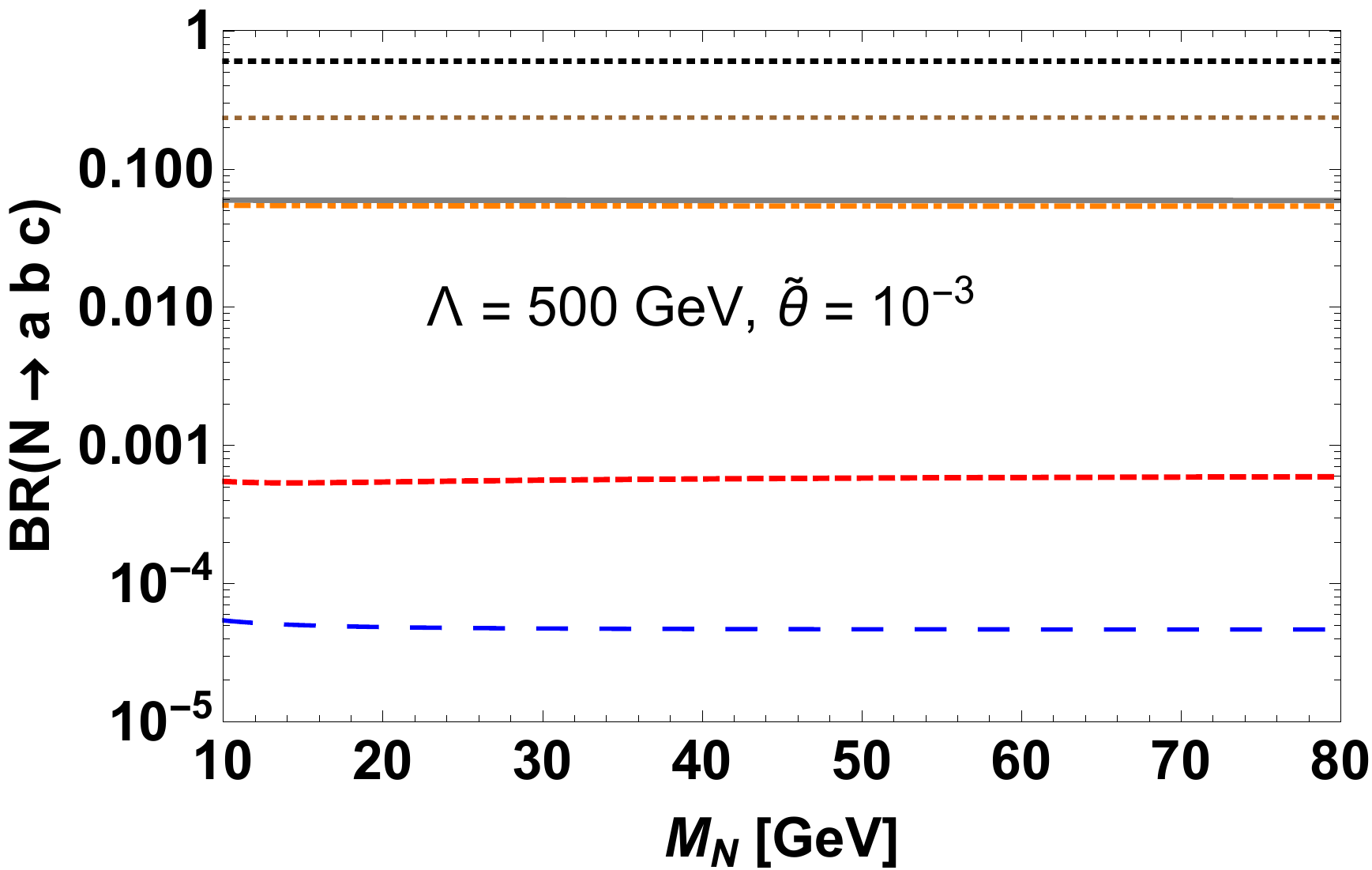}
\includegraphics[width=0.55\textwidth,height=0.25\textheight]{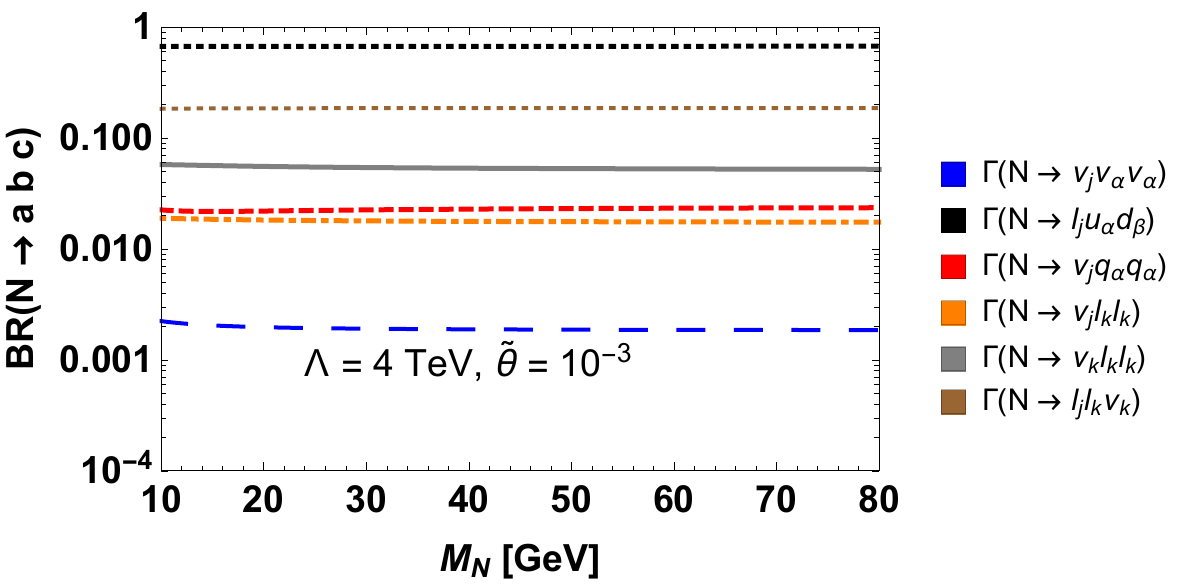}
\includegraphics[width=0.6\textwidth,height=0.25\textheight]{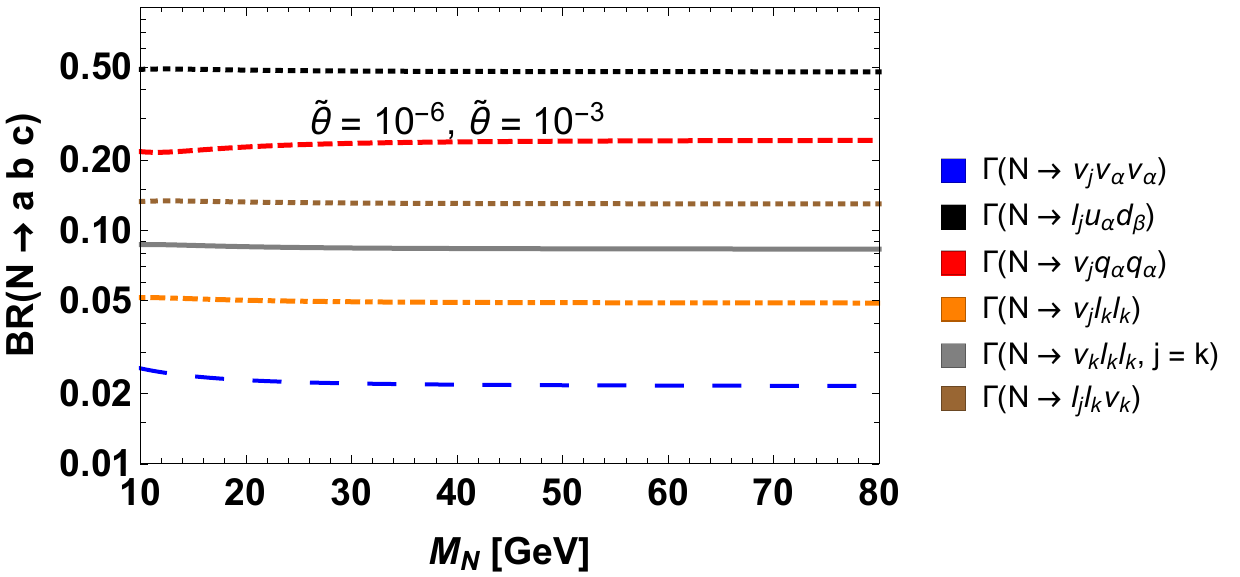}
\caption{Upper Panel: The branching ratios of  RHN mass ranging from from 10~GeV to 80~GeV in various three body modes for the benchmark points \textbf{BP1} ($\Lambda$ = 500~GeV) and \textbf{BP2} ($\Lambda$ =4~TeV). In both cases, we have considered the active sterile mixing angle $\tilde{\theta} = 10^{-3}$. Lower Panel: We calculate the branching ratio of the RHN mass ranging from 10~GeV to 80~GeV in different three body decay modes while considering renormalisable part of the Lagrangian.}
\label{Fig:3bodyBR}
\end{figure}

We now present the branching ratio for the benchmark points \textbf{BP1}~($\Lambda=500$ GeV) and \textbf{BP2}~($\Lambda=4$ TeV) while setting active-sterile mixing angle to be $\tilde{\theta} = 10^{-3}$. In upper panel of Fig.~\ref{Fig:3bodyBR}, we present the corresponding plots for these scenarios. For both these cases the BR value for the $\ell_{j}ud$ channel will be maximum for the entire mass range as shown by the black dotted curve. Similarly, the BR value for the $\nu_{j}\nu\bar{\nu}$ mode is minimum. But the relative difference between the other possible channels vary depending on the underlying benchmark choice. For comparison purpose we also present the branching ratio in the lower panel of Fig.~\ref{Fig:3bodyBR} while considering only the renormalisable part of the EFT Lagrangian. The important change in dimension six $N_{R}$-EFT is the enhancement of the BR value of pure leptonic three body modes. In renormalisable level the BR for $N_{i} \to \nu_{j}q_{\alpha}\overline{q}_{\alpha}$ channels dominate over entire range of 10~GeV$< M_{N} < M_{W}$. However, the interplay between the operators $\mathcal{O}_{LNLe}$ and $\mathcal{O}_{HNe}$ alter these result and BR of $N_{i} \to \ell_{k} \ell_{k} \nu_{k}/ \ell_{j} \ell_{k} \nu_{k}$ (Grey solid and brown dotted) dominates over $N_{j} \to \nu_{j} q_{\alpha} q_{\alpha}$ mode for both the benchmark scenarios.  
\section{Prospective  Multi-Lepton Final States}
\label{Sec:multilep}
\begin{figure}[t!]
\centering
\subfigure[]{\includegraphics[width=0.48\textwidth,height=0.3\textheight]{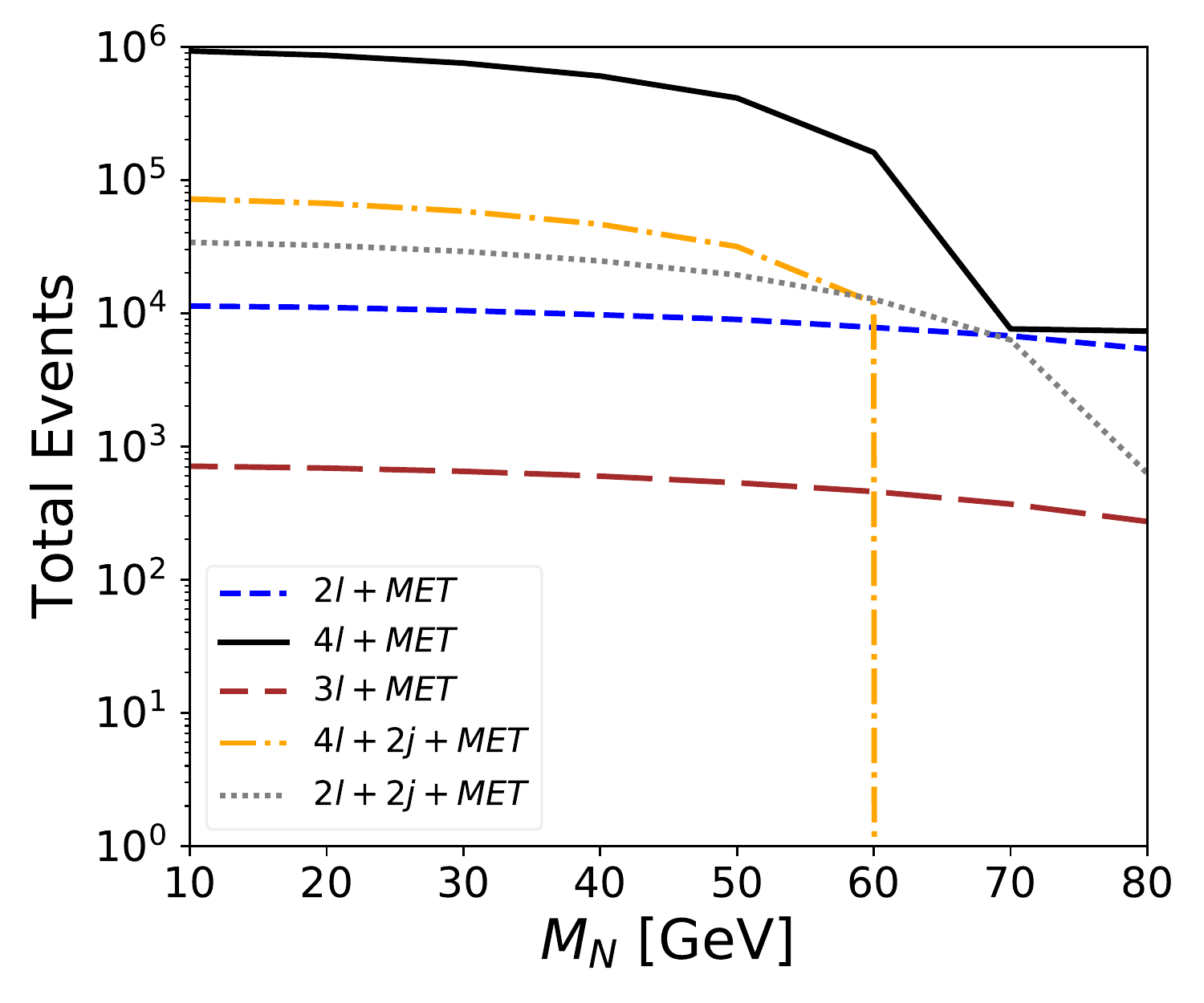}\label{Fig:ppN}}
\subfigure[]{\includegraphics[width=0.48\textwidth,height=0.3\textheight]{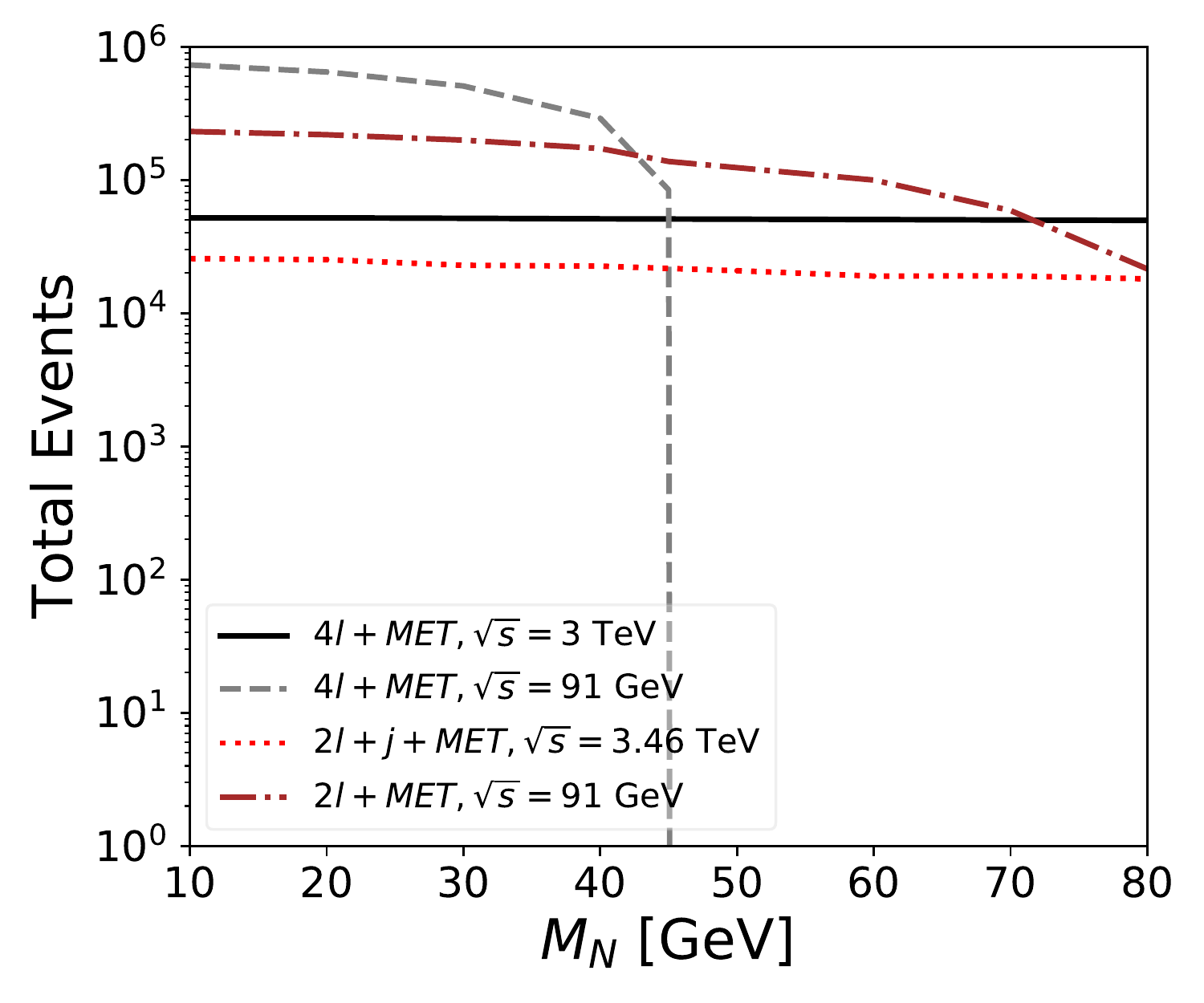}\label{Fig:eeep}}
\caption{The left panel denotes the number of events that can be obtained at HL-LHC with c.m.energy 14 TeV and  $\mathcal{L} = 3000 \, \rm{fb}^{-1}$. The right panel shows achievable  number events that can be obtained 
at $e^+e^-$ and $e^-p$ colliders for different c.m.energies.}
\label{Fig:ppep}
\end{figure}

 We focus on producing multi-lepton final states that arise from single/pair production of RHN fields and their subsequent decays. We have calculated various three-body decay modes of the RHNs. In Fig.~\ref{Fig:3bodyBR}, we show that the $N$ decaying to purely leptonic final states have branching ratios of 20$\%$ to 25$\%$, depending on the cut-off scale $\Lambda=500 $ GeV and 4 TeV. With these results, one can estimate the expected number of events in the heavy neutrino production processes at $pp$, $ep$ and $e^+e^-$ colliders, which we show in  Fig.~\ref{Fig:ppep}. We assume in all these cases that the RHN decays purely leptonically. 

Fig.~\ref{Fig:ppN} shows the number of events expected at the HL-LHC  with c.m. energy $\sqrt{s} = 14$ TeV and $\mathcal{L} = 3000 \, \rm{fb}^{-1}$. The solid black line represents the number of events for the $4\ell + MET$ final state, which results from the pair-production of $N$'s and their subsequent decay to leptonic final states. $N$-pair production is either mediated via Higgs decay, where the Higgs boson is produced through gluon fusion, or via four-fermion operators. The $Z$ boson, produced via the Drell-Yan process, can give rise to the same final state. However, this process is suppressed by the mixing angle $\tilde{\theta}$ and fails to provide any appreciable contribution. As seen from the figure, the total number of events for this channel is as large as $\mathcal{O}(10^6)$.
 
Instead, the blue dashed line shows the $2\ell + MET$ final state, resulting from single $N$ production and its subsequent leptonic decay. Similar to the previous channel, the $N$ in this case can be produced either via Higgs decay or via four-fermion operators. Additionally, there is a Drell-Yan contribution via $Z$-mediated contributions. These production modes,  however, depend on the active-sterile mixing angle. Hence, the total number of events for this process is small compared to the $4\ell + MET$ final state, where the four-fermion $\mathcal{O}^5_2$ operators can give mixing angle independent, unsuppressed contributions.
 
The $4\ell + 2j + MET$ and $2\ell + 2j + MET$ final states are represented by the orange dot-dashed line and the grey dotted line, respectively. These final states are produced via Higgs and $Z$ boson-mediated VBF production modes. The above two final states can appear depending on the rates for the decay $h/Z \to \nu N/ N N$  and subsequent leptonic decay of $N$. Amongst these two processes, the former is dominant, and many signal events can be expected $\sim \mathcal{O}(10^5)$.
 
The $3\ell + MET$ final state, shown as red dashed line in Fig.~\ref{Fig:ppN}, which arises from the $q q^{'} \to W \to \ell N$ production mode can provide signal events ranging between $\mathcal{O}(10^{4})$ to $\mathcal{O}(10^{3})$.

In \ref{Fig:eeep} we present the number of signal events one can obtain at $e^+e^-$ and $e^-p$ colliders. For the $e^+e^-$ collider, we consider a c.o.m energy of 91 GeV and 3 TeV,  respectively. On the other hand, for $e^-p$ colliders, the c.o.m energy has been chosen to be 3.46 TeV. In all these cases, we have considered the luminosity $\mathcal{L} = 1 \, \rm{ab}^{-1}$. 
The grey dashed line denotes the $4\ell + MET$ final state for $\sqrt{s} = 91$ GeV. There can be different processes contributing to the final state, among which the $Z$ mediated diagram generating two $N$s followed by $N \to \ell \ell \nu$ decay give rise to a significant number of events $N \sim \mathcal{O}(10^5)$. One can obtain $2\ell + MET$ final states via $Z$ boson decay or four-fermi operators. Although, the relevant production vertex is suppressed by the mixing angle $\tilde{\theta}$, the presence of $Z-$pole leads to substantial number of signal events. 

The number of signal events associated with $4\ell + MET$ final state is shown for  $e^+e^-$ collider with c.o.m. energy of 3 TeV. This process can arise via VBF production or four-fermion contact operators, which give rise to a sizeable number of events $N \sim \mathcal{O}(10^4)$. 

For $e^-p$ colliders, the only process which can generate a significant number of signal events is the $e^-p \to j N \to 2l+j+ MET$ channel. The number of signal events corresponding to this channel is represented by the red dotted curve and is  $N \sim \mathcal{O}(10^{4})$.

\section{Summary and Conclusion}\label{Sec:conc}
The $N_{R}$-EFT framework can provide a suitable phenomenological description of the BSM physics. In this work we have systematically established this EFT up to dimension six level and shown the detailed construction of all possible operators that one can construct at $d = 5$ and $d = 6$ respectively. At each order these higher dimensional operators either modify different $d = 4$ couplings involving RHN field or introduce new set of interactions relevant to neutrino phenomenology.

The interaction of the RHN with SM fields leads to new decay modes and hence modifies the decay width of $h/W/Z$. However, the existing experimental observations restrict such modifications and impose meaningful bounds on mixing angle $\tilde{\theta}$, Wilson coefficients and the associated cut off scale $\Lambda$. One can further tighten these constraints while incorporating different direct search limits and determine allowed range of EFT parameters. Our analysis deduce in case of cut off $\Lambda$ = 4 TeV one is allowed to fix the mixing angle $\tilde{\theta} = 10^{-3}$, the Wilson coefficients $\alpha^{(5)}_{2} \leq 0.1$  $\alpha_{HN}, \alpha_{HNe} \leq 1$, $\alpha_{LNW/B} \leq 0.1$ and coefficients correspond to four-Fermi operators to be $\alpha_{\text{four Fermi}} \leq 0.5$. Similarly, for cut off scale $\Lambda$ = 500 GeV one needs to rescale these \emph{Wilson} coefficients accordingly. 

We have presented the viable production mode of the RHN fields at 14 TeV $pp$, as well as \emph{future} $e^{+}e^{-}$ and $e^{-}p$ colliders. In case of $pp$ collider pair production of $N$ field via $gg \to h \to N N$ mode remains dominant production mode for the mass range $M_{N} < \frac{m_{h}}{2}$ with the cross section roughly around $\mathcal{O}(10)$ pb. Above this mass range RHN pair production through four-Fermi contact interaction serves as a dominant production mode with the cross section ranging between $\mathcal{O}(10^{-1})$ pb to $\mathcal{O}(10^{-3})$ pb for the mass range 100~GeV$< M_{N} <$ 1~TeV. All the operators which can generate $NN$ pair can also invoke single $N$ production. However the cross section correspond to single RHN field production is comparatively low \emph{w.r.t} RHN pair production as this is active-sterile mixing suppressed. For completeness we also present RHN production through VBF as well as Drell-Yan mechanism. Apart from RHN production via $W$ boson decay (where $W$ is produced via Drell-Yan process), for all other cases the cross section is significantly low. The cross section for RHN production via Drell-Yan $W$ mode is large (roughly around $\mathcal{O}(10^{-1})$ pb to $\mathcal{O}(10^{-3})$ pb) due to the significant contribution coming from $d = 6$ operator $\mathcal{O}_{HNe}$. In case of $ep$ collider $e p \to j N$ mode stands as a viable production channel for the RHN field. This process can occur either via t-channel $W$-mediated process or via contact interaction through different four-Fermi operators and corresponding cross section lies within the range $\mathcal{O}(10^{-1})$ pb to $\mathcal{O}(10^{-3})$ pb. In case of $e^{+} e^{-}$ collider we have studied the feasible production modes for $c.m.$ energies $\sqrt{s} =$ 91 GeV and 3 TeV respectively. For $\sqrt{s} =$ 91~GeV, RHN pair production via s-channel $Z$ boson  is the dominant production mode. The presence of $Z$ pole in the underlying process significantly increases the production cross section to 10 pb. In contrast to that for $\sqrt{s}$ = 3 TeV, this process is primarily governed by different four-Fermi operators ($\mathcal{O}_{eN}$, $\mathcal{O}_{LN}$ \emph{etc.}) and the corresponding cross section lies roughly around $\mathcal{O}(0.5)$ pb  for $M_{N}$ ranging between 100~GeV to 1~TeV.

Apart from the production mechanism, we evaluate the decay width as well as branching ratio of RHN fields while considering both renormalisable part of effective Lagrangian as well as relevant higher dimension operators. Depending upon RHN masses it can either decay via two body or three body decay modes where we assume decay among RHN fields are forbidden due to mass degeneracy.  In case of two body decay calculation, the $N$ fields can decay into $\nu h$, $\nu Z$ and $\ell W$ modes if we only consider $d = 4$ Lagrangian. The presence of EFT operators $\mathcal{O}^{(5)}_{3}$, $\mathcal{O}_{LNB}$ and $\mathcal{O}_{LNW}$ invoke an additional decay mode $\nu\gamma$ for the RHN field. The branching ratio correspond to $N \to \ell W$ mode remains dominant for entire $M_{N}$ mass range in both $d = 4$ as well as $d = 6$ set up and the contributions from $\mathcal{O}_{HNe}$ significantly enhance BR$(N \to \ell W)$ \emph{w.r.t} the $d = 4$ result. The BR correspond to $N \to \nu h$ mode remains minimum for entire RHN mass range in both these set up. 

In case of $M_N < M_W$, the RHN fields can decay via different leptonic as well as semileptonic three body modes in addition to the two body $\nu\gamma$ mode. Upto dimension five these three body decay modes can only occur due to the off shell decay of $W$ and $Z$ bosons. In dimension six the presence of different four-Fermi operators significantly alters this situation. These four-Fermi operators can contribute to the three body modes through contact interaction and enhance the corresponding partial width. For example, operator $\mathcal{O}_{LNLe}$ would participate in different leptonic channels $N_{i} \to \ell_{j}\ell_{k}\nu_{k}/\ell_{k}\ell_{k}\nu_{k}/\nu_{j}\ell_{k}\ell_{k}$ and increases partial width. The $d = 6$ operator $\mathcal{O}_{HNe}$ noticeably increases off shell $W$ mediated contribution but the off shell $Z$ boson decay does not receive any significant enhancement from the EFT operators as the relevant couplings are either mixing angle $\tilde{\theta}$ suppressed or loop suppressed. We also present comparative analysis between the branching ratio for two EFT benchmark scenarios ($\Lambda =$ 500 GeV, 4 TeV) as well as renormalisable level Lagrangian. Our analysis dictates the BR value for $\ell_{j}u_{\alpha}d_{\beta}$ mode is maximum and BR value for $\nu_{j}\nu_{\alpha}\nu_{\alpha}$ remains minimum in all these scenarios. The important aspect of dimension six $N_{R}$-EFT is the enhancement of the BR value of pure leptonic three body modes. In renormalisable level the BR for $N_{i} \to \nu_{j}q_{\alpha}\overline{q}_{\alpha}$ channels dominate over entire range of 10~GeV$< M_{N} < M_{W}$. However, the interplay between the operators $\mathcal{O}_{LNLe}$ and $\mathcal{O}_{HNe}$ alter these result and BR of $N_{i} \to \ell_{j} \ell_{k} \nu_{k}/ \ell_{k} \ell_{k} \nu_{k}$ dominates over $N_{j} \to \nu_{j} q_{\alpha} q_{\alpha}$ (red dashed line) mode for both the benchmark scenarios. 

The major advantage of the $N_{R}$-EFT set up is the enhancement in the BR value for different three body pure leptonic decay modes. This motivates us to propose few \emph{golden} channel signatures for the RHN field in multi-leptonic final states. Our study suggest $4\ell + MET$ channel can serve as an optimal discovery mode for the RHN field at both 14 TeV HL-LHC and future $e^{+} e^{-}$ (with c.m $\sqrt{s} =$ 91~GeV and 3~TeV) machine. In case of $ep$ collider the viable discovery mode would be $2\ell + j +MET$ final state. We will investigate these channels for a detailed cut-based analysis in our future studies.

\begin{acknowledgments}
The work of S.M. is supported by KIAS Individual Grants (PG086001) at Korea Institute for Advanced Study. M.M and A.S acknowledge the support from the Indo-French Centre for the Promotion of Advanced Research (Grant no: 6304-2). M.M acknowledges the support from IPPP DIVA Programme. RP acknowledges the support of Fulbright-Nehru Doctoral Research Fellowship, 2022-2023.
\end{acknowledgments}
\appendix

\section{: Useful Transformation Rules between Flavour and Mass Basis}
\label{App:Eigen}
In Eq.~\ref{Eq:EigenVV5} and Eq.~\ref{Eq:EigenV6} we established the relation between the gauge basis $\{\nu_{L}, N^{c}_{R}\}$ and the mass basis $\{\nu_{L,m}, N^{c}_{R,m}\}$ in dimension five and dimension six framework respectively. Furthermore, for the explicit coupling extraction one need other possible gauge basis definitions like $\nu^{c}_{L}$, $N_{R}$ \emph{etc.} In Eq.~\ref{Eq:AllNr}, and Eq.~\ref{Eq:RHNnewtrans}  we illustrate different relationship between the gauge basis and the mass basis states of both the active and sterile neutrinos.

\begin{equation}
\begin{split}
\nu_{L} & \simeq U_{\text{PMNS}}\nu_{L, m} + \theta N^{c}_{R, m}, ~~~~~ \overline{\nu_{L}}  \simeq  \overline{\nu}_{L, m}U^{\dagger}_{\text{PMNS}} + \overline{N}^{c}_{R, m}\theta^{\dagger} \\
\nu_{L}^{c} & \simeq U^{*}_{\text{PMNS}}\nu^{c}_{L, m} + \theta^{*}N_{R, m}, ~~~~~~~ \overline{\nu}^{c}_{L}  \simeq \overline{\nu}^{c}_{L, m}U^{\dagger}_{\text{PMNS}} + \overline{N}_{R, m}\theta^{T}.
\end{split}
\label{Eq:AllNr}
\end{equation}

\begin{equation}
\begin{split}
N^{c}_{R} & \simeq - \theta^{T}\nu_{L, m} + \kappa N^{c}_{R, m}, ~~~~~ \overline{N_{R}}^{c}  \simeq - \theta^{*}\overline{\nu}_{L, m} + \kappa^{\dagger}\overline{N}^{c}_{R, m} \\
N_{R} & \simeq - \theta^{\dagger}\nu^{c}_{L, m} + \kappa^{*}N_{R, m}, ~~~~~ \overline{N}_{R}  \simeq - \theta\overline{\nu}^{c}_{L, m} + \kappa^{T}\overline{N}_{R, m}.  
\end{split}
\label{Eq:RHNnewtrans}
\end{equation}
\noindent
In the above we present all these relations for dimension five set up. At dimension six, one should replace the mixing angle $\theta$ with $\tilde{\theta}$.
\section{: Evaluating Couplings Upto Dimension five set up}
\label{App:Dim5Couplings}
The RHN fields couple to different SM fields even in the minimal seesaw Lagrangian. At renormalisable level these couplings are governed by the mixing angle between the active and sterile neutrinos. In $d = 5$ $N_{R}$-EFT one can write three more operators $\mathcal{O}^{(5)}_{1}$, $\mathcal{O}^{(5)}_{2}$ and $\mathcal{O}^{(5)}_{3}$ which can modify the existing couplings as well as introduce new couplings involving RHN fields. Here we present the explicit expansion correspond to the relevant part of the renormalisable Lagrangian as well as dimension five operators.
\begin{equation}
\begin{split}
\bar{L}\gamma_{\mu}D_{\mu}L + h.c. & \subset \frac{g}{\sqrt{2}}\bar{\nu}_{L}\gamma^{\mu}W^{+}_{\mu}\ell_{L} + \bar{\nu}_{L}\gamma^{\mu}\left(\frac{g}{2}W^{3}_{\mu}- \frac{g^{'}}{2}B_{\mu}\right)\nu_{L} +  h.c.  \\
							 & \subset \frac{g}{\sqrt{2}}\bar{\nu}_{L}\gamma^{\mu}W^{+}_{\mu}\ell_{L} + \frac{g}{2c_{w}}\bar{\nu}_{L}\gamma^{\mu}Z_{\mu}\nu_{L} + h.c. \\
							 & \subset \frac{gU}{\sqrt{2}}[\overline{\nu}_{m}\gamma_{\mu}P_{L} + \overline{N}_{m}\gamma_{\mu}P_{L}]\ell_{m} + \frac{g}{2c_{w}}Z_{\mu}[ U^{\dagger}U\overline{\nu}_{m}\gamma^{\mu}P_{L}\nu_{m} + U^{\dagger}\theta\overline{\nu}_{m}\gamma^{\mu}P_{L}N_{m} \\
							 & + U\theta^{\dagger}\overline{N}_{m}\gamma^{\mu}P_{L}\nu_{m} + \theta^{\dagger}\theta\overline{N}_{m}\gamma^{\mu}P_{L}N_{m}] + h.c.
\end{split}
\label{Eq:NuKin}
\end{equation}
The lepton doublets are charged under the SM $SU(2)_{L}\times U(1)_{Y}$ gauge group. As a result the left-handed $\nu_{L}$ field couples to electroweak gauge bosons through the covariant derivative term as illustrated in Eq.~\ref{Eq:NuKin}. Using the transformation rule as discussed in Appendix.~\ref{App:Eigen} one can rewrite the $\nu_{L}$ fields in terms of mass basis $\{\nu_{m}, N_{m}\}$. As a result one obtain different three point couplings between the RHN fields and SM electroweak gauge bosons. One can notice, the coupling involving RHN field is mixing angle suppressed as it does not directly couple to $W/Z$ boson in the renormalisable Lagrangian.
\begin{align}
Y_{\nu}\bar{L}\tilde{H}N_{R} & = \frac{Y_{\nu}v}{\sqrt{2}}\bar{\nu}_{\ell}\left(1 + \frac{h}{v}\right)N_{R} \subset \frac{Y_{\nu}}{\sqrt{2}}h\bar{\nu}_{\ell}N_{R}  \nonumber\\
					    &  =   \frac{Y_{\nu}}{\sqrt{2}}h\left[ U^{\dagger}\theta^{\dagger}\overline{\nu}_{m}P_{R}\nu_{m} - U^{\dagger}\kappa^{*}\overline{\nu}_{m}P_{R}N_{m} + \theta^{\dagger}\theta^{\dagger}\overline{N}_{m}P_{R}\nu_{m} - \theta\kappa^{*}\overline{N}_{m}P_{R}N_{m}\right]     \label{Eq:Yukanu}  
\end{align}
In renormalisable part of the Lagrangian the neutrino fields can achieve mass through usual Yukawa term. This will invoke different couplings between both active, sterile neutrinos with Higgs boson. In Eq.~\ref{Eq:Yukanu} we present the explicit evaluation of these couplings. 
\begin{equation}
\begin{split}
\mathcal{O}^{\left(5\right)}_{1} & = \left(\frac{\alpha^{(5)}_{1}}{\Lambda}\right)\left(\overline{L}^{c}\tilde{H}^{\dagger}\tilde{H}L\right) = \left(\frac{\alpha^{(5)}_{1}}{\Lambda}\right)\overline{\nu_{\ell}}^{c}\frac{\left(v + h\right)^{2}}{2}\nu_{\ell} \\ 
						& \subset  \left(\frac{\alpha^{(5)}_{1}v}{\Lambda}\right)h\left[U^{T}U\overline{\nu}_{m}P_{L}\nu_{m} + \theta^{T}U\overline{N}_{m}P_{L}\nu_{m} + U^{T}\theta\overline{\nu}_{m}P_{L}N_{m} + \theta^{T}\theta\overline{N}_{m}P_{L}N_{m}\right] 
\end{split}	
 \label{Eq:O51}					
\end{equation} 
The Higgs-neutrino couplings receive further modification through the dimension five operators $\mathcal{O}^{(5)}_{1}$ and $\mathcal{O}^{(5)}_{2}$. In Eq.~\ref{Eq:O51} and Eq.~\ref{Eq:O52} we present these corrections in detailed manner.
\begin{equation}
\begin{split}
\mathcal{O}^{\left(5\right)}_{2} & = \left(\frac{\alpha^{(5)}_{2}}{\Lambda}\right)\left(\overline{N_{R}}^{c}N_{R}\right)\left(H^{\dagger}H\right) = \left(\frac{\alpha^{(5)}_{2}}{\Lambda}\right)\left(\overline{N_{R}}^{c}N_{R}\right)\frac{\left(v + h\right)^{2}}{2}  \\
	                                        & \subset \left(\frac{\alpha^{(5)}_{2}v}{\Lambda}\right)h\left[\theta^{*}\theta^{\dagger}\overline{\nu}_{m}P_{R}\nu_{m} - \kappa^{\dagger}\theta^{\dagger}\overline{N}_{m}P_{R}\nu_{m} - \theta^{*}\kappa^{*}\overline{\nu}_{m}P_{R}N_{m} + \kappa^{\dagger}\kappa^{*}\overline{N}_{m}P_{R}N_{m}\right] 
	                                        \end{split}
\label{Eq:O52}
\end{equation}
In the above, we only have presented the three point vertices involving RHN fields and SM Higgs field. Apart from the operators involving Higgs field, one can write another operator $\mathcal{O}^{(5)}_{3}$ involving $U(1)_{Y}$ field-strength tensor in $d = 5$ $N_{R}$-EFT.
\begin{equation}
\begin{split}
\mathcal{O}^{\left(5\right)}_{3} & = \left(\frac{\alpha^{(5)}_{3}}{\Lambda}\right)\left(\overline{N_{R}}^{c}\sigma_{\mu\nu}N_{R}\right)B_{\mu\nu} = \left(\frac{\alpha^{(5)}_{3}}{\Lambda}\right)\left(\overline{N_{R}}^{c}\sigma_{\mu\nu}N_{R}\right)\left(\partial_{\mu}B_{\nu} - \partial_{\nu}B_{\mu}\right)   \\
					     & =  \left(\frac{\alpha^{(5)}_{3}}{\Lambda}\right)\left(\overline{N_{R}}^{c}\sigma_{\mu\nu}N_{R}\right)\left(- \sin\theta_{w}\{\partial_{\mu}Z_{\nu} - \partial_{\nu}Z_{\mu}\} + \cos\theta_{w}\{\partial_{\mu}A_{\nu} - \partial_{\nu}A_{\mu}\}\right)  \\
					     & = c_{w}A_{\mu\nu}\left(\frac{\alpha^{(5)}_{3}}{\Lambda}\right)\left[\theta^{*}\theta^{\dagger}\overline{\nu}_{m}\sigma_{\mu\nu}P_{R}\nu_{m} - \kappa^{\dagger}\theta^{\dagger}\overline{N}_{m}\sigma_{\mu\nu}P_{R}\nu_{m} - \theta^{*}\kappa^{*}\overline{\nu}_{m}\sigma_{\mu\nu}P_{R}N_{m} + \kappa^{\dagger}\kappa^{*}\overline{N}_{m}\sigma_{\mu\nu}P_{R}N_{m}\right]  \\
					     & - s_{w}Z_{\mu\nu}\left(\frac{\alpha^{(5)}_{3}}{\Lambda}\right)\left[\theta^{*}\theta^{\dagger}\overline{\nu}_{m}\sigma_{\mu\nu}P_{R}\nu_{m} - \kappa^{\dagger}\theta^{\dagger}\overline{N}_{m}\sigma_{\mu\nu}P_{R}\nu_{m} - \theta^{*}\kappa^{*}\overline{\nu}_{m}\sigma_{\mu\nu}P_{R}N_{m} + \kappa^{\dagger}\kappa^{*}\overline{N}_{m}\sigma_{\mu\nu}P_{R}N_{m}\right] 
\end{split}
\label{Eq:O53}
\end{equation}
Here we define $A_{\mu\nu} = \partial_{\mu}A_{\nu} - \partial_{\nu}A_{\mu}$ and $Z_{\mu\nu} = \partial_{\mu}Z_{\nu} - \partial_{\nu}Z_{\mu} $. In above equation, we present the explicit expansion of this operator. The $B_{\mu}$ field correspond to $U(1)_{Y}$ gauge group couples to RHN field through this operator. One can re-define this $B_{\mu}$ field as linear sum of neutral gauge boson $Z_{\mu}$ and the photon field $A_{\mu}$. As a consequence the RHN fields would couple to both the $Z$ and $\gamma$. The coupling between the $N$ fields and the photon is a direct consequence of the underlying $N_{R}$-EFT.

\section{: Evaluating Coupling Using Dimension Six Operators}
\label{App:dim6}
In this section we present the explicit expansion of various operators which leads to different couplings that involves heavy $N_{R}$ fields. The dimension six set up allows one to construct different class of operators which are $\psi^{2}H^{3}$, $\psi^{2}H^{2}D$, $\psi^{2}H^{2}X$ and four fermi respectively. The Higgs to neutrino couplings get substantial modification due to the operator $\mathcal{O}_{LNH}$ which comes under $\psi^{2}H^{3}$ class. The $\tilde{H}\left(H^{\dagger}H\right)$ part of the operator invokes sufficient change in the neutrino mass matrix once the Higgs field acquires \emph{vev} and break the electroweak symmetry. As an outcome of the EWSB, this operator also introduce an Yukawa interaction between Higgs and neutral leptons. Transforming the gauge basis to mass basis using the prescriptions which is mentioned in Appendix.~~\ref{App:Eigen}, one can deduce the precise form of interaction terms. The parameter $\tilde{\theta}$ is the redefined active sterile mixing. In our presentation of different interaction terms involving Higgs field, we stick to three point vertices. 
\begin{equation}
\begin{split}
\mathcal{O}_{LNH} &= \frac{\alpha_{LNH}}{\Lambda^{2}}\left(\overline{L}N_{R}\right)\tilde{H}\left(H^{\dagger}H\right) + h.c.  \\
			      & \subset \frac{3v^{2}\alpha_{LNH}}{2\sqrt{2}\Lambda^{2}}h\left[-U^{\dagger}\tilde{\theta}^{\dagger}\overline{\nu}_{m}P_{R}\nu_{m} - \tilde{\theta}^{\dagger}\tilde{\theta}^{\dagger}\overline{N}_{m}P_{R}\nu_{m} + U^{\dagger}\kappa^{*}\overline{\nu}_{m}P_{R}N_{m} + \tilde{\theta}^{\dagger}\kappa^{*}\overline{N}_{m}P_{R}N_{m}\right] + h.c. 
\end{split}
\label{Eq:OLNH}
\end{equation}
In Eq[\ref{Eq:OLNH}], we present the precise form of $\mathcal{C}^{(6)}_{h\overline{\nu}\nu}$, $\mathcal{C}^{(6)}_{h\overline{N}N}$ and $\mathcal{C}^{(6)}_{h\left(\overline{\nu}N + \overline{N}\nu\right)}$ couplings respectively. Under the class of $\psi^{2}H^{2}D$ there are two operators $\mathcal{O}_{HNe}$ and $\mathcal{O}_{HN}$. Here also we focus only on the three point vertices involving gauge boson fields. The operator $\mathcal{O}_{HNe}$ invokes coupling between the $W$ boson and the heavy neutrinos.  
\begin{align}
\mathcal{O}_{HNe} & = \frac{\alpha_{HNe}}{\Lambda^{2}}\left(\overline{N}_{R}\gamma^{\mu}e_{R}\right)\left(\tilde{H}^{\dagger}iD_{\mu}H\right) + h.c.  \subset \frac{\sqrt{2}\alpha_{HNe}M^{2}_{W}}{g\Lambda^{2}}W^{+}_{\mu}\left[ \overline{N}_{R}\gamma^{\mu}P_{R}e_{R}\right] + h.c. \\
			& = \frac{\sqrt{2}\alpha_{HNe}M^{2}_{W}}{g\Lambda^{2}}W^{+}_{\mu}\left[- \tilde{\theta}\overline{\nu}_{m}\gamma^{\mu}P_{R}e_{m} + \kappa^{T}\overline{N}_{m}\gamma^{\mu}P_{R}e_{m}\right] + h.c.  \nonumber
\end{align}
In addition to that, this operator also generate additional term to the charged gauge boson and SM leptons. The presence of right handed chiral matrix suggest non standard interaction which indicates a distinct deviation from the SM $SU(2)_{L}$ gauge properties. Using explicit definition of the $H^{\dagger}\overleftrightarrow{D}H$ \emph{i.e.} $H^{\dagger}D_{\mu}H - \left(D_{\mu}H\right)^{\dagger}H$ one can see the $N_{R}$ fields only couple the the neutral gauge bosons via three point interactions. 
\begin{equation}
\begin{split}
\mathcal{O}_{HN} & = \frac{\alpha_{HN}}{\Lambda^{2}}\left(\overline{N}_{R}\gamma^{\mu}N_{R}\right)\left(iH^{\dagger}\overleftrightarrow{D}_{\mu}H\right)  = -\left(\frac{M_{Z}v\alpha_{HN}}{\Lambda^{2}}\right)\left(\overline{N}_{R}\gamma^{\mu}N_{R}\right)Z_{\mu}\left[v + h\right]^{2}  \\
			& \subset -\left(\frac{M_{Z}\alpha_{HN}v}{\Lambda^{2}}\right)Z_{\mu}\left[\tilde{\theta}\tilde{\theta}^{\dagger}\overline{\nu}_{m}\gamma^{\mu}P_{R}\nu_{m} - \kappa^{T}\tilde{\theta}^{\dagger}\overline{N}_{m}\gamma^{\mu}P_{R}\nu_{m} - \tilde{\theta}\kappa^{*}\overline{\nu}_{m}\gamma^{\mu}P_{R}N_{m} + \kappa^{T}\kappa^{*}\overline{N}_{m}\gamma^{\mu}P_{R}N_{m}\right]             
\end{split}
\label{Eq:OHN}
\end{equation}
In Eq[\ref{Eq:OHN}], we present the explicit form of these couplings. One can notice that, this operator modifies the coupling between the $Z$ boson and active neutrino pair. This coupling substantially change the total decay width of the $Z$ boson which is precisely measured. However, apart from the dimension six operator co-efficient this coupling is also dependent on the square of the mixing angle. Hence the bounds coming from the EWPO can be evaded by suitably fixing the value of $\frac{\alpha_{HN}}{\Lambda^{2}}$ and $\tilde{\theta}$. Similar to dimension five one can write down two more operators that include stress-energy tensors. We begin with the operator which consider the $B_{\mu\nu}$ tensor associated with the abelian $U(1)_{Y}$ group. In Eq[\ref{Eq:ONB}], we present the explicit form of this operator.        
\begin{equation}
\begin{split}
\mathcal{O}_{NB} & = \frac{\alpha_{LNB}}{\Lambda^{2}}\overline{L}\sigma_{\mu\nu}N_{R}\tilde{H}B_{\mu\nu} + h.c.  \subset \frac{\alpha_{LNB}v}{\sqrt{2}\Lambda^{2}}\left[c_{w}A_{\mu\nu} - s_{w}Z_{\mu\nu}\right]\left(\overline{\nu}_{L}\sigma_{\mu\nu}N_{R}\right) + h.c.  \\
			& \subset \frac{\alpha_{LNB}v}{\sqrt{2}\Lambda^{2}}\left[c_{w}A_{\mu\nu} - s_{w}Z_{\mu\nu}\right][-U^{\dagger}\tilde{\theta}^{\dagger}\overline{\nu}_{m}\sigma_{\mu\nu}P_{R}\nu_{m} - \tilde{\theta}^{\dagger}\tilde{\theta}^{\dagger}\overline{N}_{m}\sigma_{\mu\nu}P_{R}\nu_{m}  \\
			&+ U^{\dagger}\kappa^{*}\overline{\nu}_{m}\sigma_{\mu\nu}P_{R}N_{m} + \tilde{\theta}^{\dagger}\kappa^{*}\overline{N}_{m}\sigma_{\mu\nu}P_{R}N_{m}] 
\end{split}
\label{Eq:ONB}
\end{equation}
From SM, one can rewrite the gauge field as the linear sum of photon and $Z$ boson fields. Using that one can express $B_{\mu\nu}$ as $\left[c_{w}A_{\mu\nu} - s_{w}Z_{\mu\nu}\right]$. Hence this operator generates different three point vertices between the leptons and the neutral gauge bosons both massless and massive. The dimension six also allows us to build operator that involve non-Abelian stress energy tensor. Depending on the generators the operator $\mathcal{O}_{NW}$ can be re-expressed as the sum of $ \mathcal{O}^{\pm}_{NW}$ and $\mathcal{O}^{0}_{NW}$.   
\begin{equation}
\begin{split}
\mathcal{O}_{NW} & = \mathcal{O}^{\pm}_{NW} + \mathcal{O}^{0}_{NW}  = \left(\frac{\alpha_{LNW}}{\Lambda^{2}}\right)\left(\overline{L}\sigma_{\mu\nu}N_{R}\right)\tau^{I}\tilde{H}W^{I}_{\mu\nu} + h.c. 
\end{split}
\label{Eq:ONWpm}
\end{equation}
\begin{align}
 \mathcal{O}^{\pm}_{NW} & = \frac{v\alpha_{LNW}}{\sqrt{2}\Lambda^{2}}W^{\pm}_{\mu\nu}\left[-\tilde{\theta}^{\dagger}\overline{e}_{m}\sigma_{\mu\nu}P_{R}\nu_{m} + \kappa^{*}\overline{e}_{m}\sigma_{\mu\nu}P_{R}N_{m}\right] + h.c. 
\end{align}
\begin{equation}
\begin{split}
\mathcal{O}^{0}_{NW} & = \frac{v\alpha_{LNW}}{\sqrt{2}\Lambda^{2}}\left[s_{w}A_{\mu\nu} + c_{w}Z_{\mu\nu}\right][- U^{\dagger}\tilde{\theta}^{\dagger}\overline{\nu}_{m}\sigma_{\mu\nu}P_{R}\nu_{m} - \tilde{\theta}^{\dagger}\tilde{\theta}^{\dagger}\overline{N}_{m}\sigma_{\mu\nu}P_{R}\nu_{m} \\
				& + U^{\dagger}\kappa^{*}\overline{\nu}_{m}\sigma_{\mu\nu}P_{R}N_{m} + \tilde{\theta}\kappa^{*}\overline{N}_{m}\sigma_{\mu\nu}P_{R}N_{m}]
\end{split}
\label{Eq:ONW0}
\end{equation}
In Eq[\ref{Eq:ONWpm}] and Eq[\ref{Eq:ONW0}] we present the analytic form of these operators in the mass basis. 

Now we turn our discussion to different four fermi operators. The operator $\mathcal{O}_{QuNL}$ comes under the class of $\left(\overline{L}R\right)\left(\overline{R}L\right)$.
\begin{align}
\mathcal{O}_{QuNL} & = \frac{\alpha_{LNQ}}{\Lambda^{2}}\left(\overline{Q}u_{R}\right)\left(\overline{N}_{R}L\right) + h.c. =  \frac{\alpha_{LNQ}}{\Lambda^{2}}\begin{bmatrix}
				\overline{u}_{L}u_{R} & \overline{d}_{L}u_{R}
				\end{bmatrix} \begin{bmatrix}
				\overline{N}_{R}\nu_{\ell} \\
				\overline{N}_{R}e_{L} \end{bmatrix} + h.c. \label{Eq:OQuNL}  \\
				& = \frac{\alpha_{LNQ}}{\Lambda^{2}}\left(\overline{u}_{m}P_{R}u_{m}\right)\left(-\tilde{\theta}U\overline{\nu}_{m}P_{L}\nu_{m} + \kappa^{T}U\overline{N}_{m}P_{L}\nu_{m} - \tilde{\theta}\tilde{\theta}\overline{\nu}_{m}P_{L}N_{m} + \kappa^{T}\tilde{\theta}\overline{N}_{m}P_{L}N_{m}\right)   \nonumber \\
				& +  \frac{\alpha_{LNQ}}{\Lambda^{2}}\left(\overline{d}_{m}P_{R}u_{m}\right)\left(-\tilde{\theta}\overline{\nu}_{m}P_{L}e_{m} + \kappa^{T}\overline{N}_{m}P_{L}e_{m}\right) + h.c. \nonumber
\end{align}
The operator $\mathcal{O}_{fN}$ that comes under $\left(\overline{R}R\right)\left(\overline{R}R\right)$ class signifies different scenario for different $f$. The label $f$ stands for various right handed gauge singlet SM quarks or charged leptons.  
\begin{align}
\left(\overline{R}R\right)\left(\overline{R}R\right) & \subset \mathcal{O}_{fN} = \frac{\alpha_{fN}}{\Lambda^{2}}\left(\overline{f}_{R}\gamma^{\mu}f_{R}\right)\left(\overline{N}_{R}\gamma_{\mu}N_{R}\right)   \\
					& =  \frac{\alpha_{fN}}{\Lambda^{2}}\left(\overline{f}_{m}\gamma^{\mu}P_{R}f_{m}\right)\left\{\tilde{\theta}\tilde{\theta}^{\dagger}\overline{\nu}_{m}\gamma^{\mu}P_{R}\nu_{m} - \kappa^{T}\tilde{\theta}\overline{N}_{m}\gamma^{\mu}P_{R}\nu_{m} - \tilde{\theta}\kappa^{*}\overline{\nu}_{m}\gamma^{\mu}P_{R}N_{m} + \kappa^{T}\kappa^{*}\overline{N}_{m}\gamma^{\mu}P_{R}N_{m} \right\}\nonumber
\end{align}
In addition to that one can construct another operator $\mathcal{O}_{duNe}$ which comes under $\left(\overline{R}R\right)\left(\overline{R}R\right)$ which takes the following explicit form. 
\begin{align}
\mathcal{O}_{duNe} & = \frac{\alpha_{duNe}}{\Lambda^{2}}\left(\overline{d}_{R}\gamma^{\mu}u_{R}\right)\left(\overline{N}_{R}\gamma_{\mu}e_{R}\right) + h.c. \nonumber \\
				& = \frac{\alpha_{duNe}}{\Lambda^{2}}\left(\overline{d}_{m}\gamma^{\mu}P_{R}u_{m}\right)\left[-\tilde{\theta}\overline{\nu}_{m}\gamma_{\mu}P_{R}e_{m} + \kappa^{T}\overline{N}_{m}\gamma^{\mu}P_{R}e_{m}\right] + h.c. 
\end{align}
The $\mathcal{O}_{FN}$ stands for two different operators which takes the generic form $\left(\overline{L}L\right)\left(\overline{R}R\right)$. The $F$ represents various SM-like quarks and lepton fields which transforms as a doublet under $SU(2)_{L}$ gauge group.
\begin{equation}   
\begin{split}
\left(\overline{L}L\right)\left(\overline{R}R\right) & \subset \mathcal{O}_{FN} = \frac{\alpha_{FN}}{\Lambda^{2}}\left(\overline{f}^{(1)}_{L}\gamma^{\mu}f^{(1)}_{L} + \overline{f}^{(2)}_{L}\gamma^{\mu}f^{(2)}_{L}\right)\left(\overline{N}_{R}\gamma_{\mu}N_{R}\right)  \\
									& = \frac{\alpha_{FN}}{\Lambda^{2}}\left(\overline{f}^{(1)}_{m}\gamma^{\mu}P_{L}f^{(1)}_{m} + \overline{f}^{(2)}_{m}\gamma^{\mu}P_{L}f^{(2)}_{m}\right)\{\tilde{\theta}\tilde{\theta}^{\dagger}\overline{\nu}_{m}\gamma^{\mu}P_{R}\nu_{m} - \kappa^{T}\tilde{\theta}\overline{N}_{m}\gamma^{\mu}P_{R}\nu_{m} \\
								&	- \tilde{\theta}\kappa^{*}\overline{\nu}_{m}\gamma^{\mu}P_{R}N_{m} + \kappa^{T}\kappa^{*}\overline{N}_{m}\gamma^{\mu}P_{R}N_{m} \} 
\end{split}
\label{Eq:OFN}
\end{equation}
The $f^{(1)}_{L}$ and $f^{(2)}_{L}$ denotes the up and down components of the left-handed doublet fermions. We conclude this section with the remaining three operators that comes under $\left(\overline{L}R\right)\left(\overline{L}R\right)$. The $\epsilon_{ij}$ in Eq[\ref{Eq:OLNLe}] to Eq[\ref{Eq:OLdQN}] is the $2\times2$ matrix which is equal to $i\sigma_{2}$, where $\sigma_{2}$ is the second Pauli matrix. The operator $\mathcal{O}_{LNLe}$ is constructed only the lepton fields whereas other two take both quarks and leptons into account.     
\begin{equation}
\begin{split}
\mathcal{O}_{LNLe} & = \left(\frac{\alpha_{LNLe}}{\Lambda^{2}}\right)\left[\overline{L}_{i}N_{R}\right]\epsilon_{ij}\left[\overline{L}_{j}e_{R}\right] + h.c.  \\
			        & = \left(\frac{\alpha_{LNLe}}{\Lambda^{2}}\right)\left[\overline{\nu}_{L}N_{R}\right]\left[\overline{e}_{L}e_{R}\right] - \left(\frac{\alpha_{LNLe}}{\Lambda^{2}}\right)\left[\overline{e}_{L}N_{R}\right]\left[\overline{\nu}_{L}e_{R}\right] + h.c.  \\
			        & =  \left(\frac{\alpha_{LNLe}}{\Lambda^{2}}\right)\left[-U^{\dagger}\tilde{\theta}^{\dagger}\overline{\nu}_{m}P_{R}\nu_{m} - \tilde{\theta}\tilde{\theta}\overline{N}_{m}P_{R}\nu_{m} + U^{\dagger}\kappa^{*}\overline{\nu}_{m}P_{R}N_{m} + \tilde{\theta}\kappa^{*}\overline{N}_{m}P_{R}N_{m}\right]\left(\overline{e}_{m}P_{R}e_{m}\right)  \\
			        & - \left(\frac{\alpha_{LNLe}}{\Lambda^{2}}\right)\left[- \tilde{\theta}^{\dagger}\overline{e}_{m}P_{R}\nu_{m} + \kappa^{*}\overline{e}_{m}P_{R}N_{m}\right]\left[U^{\dagger}\overline{\nu}_{m}P_{R}e_{m} + \tilde{\theta}\overline{N}_{m}P_{R}e_{m}\right] +  h.c. 
\end{split}
\label{Eq:OLNLe}
\end{equation}
\begin{equation}
\begin{split}
\mathcal{O}_{LNQd} & = \left(\frac{\alpha_{LNQd}}{\Lambda^{2}}\right)\left[\overline{L}_{i}N_{R}\right]\epsilon_{ij}\left[\overline{Q}_{j}d_{R}\right] + h.c. \\
				& = \left(\frac{\alpha_{LNQd}}{\Lambda^{2}}\right)\left[\overline{\nu}_{L}N_{R}\right]\left[\overline{d}_{L}d_{R}\right] - \left(\frac{\alpha_{LNQd}}{\Lambda^{2}}\right)\left[\overline{e}_{L}N_{R}\right]\left[\overline{u}_{L}d_{R}\right] + h.c. \\
				& =  \left(\frac{\alpha_{LNQd}}{\Lambda^{2}}\right)\left[-U^{\dagger}\tilde{\theta}\overline{\nu}_{m}P_{R}\nu_{m} - \tilde{\theta}^{\dagger}\tilde{\theta}^{\dagger}\overline{N}_{m}P_{R}\nu_{m} + U^{\dagger}\kappa^{*}\overline{\nu}_{m}P_{R}N_{m} + \tilde{\theta}\kappa^{*}\overline{N}_{m}P_{R}N_{m}\right]\left(\overline{d}_{m}P_{R}d_{m}\right) \\
				& -  \left(\frac{\alpha_{LNQd}}{\Lambda^{2}}\right)\left[\kappa^{*}\overline{e}_{m}P_{R}e_{m} - \tilde{\theta}\overline{e}_{m}P_{R}e_{m}\right]\left(\overline{u}_{m}P_{R}d_{m}\right) 
\end{split}
\label{Eq:OLNQd}
\end{equation}
\begin{equation}
\begin{split}
\mathcal{O}_{LdQN} & = \left(\frac{\alpha_{LdQN}}{\Lambda^{2}}\right)\left[\overline{L}_{i}d_{R}\right]\epsilon_{ij}\left[\overline{Q}_{j}N_{R}\right] + h.c.  \\
				& =  \left(\frac{\alpha_{LdQN}}{\Lambda^{2}}\right)\left[\overline{\nu}_{L}d_{R}\right]\left[\overline{d}_{L}N_{R}\right] - \left(\frac{\alpha_{LdQN}}{\Lambda^{2}}\right)\left[\overline{e}_{L}d_{R}\right]\left[\overline{u}_{L}N_{R}\right] + h.c. \\
				& = \left(\frac{\alpha_{LdQN}}{\Lambda^{2}}\right)\left[U^{\dagger}\overline{\nu}_{m}P_{R}d_{m} + \tilde{\theta}^{\dagger}\overline{N}_{m}P_{R}d_{m}\right]\left[- \tilde{\theta}\overline{d}_{m}P_{R}\nu_{m} + \kappa^{*}\overline{d}_{m}P_{R}N_{m}\right]\ \\
				& - \left(\frac{\alpha_{LdQN}}{\Lambda^{2}}\right)\left[\overline{e}_{m}P_{R}d_{m}\right]\left[- \tilde{\theta}^{\dagger}\overline{u}_{m}P_{R}\nu_{m} + \kappa^{*}\overline{u}_{m}P_{R}N_{m}\right] 
\end{split}
\label{Eq:OLdQN}
\end{equation}

\section{Integration Formula}
\label{App:Int}
The phase space integrals which are required to evaluate the three body decay modes of RHN field. The $\lambda$ here stands for the usual Kellen function of the form $\lambda(x, y, z) = x^{2} + y^{2} + z^{2} - 2xy -2 yz -2xz$.
\begin{align}
\mathcal{I}_{1}\left(x_{a}, x_{b}, x_{c}\right) & =  \int^{\left(1 - x_{c}\right)^{2}}_{\left(x_{a} + x_{b}\right)^{2}}\frac{dz}{z}\left(z - x^{2}_{a} - x^{2}_{b}\right)\left(1 + x^{2}_{c} - z\right)\lambda^{\frac{1}{2}}\left(1, z, x^{2}_{c}\right)\lambda^{\frac{1}{2}}\left(1, x^{2}_{a}, x^{2}_{b}\right)
\nonumber \\
\mathcal{I}_{2}\left(x_{a}, x_{b}, x_{c}\right) & = - \int^{\left(1 - x_{c}\right)^{2}}_{\left(x_{a} + x_{b}\right)^{2}}\frac{dz}{z}x_{c}\left(z - x^{2}_{a} - x^{2}_{b}\right)\lambda^{\frac{1}{2}}\left(1, z, x^{2}_{c}\right)\lambda^{\frac{1}{2}}\left(1, x^{2}_{a}, x^{2}_{b}\right) \nonumber \\
\mathcal{I}_{3}\left(x_{a}, x_{b}, x_{c}\right) & = \int^{\left(1 - x_{c}\right)^{2}}_{\left(x_{a} + x_{b}\right)^{2}}\frac{dz}{z^{2}}\{x_{b}x_{c}\left(z + x^{2}_{a} - x^{2}_{b}\right)\left(1 - x^{2}_{c} + z\right) - \frac{3}{2}x_{a}x_{a}\left(z - x^{2}_{a} + x^{2}_{b}\right)\left(1 - x^{2}_{c} - z\right)\}\nonumber \\
								&~~~~\times\lambda^{\frac{1}{2}}\left(1, z, x^{2}_{c}\right)\lambda^{\frac{1}{2}}\left(1, x^{2}_{a}, x^{2}_{b}\right) \nonumber \\
\mathcal{I}_{4}\left(x_{a}, x_{b}, x_{c}\right) & = \int^{\left(1 - x_{c}\right)^{2}}_{\left(x_{a} + x_{b}\right)^{2}}\frac{dz}{z^{2}}\{\frac{3}{2}x_{c}x_{a}\left(z + x^{2}_{b} - x^{2}_{a}\right)\left(1 - x^{2}_{c} + z\right) - 2x_{b}\left(z + x^{2}_{a} - x^{2}_{b}\right)\left(1 - x^{2}_{c} - z\right)\}  \nonumber \\
                            &~~~~\times\lambda^{\frac{1}{2}}\left(1, z, x^{2}_{c}\right)\lambda^{\frac{1}{2}}\left(1, x^{2}_{a}, x^{2}_{b}\right)      \nonumber \\
\mathcal{I}_{5}\left(x_{a}, x_{b}, x_{c}\right) & = - \int^{\left(1 - x_{c}\right)^{2}}_{\left(x_{a} + x_{b}\right)^{2}}\frac{dz}{z}x_{c}x_{a}x_{b}\lambda^{\frac{1}{2}}\left(1, z, x^{2}_{c}\right)\lambda^{\frac{1}{2}}\left(1, x^{2}_{a}, x^{2}_{b}\right) \nonumber \\
\mathcal{H}_{1}\left(x_{a}, x_{b}\right) & = \int^{4x^{2}_{b}}_{\left(1 - x_{a}\right)^{2}}\frac{dz}{z}x^{2}_{b}\left(1 + x^{2}_{a} - z\right)\lambda^{\frac{1}{2}}\left(1, x^{2}_{a}, z\right)\lambda^{\frac{1}{2}}\left(z, x^{2}_{b}, x^{2}_{b}\right) \nonumber \\
\mathcal{H}_{2}\left(x_{a}, x_{b}\right) & = \int^{4x^{2}_{b}}_{\left(1 - x_{a}\right)^{2}}\frac{dz}{z}x_{b}\left(1 - x^{2}_{a} - z\right)\lambda^{\frac{1}{2}}\left(1, x^{2}_{a}, z\right)\lambda^{\frac{1}{2}}\left(z, x^{2}_{b}, x^{2}_{b}\right) \nonumber \\
\mathcal{H}_{3}\left(x_{a}, x_{b}\right) & = - \int^{4x^{2}_{b}}_{\left(1 - x_{a}\right)^{2}}\frac{dz}{z}x_{a}x_{b}\left(1 - x^{2}_{a} + z\right)\lambda^{\frac{1}{2}}\left(1, x^{2}_{a}, z\right)\lambda^{\frac{1}{2}}\left(z, x^{2}_{b}, x^{2}_{b}\right) \nonumber \\
\mathcal{G}_{1}\left(x_{a}, x_{b}, x_{c}\right) & = \int^{\left(1 - x_{c}\right)^{2}}_{\left(x_{a} + x_{b}\right)^{2}}\frac{dz}{z^{2}}x_{a}\left(1 - x^{2}_{c} - z\right)\left(z - x^{2}_{a} + x^{2}_{b}\right)\lambda^{\frac{1}{2}}\left(1, x^{2}_{c}, z\right)\lambda^{\frac{1}{2}}\left(z, x^{2}_{a}, x^{2}_{b}\right)                       \nonumber \\
\mathcal{G}_{2}\left(x_{a}, x_{b}, x_{c}\right) & = - \int^{\left(1 - x_{c}\right)^{2}}_{\left(x_{a} + x_{b}\right)^{2}}\frac{dz}{z^{2}}x_{c}x_{a}\left(1 - x^{2}_{c} + z\right)\left(z - x^{2}_{a} + x^{2}_{b}\right)\lambda^{\frac{1}{2}}\left(1, x^{2}_{c}, z\right)\lambda^{\frac{1}{2}}\left(z, x^{2}_{a}, x^{2}_{b}\right)     \nonumber      \\
\mathcal{G}_{3}\left(x_{a}, x_{b}, x_{c}\right) & = - \int^{\left(1 - x_{c}\right)^{2}}_{\left(x_{a} + x_{b}\right)^{2}}\frac{dz}{z}x_{a}x_{b}\left(1 + x^{2}_{c} - z\right)\lambda^{\frac{1}{2}}\left(1, x^{2}_{c}, z\right)\lambda^{\frac{1}{2}}\left(z, x^{2}_{a}, x^{2}_{b}\right)      \nonumber
\end{align}

\section{Collider Signatures For RHN fields} 
In this section we will present different possible signatures involving RHN field for $pp$, $e^{+} e^{-}$ and $e^{-}p$ colliders. In Section.~\ref{Sec:multilep}, we discussed few of the multi-lepton final states where the RHN fields produce either via single or via pair production mechanism and then subsequently decay to three body pure leptonic channels. In addition to those channels one can also look for the RHN fields in other final states. In Table.~\ref{Tab:ppsig}, we present different production mode for the RHN fields for $pp$ colliders. The first row and first column of the Table.~\ref{Tab:ppsig} illustrate the flavour label of the final state leptons and quarks.     

\begin{table}[!ht]
\centering
\begin{tabular}{|c|c|c|c|c|c|c|}
\hline
\diagbox[width=11em]{Production}{Decay} & $N_{i} \to \ell_{j} \ell_{k} \nu_{k}$ & $N_{i} \to \ell_{k} \ell_{k} \nu_{k}$ & $N_{i} \to \nu_{j} \ell_{k} \ell_{k}$ & $N_{i} \to \ell_{j} u_{\alpha} d_{\beta}$ & $N_{i} \to \nu_{j} q_{\alpha} q_{\alpha}$ & $N_{i} \to \nu_{j} \nu_{k} \nu_{k}$  \\
\hline
$gg \to h \to \nu_{\delta} N_{i}$ & $2\ell +\slashed{E}_{T}$ & $2\ell +\slashed{E}_{T}$ & $2\ell +\slashed{E}_{T}$ & $\ell + 2q +\slashed{E}_{T}$ & $2q +\slashed{E}_{T}$ & $\slashed{E}_{T}$   \\
\hline
$q q' \to W \to \ell_{\delta}N_{i}$ & $3\ell +\slashed{E}_{T}$ & $3\ell +\slashed{E}_{T}$ & $3\ell +\slashed{E}_{T}$ & $2\ell + 2q$ & $\ell + 2q + \slashed{E}_{T}$ & $\ell + \slashed{E}_{T}$  \\
\hline
$q \bar{q} \to Z \to \nu_{\delta} N_{i}$ & $2\ell +\slashed{E}_{T}$ & $2\ell +\slashed{E}_{T}$ & $2\ell +\slashed{E}_{T}$ & $\ell + 2q +\slashed{E}_{T}$ & $2q +\slashed{E}_{T}$ & $\slashed{E}_{T}$   \\
\hline
$p p \to Z h$ & $4\ell + \slashed{E}_{T}$ & $4\ell + \slashed{E}_{T}$ & $4\ell + \slashed{E}_{T}$ & $3\ell + 2q$ & $2\ell + 2q$ & $2\ell + \slashed{E}_{T}$  \\
~~~~$\to \ell_{\rho} \ell_{\rho} \nu_{\delta} N_{i}$ & & & & $\slashed{E}_{T}$ & $\slashed{E}_{T}$ &   \\
\hline
$p p \to Z h$ &  $2\ell + 2b$  &  $2\ell + 2b$ &  $2\ell + 2b$ & $\ell + 2b$ & $2b + 2q$ & $2b + \slashed{E}_{T}$   \\
~~~~$\to \nu_{\delta}N_{i} 2b$ & $+ \slashed{E}_{T}$ & $+ \slashed{E}_{T}$ & $+ \slashed{E}_{T}$ & $2q + \slashed{E}_{T}$ & $\slashed{E}_{T}$ &  \\
\hline
$p p \to \nu_{\delta} N_{i}$ & $2\ell $ &  $2\ell $ &  $2\ell $ & $\ell + 2q $ & $2q $ & $\ell $  \\
via Four-Fermi & $+ \slashed{E}_{T}$ & $+ \slashed{E}_{T}$ & $+ \slashed{E}_{T}$ & $ + \slashed{E}_{T}$ & $+ \slashed{E}_{T}$ & $+ \slashed{E}_{T}$  \\
\hline
VBF $\to W q q' \to \ell_{\delta} N_{i} q q'$ & $3\ell + 2q + \slashed{E}_{T}$ & $3\ell + 2q + \slashed{E}_{T}$ & $3\ell + 2q + \slashed{E}_{T}$ & $2\ell + 4q$ & $\ell + 4q +\slashed{E}_{T}$ & $\ell + 2q + \slashed{E}_{T}$  \\
\hline  
VBF $\to Z q q' \to \nu_{\delta} N_{i} q q'$ & $2\ell + 2q + \slashed{E}_{T}$ & $2\ell + 2q + \slashed{E}_{T}$ & $2\ell + 2q + \slashed{E}_{T}$ & $\ell + 4q +\slashed{E}_{T}$ & $4q + \slashed{E}_{T}$ & $2q + \slashed{E}_{T}$  \\ 
\hline
VBF $\to h q q' \to \nu_{\delta} N_{i} q q'$  & $2\ell + 2q + \slashed{E}_{T}$ & $2\ell + 2q + \slashed{E}_{T}$ & $2\ell + 2q + \slashed{E}_{T}$ & $\ell + 4q +\slashed{E}_{T}$ & $4q + \slashed{E}_{T}$ & $2q + \slashed{E}_{T}$   \\
\hline
\end{tabular}
\caption{Different collider signatures for the single RHN production at $pp$ collider. Apart from the $\text{VBF} \to W q q' \to \ell_{\delta} N_{i} q q'$ and Drell-Yan $q q' \to W \to \ell_{\delta}N_{i}$ process all the other channels can be used for the pair production of the RHN fields.  For details regarding the final states correspond to $N$-pair production see the text.}
\label{Tab:ppsig}
\end{table}

Similarly in Table.~\ref{Tab:eesig} we tabulate different single production of $N$ fields at $e^{+} e^{-}$ colliders. The commonalities between Table.~\ref{Tab:ppsig} and Table.~\ref{Tab:eesig} is that we only highlight the single $N$ field production. Majority of the channels which are presented in these two Tables can serve as the pair production mode for the RHN fields. For example in case of $pp$ collider apart from  $\text{VBF} \to W q q' \to \ell_{\delta} N_{i} q q'$ and Drell-Yan $q q' \to W \to \ell_{\delta}N_{i}$ process all the other modes can potentially produce pair of RHN fields. On other hand in case of $e^{+} e^{-}$ collider apart from the t-channel $W$ boson mediated process all the other process can generate pair of $N$ fields. Using Table.~\ref{Tab:NNpairsig} one can evaluate the final state signature for these modes. In Table.~\ref{Tab:NNpairsig} we only illustrate the final states that can arise due to the subsequent decay of both $N_{i}$ and $N_{j}$. Moreover, at the production level other SM particle can arise in association with the RHN fields. Hence one need to put suitable particles in place of $X$ as mentioned in the first row of Table.~\ref{Tab:NNpairsig}.

\begin{table}[!ht]
\centering
\begin{tabular}{|c|c|c|c|c|c|c|}
\hline
\diagbox[width=11em]{Production}{Decay} & $N_{i} \to \ell_{j} \ell_{k} \nu_{k}$ & $N_{i} \to \ell_{k} \ell_{k} \nu_{k}$ & $N_{i} \to \nu_{j} \ell_{k} \ell_{k}$ & $N_{i} \to \ell_{j} u_{\alpha} d_{\beta}$ & $N_{i} \to \nu_{j} q_{\alpha} q_{\alpha}$ & $N_{i} \to \nu_{j} \nu_{k} \nu_{k}$  \\
\hline
$e^{+} e^{-} \to \nu_{\delta} N_{i}$ & $2\ell $ &  $2\ell $ &  $2\ell $ & $\ell + 2q $ & $2q $ & $\ell $  \\
via Four-Fermi & $+ \slashed{E}_{T}$ & $+ \slashed{E}_{T}$ & $+ \slashed{E}_{T}$ & $ + \slashed{E}_{T}$ & $+ \slashed{E}_{T}$ & $+ \slashed{E}_{T}$  \\
\hline
$e^{+} e^{-}  \to Z h$ & $4\ell + \slashed{E}_{T}$ & $4\ell + \slashed{E}_{T}$ & $4\ell + \slashed{E}_{T}$ & $3\ell + 2q$ & $2\ell + 2q$ & $2\ell + \slashed{E}_{T}$  \\
~~~~$\to \ell_{\rho} \ell_{\rho} \nu_{\delta} N_{i}$ & & & & $\slashed{E}_{T}$ & $\slashed{E}_{T}$ &   \\
\hline
$e^{+} e^{-}  \to Z h$ &  $2\ell + 2b$  &  $2\ell + 2b$ &  $2\ell + 2b$ & $\ell + 2b$ & $2b + 2q$ & $2b + \slashed{E}_{T}$   \\
~~~~$\to \nu_{\delta}N_{i} 2b$ & $+ \slashed{E}_{T}$ & $+ \slashed{E}_{T}$ & $+ \slashed{E}_{T}$ & $2q + \slashed{E}_{T}$ & $\slashed{E}_{T}$ &  \\
\hline
$e^{+} e^{-}  \to Z \to \nu_{\delta} N_{i}$ & $2\ell + \slashed{E}_{T}$ & $2\ell + \slashed{E}_{T}$ & $2\ell + \slashed{E}_{T}$ & $\ell + 2q + \slashed{E}_{T}$ & $2q + \slashed{E}_{T}$ & $\slashed{E}_{T}$  \\
\hline
VBF $\to Z \nu_{\rho} \nu_{\delta}$ & $2\ell +\slashed{E}_{T}$ & $2\ell +\slashed{E}_{T}$ & $2\ell +\slashed{E}_{T}$ & $\ell + 2q$ & $2q + \slashed{E}_{T}$ & $\slashed{E}_{T}$  \\
~~~~ $\to N_{i} \nu_{\sigma} \nu_{\rho} \nu_{\delta}$ & & & & $+ \slashed{E}_{T}$ & &  \\
\hline
VBF $\to W \ell_{\delta} \nu_{\rho}$ & $4\ell + \slashed{E}_{T}$ & $4\ell + \slashed{E}_{T}$ & $4\ell + \slashed{E}_{T}$ & $3\ell + 2q$ & $2\ell + 2q$ & $2\ell + \slashed{E}_{T}$ \\
~~~~$\to \ell_{\delta} \nu_{\rho} N_{i} \ell_{\sigma}$ & & &  & $\slashed{E}_{T}$ & $\slashed{E}_{T}$ &  \\
\hline
$e^{+} e^{-}  \to W \to \ell_{\delta} N_{i}$ & $3\ell + \slashed{E}_{T}$ & $3\ell + \slashed{E}_{T}$ & $3\ell + \slashed{E}_{T}$ & $2\ell + 2q$ & $\ell + 2q$ & $\ell + \slashed{E}_{T}$  \\
t channel & & & & + $\slashed{E}_{T}$ & + $\slashed{E}_{T}$ &  \\
\hline
\end{tabular}
\caption{Different collider signatures for the single RHN production at $e^{+} e^{-}$ collider. Apart from the t-channel $W$ boson mediated process all the other modes can serve as the pair production mode for RHN fields. For details regarding the final states correspond to $N$-pair production see the text.}
\label{Tab:eesig}
\end{table}
We like to elaborate this point with two suitable examples. For the process $gg \to h \to N_{i} N_{j}$ the possible final states are the states that are mentioned in this Table and one does not need to put any SM fields in place of $X$. On the other hand, in case of $pp \to Z h \to \ell_{\rho} \ell_{\rho} N_{i} N_{j}$, the $X$ would be replaced with $\ell_{\rho} \ell_{\rho}$ and to obtain the final states correspond to this process one needs to add $2\ell$ in each entires of Table.\ref{Tab:NNpairsig}.
\begin{table}[!ht]
\centering
\begin{tabular}{|c|c|c|c|c|c|c|}
\hline
 Possible Production Mode & \multicolumn{6}{ |c| }{$pp \to N_{i} N_{j} X$ and $e^{+} e^{-} \to N_{i} N_{j} X$ }  \\
 \hline
\diagbox[width=11em]{$N_{j}$ Decay}{$N_{i}$ Decay}      & $N_{i} \to \ell_{j} \ell_{k} \nu_{k}$ & $N_{i} \to \ell_{k} \ell_{k} \nu_{k}$ & $N_{i} \to \nu_{j} \ell_{k} \ell_{k}$ & $N_{i} \to \ell_{j} u_{\alpha} d_{\beta}$ & $N_{i} \to \nu_{j} q_{\alpha} q_{\alpha}$ & $N_{i} \to \nu_{j} \nu_{k} \nu_{k}$ \\
\hline
$N_{j} \to \ell_{a} \ell_{b} \nu_{b}$ & $4\ell + \slashed{E}_{T} $ & $4\ell + \slashed{E}_{T} $  & $4\ell + \slashed{E}_{T} $  & $3\ell + 2q + \slashed{E}_{T} $ & $2\ell + 2q +\slashed{E}_{T} $ & $2\ell + \slashed{E}_{T} $ \\
\hline
$N_{j} \to \ell_{b} \ell_{b} \nu_{b}$ & $4\ell + \slashed{E}_{T} $ & $4\ell + \slashed{E}_{T} $  & $4\ell + \slashed{E}_{T} $  & $3\ell + 2q + \slashed{E}_{T} $ & $2\ell + 2q +\slashed{E}_{T} $ & $2\ell + \slashed{E}_{T} $ \\
\hline
$N_{j} \to \ell_{a} \ell_{a} \nu_{b}$ & $4\ell + \slashed{E}_{T} $ & $4\ell + \slashed{E}_{T} $  & $4\ell + \slashed{E}_{T} $  & $3\ell + 2q + \slashed{E}_{T} $ & $2\ell + 2q +\slashed{E}_{T} $ & $2\ell + \slashed{E}_{T} $ \\
\hline
$N_{j} \to \ell_{a} u_{\rho} d_{\delta}$ & $3\ell + 2q + \slashed{E}_{T} $ & $3\ell + 2q + \slashed{E}_{T} $ & $3\ell + 2q + \slashed{E}_{T} $ & $2\ell + 4q + \slashed{E}_{T} $ & $\ell + 4q + \slashed{E}_{T} $ & $\ell + 2q + \slashed{E}_{T} $  \\
\hline
$N_{j} \to \nu_{a} q_{\rho} q_{\rho}$ & $2\ell + 2q + \slashed{E}_{T}$ & $2\ell + 2q + \slashed{E}_{T}$ & $2\ell + 2q + \slashed{E}_{T}$ & $\ell + 4q + \slashed{E}_{T}$ & $4q + \slashed{E}_{T}$ & $2q + \slashed{E}_{T}$ \\
\hline
$N_{j} \to \nu_{a} \nu_{b} \nu_{b}$ & $2\ell + \slashed{E}_{T}$ & $2\ell + \slashed{E}_{T}$ & $2\ell + \slashed{E}_{T}$ & $\ell + 2q + \slashed{E}_{T}$ &  $2q + \slashed{E}_{T}$ & $\slashed{E}_{T}$ \\
\hline 
\end{tabular}
\caption{Different possible final states that can arise due to the subsequent three body decays of RHN pair. The $X$ in the first row signifies the SM fields that can generate due the underlying production mechanism in association with RHN pair. For explicit evaluation one should replace $X$ with appropriated field content.}
\label{Tab:NNpairsig}
\end{table}
In Table.~\ref{Tab:epsig}, we present different RHN production modes relevant for $e^{-}p$ collider. Here we will restrict ourselves to single production of the $N$ fields. In principle one can generate more than one RHN fields in this collider. However, the processes related to these would involve more than one EFT vertices. Hence the cross section would receive higher order cut off scale suppression (see Fig.~\ref{fig:eph} for details).
\begin{table}[!ht]
\centering
\begin{tabular}{|c|c|c|c|c|c|c|}
\hline
\diagbox[width=11em]{Production}{Decay} & $N_{i} \to \ell_{j} \ell_{k} \nu_{k}$ & $N_{i} \to \ell_{k} \ell_{k} \nu_{k}$ & $N_{i} \to \nu_{j} \ell_{k} \ell_{k}$ & $N_{i} \to \ell_{j} u_{\alpha} d_{\beta}$ & $N_{i} \to \nu_{j} q_{\alpha} q_{\alpha}$ & $N_{i} \to \nu_{j} \nu_{k} \nu_{k}$  \\
\hline
$e p \to W \to q_{\delta} N_{i}$ & $2\ell + q + \slashed{E}_{T}$ & $2\ell + q + \slashed{E}_{T}$ & $2\ell + q + \slashed{E}_{T}$ & $\ell + 3q$ & $3q + \slashed{E}_{T}$ & $q + \slashed{E}_{T}$  \\
via t- channel & & & & & &    \\
\hline
$e p \to q_{\delta} N_{i}$ & $2\ell + q + \slashed{E}_{T}$ & $2\ell + q + \slashed{E}_{T}$ & $2\ell + q + \slashed{E}_{T}$ & $\ell + 3q$ & $3q + \slashed{E}_{T}$ & $q + \slashed{E}_{T}$  \\
via Four-Fermi & & & & & &  \\
\hline
VBF $\to h \nu_{\rho} q_{\delta} $ & $2\ell + q + \slashed{E}_{T}$ & $2\ell + q + \slashed{E}_{T}$ & $2\ell + q + \slashed{E}_{T}$ & $\ell + 3q + \slashed{E}_{T}$ & $3q + \slashed{E}_{T}$ & $q + \slashed{E}_{T}$  \\
$\to \nu_{\sigma} N_{i} \nu_{\rho} q_{\delta}$ & & & & & &  \\
\hline
$e p \to t b N_{i} \to N_{i} + 2b$ & $2\ell + 2b$  & $2\ell + 2b $  & $2\ell + 2b$  & $2\ell + 2b $  & $\ell + 2b $  & $\ell +2b $  \\
$+ q_{\delta} + q_{\sigma}$ &  $+ 2q + \slashed{E}_{T}$  & $ + 2q + \slashed{E}_{T}$ & $ + 2q + \slashed{E}_{T}$ & $ + 4q + \slashed{E}_{T}$ & $2q + \slashed{E}_{T}$ & $+ \slashed{E}_{T}$   \\
\hline 
$e p \to t b N_{i} \to N_{i} + 2b $ & $3\ell + 2b$  & $3\ell + 2b$ & $3\ell + 2b$ & $2\ell + 2q$ & $\ell + 2q$ & $\ell + 2b$  \\
$ + \ell_{\delta} \nu_{\rho} $ &   +$\slashed{E}_{T}$    & +$\slashed{E}_{T}$ & + $\slashed{E}_{T}$ & $2b + \slashed{E}_{T}$ & $2b + \slashed{E}_{T}$ & $\slashed{E}_{T}$    \\
\hline
\end{tabular}
\caption{Different collider signatures for the single RHN production at $e^{-}p$ collider.}
\label{Tab:epsig}
\end{table}


\bibliographystyle{utphys}
\bibliography{bibitem}  
\end{document}